\documentclass[prd,showkeys,floatfix,nofootinbib,
               preprint,12pt,
               tightenlines,fleqn]{revtex4-1} 

\usepackage{amsmath,amssymb,revsymb,graphicx,dcolumn}
\usepackage{leftidx}
\usepackage{slashed,physics}
\usepackage{marginnote}
\usepackage[normalem]{ulem}

\newcommand{\stkout}[1]{\ifmmode\text{\sout{\ensuremath{#1}}}\else\sout{#1}\fi}

\usepackage{xcolor}
\definecolor{mypink}{rgb}{0.9, 0.1, 0.3}
\usepackage{mathtools}
\usepackage{framed}

\newcommand{\beq}{\begin{equation}}
\newcommand{\eeq}{\end{equation}}
\newcommand{\beqa}{\begin{eqnarray}}
\newcommand{\eeqa}{\end{eqnarray}}
\newcommand{\bsubeqs}{\begin{subequations}}
\newcommand{\esubeqs}{\end{subequations}}

\begin{document}
\title[]
      {Field Quantisations in Schwarzschild Spacetime: \\ Theory versus Low-Energy Experiments}
\author{Viacheslav\;A.\;Emelyanov}
\email{viacheslav.emelianov@rwth-aachen.de}
\affiliation{Institute\;for\;Theoretical\;Physics\\
Karlsruhe\;Institute\;of\;Technology\\
76131\;Karlsruhe,\;Germany\\ \vspace{-4mm}}
\affiliation{Department\;of\;Mathematics\\
RWTH\;Aachen\;University\\
52056\;Aachen,\;Germany}

\begin{abstract}
\vspace*{2.5mm}\noindent
Non-relativistic quantum particles in the Earth's gravitational field are successfully described~by~the
Schr\"{o}dinger equation with Newton's gravitational potential.~Particularly,~quantum~mechanics is in
agreement with such experiments as free fall and quantum interference induced by gravity.~However,
quantum mechanics is a low-energy approximation to quantum field theory.\,The latter~is~successful
by the description of high-energy experiments.\,Gravity is embedded in quantum field~theory~through
the general-covariance principle.~This framework is known in the literature as quantum~field theory
in curved spacetime, where~the concept of a quantum particle
is,~though,~ambiguous.~In~this~article,
we~study in this framework~how a Hawking particle moves in the far-horizon region of
Schwarzschild spacetime~by computing its propagator.~We find this propagator
differs from that which follows~from the path-integral formalism -- the formalism which
adequately describes both free fall and~quantum interference induced by gravity.
\end{abstract}


\maketitle

\section{Introduction}
\label{sec:introduction}

Classical mechanics allows multiple equivalent formulations,~one of 
which is based~on the Hamilton-Jacobi equation~\cite{Landau&Lifshitz-1}:
\beqa\label{eq:hjeq}
\partial_t S(t,\boldsymbol{x}) + H(t,\boldsymbol{x},\boldsymbol{p}) &=& 0
\quad \textrm{with} \quad \boldsymbol{p} \;=\; \partial_{\boldsymbol{x}} S(t,\boldsymbol{x})\,,
\eeqa
where $S(t,\boldsymbol{x})$ is Hamilton's principle function and $H(t,\boldsymbol{x},\boldsymbol{p})$ 
is Hamiltonian.~The~elementary system we intend to deal with is
a non-relativistic particle of mass $M$ placed in the vicinity~of the Earth's surface.~Such a particle is 
accordingly described relative to the Earth's surface~by
\beqa
S(x,X) &=& 
M\Delta{t}\left(
\frac{\Delta\boldsymbol{x}^2}{2\Delta{t}^2} - \phi_\oplus -  \frac{g_\oplus(z+Z)}{2}
- \frac{(g_\oplus\Delta{t})^2}{24}\right),
\eeqa
where $\Delta{x} = x - X$,
$x = (t,\boldsymbol{x})$ and $X = (T,\boldsymbol{X})$ parametrises
initial time and position of the particle,
$\phi_\oplus$ and $g_\oplus$
stand for,~respectively,~Newton's gravitational potential and the free-fall acceleration
at the Earth's surface.

Classical mechanics is,~however,~incapable of describing the wave-like aspects of particles,
requiring $\hbar > 0$.~Quantum mechanics in this respect surpasses classical mechanics~in~the~non-relativistic regime.~The particle's dynamics is described by the particle's propagator~to~equal
the probability amplitude for $|X\rangle$ to evolve into $|x\rangle$.~The
path-integral~formalism~\cite{Dirac,Feynman}~gives
for a quantum particle of mass $M$ placed in the vicinity~of~the Earth's surface that
\beqa\label{eq:qm-propagator}
\langle x|X\rangle &=&
\left(\frac{M/\hbar}{2\pi i \Delta{t}}\right)^\frac{3}{2}
{\exp}\big(iS(x,X)/\hbar\big)\,.
\eeqa
Such a description is consistent not only with free fall,~but also with gravity-induced quantum
interference  \cite{Colella&Overhauser,Colella&Overhauser&Werner}.~The propagator
regarded as a non-normalisable wave function is a solution of Schr\"{o}dinger's equation
which includes the Hamilton-Jacobi equation~\eqref{eq:hjeq}~in~the limit $\hbar \to 0$
\cite{Schroedinger,Sakurai&Napolitano}.~Quantum mechanics coherently reduces to
classical mechanics
in the classical limit.\;\;\;\;

Quantum mechanics is,~however, a low-energy approximation to quantum field theory~\cite{Weinberg}.
The basic concept there is a quantum field,~e.g.,~$\hat{\Phi}(x)$.~This is a distribution-valued operator. 
The theory accordingly needs the consideration of a Hilbert-space representation~of~the~field
operator algebra.~This is constructed by adopting the concept of quantum vacuum which~is~a
state with no particles present.~The choice of quantum vacuum thus determines~the~concept
of~quantum particles~in theory.~This choice is,~in general,~non-unique~\cite{Haag}.

In theoretical particle physics,~there~exists~a unique state, $|\Omega\rangle$,
which is invariant under~the Minkowski-spacetime isometries -- the
Poincar\'{e} group.~Particles' states built on top~of~$| \Omega \rangle$~are
related this way with irreducible unitary representations of the Poincar\'{e}\,group,\,in~accord~with
the Wigner classification~\cite{Weinberg,Haag}.~This~implies
\bsubeqs
\beqa\label{eq:minkowski-quantum-field-rep}
\hat{\Phi}(x) &=& \hat{a}(x) + \hat{a}^\dagger(x)\,,
\eeqa
where $\hat{a}(x)$ annihilates $|\Omega\rangle$,~meaning $\hat{a}(x)|\Omega\rangle = 0$,~and
a one-particle state reads
\beqa
|a(x)\rangle &\equiv& \hat{a}^\dagger(x)|\Omega \rangle\,.
\eeqa
\esubeqs
The invariance under unitary trasformations implementing 
the Poincar\'{e} group gives\;\;\;\;
\beqa\label{eq:a-minkowski}
\langle a(x)|\boldsymbol{p}\rangle &=& e^{-ip{\cdot}x/\hbar}
\quad \textrm{with} \quad
p \;\equiv\; \big(\sqrt{(Mc)^2 + \boldsymbol{p}^2},\boldsymbol{p}\big)\,,
\eeqa
where $|\boldsymbol{p}\rangle$ is a state with a single particle of momentum
$\boldsymbol{p}$.~This is the defining characteristic~of $|a(x)\rangle$, which
allows to interpret it as an asymptotic state appearing in scattering processes~in collider
physics.~Such processes are described in quantum field theory by $S$-matrix elements
which equal probability 
amplitudes for some initial $N$-particle states to evolve into some~final 
$n$-particle states.~The
$S$-matrix in turn is linked to time-ordered $n+N$-point functions~via~the
Lehmann-Symanzik-Zimmermann reduction formula~\cite{Weinberg}.~Asymptotic states
have~accordingly
to satisfy~\eqref{eq:a-minkowski} in order to give a non-zero $S$-matrix
element~\cite{Schwartz}.~This assertion follows~from~the reduction formula and
the completeness relation for the one-particle states, namely
\beqa\label{eq:rcr}
{\int}{\frac{d^3\boldsymbol{p}}{(2\pi\hbar)^3}}\,
\frac{c}{2\omega_{\boldsymbol{p}}}\,
|\boldsymbol{p}\rangle \langle\boldsymbol{p}|  &=& \hat{1}
\quad \textrm{with} \quad
 \omega_{\boldsymbol{p}} \;\equiv\; (c/\hbar)\sqrt{(Mc)^2 + \boldsymbol{p}^2}\,,
\eeqa
ensuring with~\eqref{eq:a-minkowski} 
that the time-ordered probability amplitude~--~Feynman's propagator~--~has
a pole on mass shell in momentum space,~making a non-zero contribution to the
$S$-matrix.~In addition, the completeness relation and the
Poincar\'{e} invariance provide
\beqa
\langle a(x) | a(X) \rangle &\xrightarrow[c \,\to\, \infty]{}& \frac{e^{-iMc^2\Delta{t}/\hbar}}{2Mc/\hbar}\,
\langle x|X\rangle|_{G \,\to\, 0}\,.
\eeqa
Accordingly,~$|a(x)\rangle$ is a one-particle state which
reduces~to~the quantum-mechanics~state~$|x\rangle$ in the non-relativistic
limit in Minkowski spacetime (Newton's constant $G \to 0$).

The observable Universe is a non-Minkowski spacetime in general relativity.~It~is~taken~into
account in quantum mechanics by adding $M\phi(\boldsymbol{x})$ to Hamiltonian, where
$\phi(\boldsymbol{x})$ is the Newton gravitational potential.~This modification
is~enough to explain the gravity-induced~quantum
interference of thermal neutrons~\cite{Colella&Overhauser,Colella&Overhauser&Werner}.~Another
example is the appearance of bound states~of neutrons above a
reflecting plate held parallelly to the Earth's 
surface~\cite{Luschikov&Frank,Nesvizhevsky&etal-2002}.~These~effects were predicted by
quantum mechanics before their observations.\,Quantum~field~theory~adopts gravity
through the principle of general covariance.~This framework is known~in~the literature 
as quantum field theory in curved spacetime~\cite{DeWitt-1975,Birrell&Davies}.

The state $|x\rangle = |t,\boldsymbol{x} \rangle$ depends on $\phi(\boldsymbol{x})$,~as the gravitational
potential alters Hamiltonian. However,~$\phi(\boldsymbol{x}) \neq 0$ leaves the
concept of a quantum particle unaltered in quantum mechanics, because~$|\boldsymbol{x} \rangle$ is
oblivious to gravity.~In contrast,~the concept of a quantum particle is generally agreed to depend on
observer's notion of time in quantum field theory~in~curved spacetime.
This hypothesis leads to quantum particles'~ambiguity in theory.~In Schwarzschild spacetime,
approximately describing the Earth's gravitational field,~this implies the doubling of particle types:
\beqa\label{eq:schwarzschild-quantum-field-rep}
\hat{\Phi}(x) &=& \hat{n}(x) + \hat{n}^\dagger(x) + \hat{h}(x) + \hat{h}^\dagger(x)\,,
\eeqa
as there are two independent types of radial-mode solutions~\cite{DeWitt-1975}.~While particles
related~with $\hat{n}^\dagger(x)$ are,~to our knowledge,~nameless,~particles associated with
$\hat{h}^\dagger(x)$ are known as~Hawking 
particles~\cite{Hawking1974,Hawking1975,Hawking1976}.~A few states are defined this way in
Schwarzschild spacetime~\cite{Boulware,Unruh,Hartle&Hawking,Sciama&Candelas&Deutsch}.~In~the 
Earth's case,~Boulware's state $|B\rangle$~yields quantum
vacuum,~i.e.~$\hat{n}(x) |B\rangle = 0$~and~$\hat{h}(x) |B\rangle = 0$,
and,~correspondingly,~there are two types of one-particle states:
\bsubeqs
\beqa\label{eq:onps}
|n(x)\rangle &\equiv& \hat{n}^\dagger(x) |B\rangle\,,
\\[2mm]\label{eq:ohps}
|h(x)\rangle &\equiv& \hat{h}^\dagger(x) |B\rangle\,,
\eeqa
\esubeqs
with the propagators $\langle n(x)|n(X) \rangle$ and $\langle h(x)|h(X)\rangle$, respectively.

Schwarzschild spacetime turns into Minkowski spacetime at spatial infinity.~In~Minkowski
spacetime,~$|a(x)\rangle$ is a one-particle state which corresponds to an asymptotic
state in collider-physics experiments.~For $|n(x)\rangle$ and $|h(x)\rangle$ 
to be asymptotic states,~these~must satisfy
\bsubeqs\label{eq:n-and-h-infinity}
\beqa
\langle n(x)|\boldsymbol{p}\rangle|_\textrm{spatial infinity} &\propto& e^{-ip{\cdot}x/\hbar}\,,
\\[2mm]
\langle h(x)|\boldsymbol{p}\rangle|_\textrm{spatial infinity} &\propto& e^{-ip{\cdot}x/\hbar}\,.
\eeqa
\esubeqs
These would,~however,~imply that $\langle n(x)|h(X) \rangle \neq 0$ which contradicts
the circumstance that $[\hat{n}(x),\hat{h}^\dagger(X)] = 0$~\cite{Hawking1976}.~This
implies in turn that either $|n(x)\rangle$~or~$|h(x)\rangle$,~or~both~give~only~zero $S$-matrix
elements,\,and,\,correspondingly,\,one of $|n(x)\rangle$ and $|h(x)\rangle$,\,or~both~are
unobservable~in collider-physics experiments.

The spatial infinity of Schwarzschild spacetime is a mathematical abstract.~The~observable
Universe is a non-Schwarzschild spacetime.~The Schwarzschild geometry is an approximation
to local geometry of spherically symmetric compact objects in the observable Universe.~The
spatial infinity is,~in practice,~a region starting from the distance being much bigger~than~the
Schwarzschild radius $R_S$ of a compact object up to the distance at which gravitational field~of
rest matter is negligible.~In the Earth's case,~such a region is available~at~the Earth's surface
at~$R_\oplus$.~The Earth's gravitational field is characterised at $R_\oplus$~by
the free-fall acceleration $g_\oplus$.~It can,~however,~be locally eliminated by introducing
local Minkowski~coordinates,~such~as,~for example,~Riemann normal coordinates~\cite{Petrov}.\,It
is~due to Einstein's equivalence principle~built into general relativity~\cite{Einstein-1916}.~In
general,~the Minkowski-spacetime structure locally emerges~in Schwarzschild spacetime by
treating Riemann normal coordinates and neglecting space-time curvature.~Hence,~instead
of asymptotically Minkowski spacetime,~we have,~in practice,~local Minkowski frames at
$R_\oplus \leq R < \infty$ with the extent much less than $R\sqrt{R/R_S}$.

In theory, particle physics employs Minkowski spacetime as a basic space-time background in
which scattering processes take place.~In practice,~particle physics employs the
Minkowski-spacetime approximation in the vicinity of a given point in Schwarzschild spacetime,~which
approximates local geometry of the observable Universe in the vicinity of the Earth.~It~means
that particle physics deals with $|a(y)\rangle$ with space-time curvature neglected, where $y$
denotes Riemann normal coordimates.~In
Section~\ref{sec:qm},~we first show that
\beqa\label{eq:result-1}
\langle a(y)|a(Y) \rangle &\xrightarrow[c \,\to\, \infty]{}& \frac{e^{-iMc^2\Delta{t}/\hbar}}{2Mc/\hbar}\,
\langle x|X\rangle\,,
\eeqa
where Riemann normal coordinates $y$ are defined at $R_\oplus$, such that $y = y(x)$ and $Y = y(X)$,
and $y \to x$ in the $G \to 0$ limit.~This agrees not only with high-energy experiments~in~colliders in the
presence of the Earth's gravitational field, but also with free-fall experiments and the
gravity-induced quantum interference.~This
implies that the field quantisation in theoretical particle physics is adequate also for quantum effects
in gravity whenever space-time~curvature
plays no role.~This field quantisation allows thus to coherently~reduce~quantum~field
theory~to quantum mechanics if $c \to \infty$ and to classical mechanics if additionally $\hbar \to 0$.

In Section~\ref{sec:qft},~we second study the
probability amplitudes $\langle n(x)|n(X) \rangle$ and $\langle h(x)|h(X)\rangle$.
The study is based, first, on analytic and numerical computations of the amplitudes at large distances
($|\boldsymbol{X}| \gg R_S$ and 
$|\boldsymbol{X}| \gg |\Delta\boldsymbol{x}|$)~and,~second,~on~comparison of the 
amplitudes~with $\langle a(x)|a(X) \rangle$~afterwards.~The latter probability
amplitude adequately describes~the~high-~and
low-energy phenomena,~even in gravity.~It is due to
the first term~on the right-hand side~of\;\;\;
\beqa\label{eq:axaX}
\langle a(x)|a(X) \rangle &=& {\int}{\frac{d^3\boldsymbol{p}}{(2\pi\hbar)^3}}\,
\frac{c}{2\omega_{\boldsymbol{p}}}\,e^{-ip{\cdot}(y(x)-y(X))/\hbar}
+ \textrm{curvature-dependent corrections}\,.
\eeqa
However,~$\langle n(x)|n(X) \rangle$ and $\langle h(x)|h(X)\rangle$ cannot simultaneously
have such a representation,~as otherwise this would contradict the
canonical commutation relation~of~the quantum field and its canonical
conjugate and $\langle n(x)|h(X) \rangle = 0$.~This means that at least
one of the amplitudes 
is irreducible to $\langle x | X \rangle$ given in~\eqref{eq:qm-propagator}
in the non-relativistic limit.~We~find that it is $\langle h(x)|h(X)\rangle$ 
which differs from~\eqref{eq:qm-propagator} in the non-relativistic limit in the far-horizon
region,~while~$\langle n(x)|n(X)\rangle$ approaches
$\langle a(x)|a(X)\rangle$, at least if $G \to 0$.

In what follows, we use natural units $c = G = \hbar = 1$, unless otherwise stated.

\section{Quantum mechanics in Schwarzschild spacetime}
\label{sec:qm}

\subsection{Schr\"{o}dinger equation with Newton's gravitational potential}

The starting point of doing quantum field theory in curved spacetime is to pick~a~quantum
field and a space-time geometry in order to provide a concrete expansion of the quantum~field
over annihilation and creation operators.~This is necessary for a solid physical interpretation
of the quantum field as we directly observe only particles in real-world experiments.~The~basic
equation we wish to consider reads
\beqa\label{eq:scalar-field-equation}
\left(\Box_x + M^2 - \frac{1}{6}\,R(x)\right) \hat{\Phi}(x) &=& 0\,,
\eeqa
where $R(x)$ stands for the Ricci scalar,~and $\hat{\Phi}(x)$ is the quantum scalar
field satisfying~with~its conjugate $\hat{\Pi}(x)$ the standard canonical commutation
relation.~Furthermore,~we consider~the Schwarzschild geometry which~is~given~in
isotropic coordinates by the following line~element:
\beqa\label{eq:schwarzschild-spacetime}
ds^2 &=&
{\left(1+2\,\frac{\phi(\boldsymbol{x})}{c^2}\right)}(cdt)^2 
- {\left(1-2\,\frac{\psi(\boldsymbol{x})}{c^2}\right)}d\boldsymbol{x}^2\,,
\eeqa
where we keep track of the $c$-factors entering the line element, as we wish to consider~the~non-
relativistic approximation shortly, and the gravitational potentials read
\bsubeqs\label{eq:gravitational-potentials}
\beqa
\phi(\boldsymbol{x}) &=& 
+\frac{c^2}{2}\left(\frac{r-\frac{1}{4}R_S}{r + \frac{1}{4}R_S}\right)^2-\frac{c^2}{2}\,,
\\[2mm]
\psi(\boldsymbol{x}) &=& 
-\frac{c^2}{2}\left(\frac{r+\frac{1}{4}R_S}{r}\right)^4 + \frac{c^2}{2}\,,
\eeqa
\esubeqs
where $r \equiv \sqrt{\boldsymbol{x}{\cdot}\boldsymbol{x}}$
and $R_S$\,is the Schwarzschild radius.\,The reason of dealing with this\,geometry consists
in the fact that the Earth's gravitational field is approximately described~by~this~line element
with the Schwarzschild radius
\beqa\label{eq:earth-schwarzschild-radius}
R_{S,\,\oplus} &\approx& 8.87{\times}10^{-3}\,\text{m}\,,
\eeqa
provided the Earth's rotation can be neglected, which we assume in what follows.

To model a quantum particle in the framework of quantum field theory,~we must~define~an
operator which creates the particle's state by applying that on quantum vacuum, 
$|\Omega\rangle$,~namely
\bsubeqs
\beqa\label{eq:particle-state}
|\varphi\rangle &\equiv& \hat{a}^\dagger(\varphi)|\Omega\rangle\,,
\eeqa
where $\varphi(x)$ describes the particle's state.~For this purpose,~we
generalise~the~operator~used~for the definition of asymptotic states in
theoretical particle physics to curved spacetime~\cite{LSZ,Srednicki}:
\beqa\label{eq:creation-operator}
\hat{a}^\dagger(\varphi) &\equiv& - i{\int_\Sigma}d\Sigma(x)\, n^\mu(x)
\Big(\varphi(x)\nabla_\mu \hat{\Phi}^\dagger(x) - \hat{\Phi}^\dagger(x)\nabla_\mu\varphi(x)\Big)\,,
\eeqa
\esubeqs
where $n^{\mu}(x)$ is a future-directed unit four-vector orthogonal to a Cauchy surface $\Sigma$
and~$d\Sigma(x)$ is the volume element in $\Sigma$, and $\nabla_\mu$ is the covariant
derivative.~It follows from
\bsubeqs
\beqa\label{eq:relativistic-schroedinger-equation}
\left(\Box_x + M^2 - \frac{1}{6}\,R(x)\right)\varphi(x) &=& 0\,,
\eeqa
that~\eqref{eq:creation-operator} is independent of the choice of a Cauchy surface.~We
obtain from~the
normalisation condition $\langle\varphi|\varphi\rangle = 1$ that
\beqa\label{eq:normalisation-condition}
-i{\int_\Sigma}d\Sigma(x)\, n^\mu(x)
\Big(\varphi(x)\nabla_\mu\overline{\varphi(x)}-
\overline{\varphi(x)}\nabla_\mu\varphi(x)\Big) &=& 1\,,
\eeqa
\esubeqs
where bar stands for the complex conjugation.~The normalisation 
condition~$\langle\varphi|\varphi\rangle = 1$ implies
that $\varphi(x)$ must be a zero-rank tensor -- scalar, -- as the right-hand side
of~\eqref{eq:normalisation-condition} is independent
of coordinates utilised.~In other words,~the quantum state $|\varphi\rangle$ cannot appear or
disappear~by
going from one reference frame to another one.~This is in accord with the general principle~of
relativity~--~general covariance.~Einstein's equations need matter~be~modelled~covariantly,\,i.e.
through energy-momentum tensor.~Since matter itself is best described~by~quantum
theory, this description should~be~invariant~under~general~coordinate~transformations,
at least in the framework of classical gravity.~It is achieved in our case if $\varphi(x)$ is a zero-rank
tensor~\cite{Emelyanov-2020,Emelyanov-2021,Emelyanov-2022a,Emelyanov-2022b,Emelyanov&Robertz}.\;\;\;

Our purpose is to model a quantum particle of mass $M$ which freely moves~near~the~Earth's
surface.~Therefore,~taking into account that the Earth's radius
\beqa\label{eq:earth-radius}
R_\oplus &\approx& 6.37{\times}10^{6}\,\text{m}
\eeqa
is much bigger than its Schwarzschild radius~\eqref{eq:earth-schwarzschild-radius},~we
find~from~\eqref{eq:relativistic-schroedinger-equation} that
\beqa\label{eq:scalar-field-equation-approx}
\left({\left(\frac{1}{c^2}-\frac{2\phi(\boldsymbol{x})}{c^4}\right)}\partial_t^2
- {\left(1+\frac{2\psi(\boldsymbol{x})}{c^2}\right)}\partial_{\boldsymbol{x}}^2
- \partial_{\boldsymbol{x}}{\frac{\phi(\boldsymbol{x})-\psi(\boldsymbol{x})}{c^2}}{\cdot}
\partial_{\boldsymbol{x}} + (Mc)^2
\right)\varphi(x)
&\approx& 0\,,
\eeqa 
where we have considered only terms linearly depending on the gravitational potentials,~and 
have taken into consideration the~fact that $\psi(\boldsymbol{x}) \approx \phi(\boldsymbol{x})$ for
$r \gg R_S$.~Introducing
\beqa\label{eq:non-relativistic-wave-function}
\tilde{\varphi}(x) & \equiv & \sqrt{2Mc}\;e^{+iMc^2 t}\,\varphi(x)\,,
\eeqa
we obtain from~\eqref{eq:scalar-field-equation-approx} in the non-relativistic limit, i.e.
$c \rightarrow \infty$, that
\bsubeqs
\beqa\label{eq:schroedinger-equation}
i\partial_t\tilde{\varphi}(x) &\approx& {\left({-}\frac{1}{2M}\,
\partial_{\boldsymbol{x}}^2 + M\phi(\boldsymbol{x})\right)}\tilde{\varphi}(x)\,.
\eeqa
This is the Schr\"{o}dinger equation with Newton's gravitational potential~\cite{Sakurai&Napolitano}.~The
normalisation
condition~\eqref{eq:normalisation-condition} turns in this limit into
the standard quantum-mechanics one:
\beqa\label{eq:non-relativistic-normalisation-condition}
{\int_t}d^3\boldsymbol{x}\,
\overline{\tilde{\varphi}(t,\boldsymbol{x})}\,
\tilde{\varphi}(t,\boldsymbol{x})
 &\approx& 1\,.
\eeqa
\esubeqs
The approximation signs in~\eqref{eq:schroedinger-equation}
and~\eqref{eq:non-relativistic-normalisation-condition} turn into equality~signs in
quantum mechanics. 

\subsection{Stationary solution at the Earth's surface}
\label{eq:qm:mode-solution}

We wish to consider a non-inertial reference frame at the point
\beqa\label{eq-x-oplus}
\boldsymbol{X}_\oplus &\equiv& R_\oplus\,\boldsymbol{e}_z
\quad \textrm{with} \quad
\boldsymbol{e}_z \;\equiv\; (0,0,1)\,.
\eeqa
Considering $\boldsymbol{x} \to \boldsymbol{x} + \boldsymbol{X}_\oplus$
until the end of this section
and then assuming $|\boldsymbol{x}| \ll R_\oplus$,~we~have
from~\eqref{eq:gravitational-potentials} at
the leading order of approximation in $\boldsymbol{x}/R_\oplus$ that
\bsubeqs\label{eq:approximate-potentials}
\beqa\label{eq:approximate-newton-potential}
\phi(\boldsymbol{x}) &\approx& \phi_\oplus + g_\oplus z\,,
\\[2mm]\label{eq:approximate-second-potential}
\psi(\boldsymbol{x}) &\approx& \psi_\oplus + g_\oplus z\,,
\eeqa
\esubeqs
where $\phi_\oplus$ and $\psi_\oplus$ are the corresponding gravitational potentials at
$\boldsymbol{X}_\oplus$, and
\beqa
g_\oplus &\equiv& \frac{c^2R_{S,\,\oplus}}{2(R_\oplus)^2} \;\approx\; 9.81\,\text{m}/\text{s}^2
\eeqa
is the free-fall acceleration at the Earth's surface,~and $z$ accordingly measures
altitude above the Earth's surface.

The Schr\"{o}dinger equation \eqref{eq:schroedinger-equation} in the homogeneous gravitational
field~\eqref{eq:approximate-newton-potential} can be solved by applying the method of
separation of variables, also known as Fourier's method.~Namely,
following~\cite{Wheeler},~we obtain
\beqa\label{eq:qm-mode-solution}
\tilde{\varphi}_{\mathcal{E},\boldsymbol{\mathcal{K}}}(x) &\equiv&
\frac{{\exp}{\left({-}i{\left(\mathcal{E} + 
\scalebox{0.8}{$\dfrac{\boldsymbol{\mathcal{K}}^2}{2M}$}\right)}t
+i\boldsymbol{\mathcal{K}}{\cdot}\boldsymbol{x}_\perp\right)}}{\left(2g_\oplus/M\right)^\frac{1}{6}/\sqrt{2}}\,
{\text{Ai}}{\left((2M^2g_\oplus)^\frac{1}{3}
{\left(z - \frac{\mathcal{E} - M\phi_\oplus}{Mg_\oplus}\right)}\right)},
\eeqa
where $\text{Ai}(z)$ is Airy's
function,~$(\mathcal{E},\boldsymbol{\mathcal{K}}) = (\mathcal{E},\mathcal{K}_x,\mathcal{K}_y)$
are Fourier parameters,~and $\boldsymbol{x}_\perp = (x,y)$.
Making use of the completeness relation 
\beqa
{\int_{\mathbf{R}^3}} \frac{d\mathcal{E} d\boldsymbol{\mathcal{K}}}{(2\pi)^2}\,
|\mathcal{E},\boldsymbol{\mathcal{K}}\rangle \langle\mathcal{E},\boldsymbol{\mathcal{K}}|
&=& \hat{1}
\eeqa
and 
$\tilde{\varphi}_{\mathcal{E},\boldsymbol{\mathcal{K}}}(x) = \langle x|\mathcal{E},\boldsymbol{\mathcal{K}}\rangle$,~we obtain
\beqa\label{eq:qmp-1}
\langle x| X\rangle &=& {\int_{\mathbf{R}^3}} \frac{d\mathcal{E}d^2\boldsymbol{\mathcal{K}}}{(2\pi)^2} \,
\tilde{\varphi}_{\mathcal{E},\boldsymbol{\mathcal{K}}}(x)\,
\overline{\tilde{\varphi}_{\mathcal{E},\boldsymbol{\mathcal{K}}}(X)}
\;=\; \left(\frac{M}{2\pi i \Delta{t}}\right)^\frac{3}{2}
{\exp}\big(iS(x,X)\big)\,.
\eeqa
This
agrees with the propagator~\eqref{eq:qm-propagator} following from the 
path-integral formalism\,($\hbar \equiv 1$).

A few remarks are in order.~First,~the derivation based on the stationary modes $\tilde{\varphi}_{\mathcal{E},\boldsymbol{\mathcal{K}}}(x)$,~in particular,~needs $\mathcal{E} \in (-\infty,+\infty)$.~The Fourier parameter $\mathcal{E} $ cannot,~therefore,~be interpreted as the particle's
energy.~Second,~the modes~$\tilde{\varphi}_{\mathcal{E},\boldsymbol{\mathcal{K}}}(x)$ are implicitly defined
relative to $\boldsymbol{X}_\oplus$.~The gravitational
potential~$\phi(\boldsymbol{x})$ has the
form~\eqref{eq:approximate-newton-potential}~at~the~Earth's surface
only.~Thereby,~$\tilde{\varphi}_{\mathcal{E},\boldsymbol{\mathcal{K}}}(x)$
approximately solve the scalar-field equation if $|\boldsymbol{x}| \ll R_\oplus$.\,Although this
assumption rules~out the orthonormality condition requiring $z \in (-\infty,+\infty)$
in theory,~$\tilde{\varphi}_{\mathcal{E},\boldsymbol{\mathcal{K}}}(x)$ adequately model
local quantum dynamics in
practice if applied to determine the propagator $\langle x| X\rangle$.~Third,~the
Fourier method is also applied in quantum field theory in Schwarzschild spacetime,~wherein,
however, stationary modes are supposed to be positive-frequency ones, see Sec.~\ref{sec:qft}.

\subsection{Covariant solution at the Earth's surface}

We now intend to gain $\langle x| X\rangle$ by applying the principle of general covariance.~In
theoretical particle physics,~an asymptotic state models a particle moving at a constant velocity.~Such~a
state is characterised by a wave packet being a superposition of plane waves
\beqa\label{eq:qft-mode-solution-y}
\varphi_{Y,K}(y) &\equiv& {\exp}{\big({-}iK{\cdot}(y-Y)\big)}\,,
\eeqa
where we have used Riemann normal coordinates $y$, such those $y = y(x)$ and $Y= y(X)$,~and
the Fourier parameters $K$ can be interpreted as four-momentum.~It,~accordingly,~satisfies~the
on-mass-shell condition:
\beqa
K{\cdot}K &=& (Mc)^2\,.
\eeqa

The~plane
waves can be expressed through the general coordinates $x$ as follows:
\beqa\label{eq:qft-mode-solution-x}
\varphi_{X,K}(x) &=& {\exp}{\big({-}iK{\cdot}(y(x)-y(X))\big)}\,,
\eeqa
where $K$ is now understood as a four-vector belonging to the cotangent space at $X$.~Deriving
geodesic distance~\cite{DeWitt1965} and afterwards using the relation between
Riemann normal coordinates and geodesic distance~\cite{Ruse},~we find in the homogeneous
gravitational field~\eqref{eq:approximate-newton-potential}~that
Riemann normal coordinates depend on the isotropic coordinates as follows:
\bsubeqs
\beqa
y^0(x) &\approx& y^0(X) + c\Delta{t}
\left(1 + \frac{\phi_\oplus}{c^2} + \frac{g_\oplus z}{c^2} + \frac{(g_\oplus\Delta{t})^2}{6c^2}\right),
\\[1mm]
y^1(x) &\approx& y^1(X) + \Delta{x}\,,
\\[3mm]
y^2(x) &\approx& y^2(X) + \Delta{y}\,,
\\[1mm]
y^3(x) &\approx& y^3(X) + \Delta{z} + \frac{g_\oplus \Delta{t}^2}{2}\,,
\eeqa
\esubeqs
where we have omitted higher-order terms being negligible in the non-relativistic limit.~Note
that terms in $y^a(x)$ linearly depending on $g_\oplus$ follow from the same term in
geodesic distance. Using this result,~we obtain in the non-relativistic limit that
\beqa\label{eq:non-relativistic-wave-function-phase}
K{\cdot}(y(x)-y(X)) &\approx& Mc^2\Delta{t} + \frac{\boldsymbol{K}^2}{2M}\,\Delta{t}
- \boldsymbol{K}{\cdot}\Delta\boldsymbol{x}
\nonumber\\[1mm]
&&+\,M\Delta{t}\left(\phi_\oplus+g_\oplus z + \frac{(g_\oplus\Delta{t})^2}{6}\right)
- K_z\,\frac{g_\oplus \Delta{t}^2}{2}\,,
\eeqa
where we have also taken into account the on-mass-shell condition.~Redefining the
covariant modes according to the non-relativistic approximation, we obtain that
\beqa\label{eq:qft-mode-solution-x-non-rel}
\tilde{\varphi}_{X,K}(x) &\equiv& \lim_{c \,\to\, \infty}e^{+iMc^2\Delta{t}}\,\varphi_{X,K}(x)
\eeqa
exactly solves the 
Schr\"{o}dinger equation~\eqref{eq:schroedinger-equation}~with~the
Newtonian potential~\eqref{eq:approximate-newton-potential}.

Such a solution has been previously obtained in~\cite{Nauenberg} by relying,~however,~on the Einstein
principle,~stating the equivalence between uniform acceleration and homogenous~gravity~\cite{Einstein-1908}.
Specifically,~this principle being applied to the motion of a quantum particle~assumes~that~its
wave function is invariant up to a phase factor under the coordinate transformation changing
a reference frame with uniform gravity to another one with no uniform gravity.

In theoretical particle physics,~we have
from the relativistic completeness relation \eqref{eq:rcr} and Poincar\'{e} symmetry that
\beqa
\langle a(y)|a(Y)\rangle &=&
{\int}\frac{d^3\boldsymbol{K}}{(2\pi)^3}\,\frac{1}{2\omega_{\boldsymbol{K}}}\, e^{-iK{\cdot}(y-Y)}
\;\;\xrightarrow[c \,\to\, \infty]{}\;\;
\frac{e^{-iMc^2\Delta{t}}}{2Mc}
{\int}\frac{d^3\boldsymbol{K}}{(2\pi)^3}\,\tilde{\varphi}_{X,K}(x)\,,
\eeqa
where we directly find
\beqa\label{eq:qmp-2}
{\int}\frac{d^3\boldsymbol{K}}{(2\pi)^3}\,\tilde{\varphi}_{X,K}(x) &=& \langle x| X\rangle\,,
\eeqa
which thus proves~\eqref{eq:result-1}, assuming $\hbar \equiv 1$.

A few remarks are in order.~First,~in contrast to $\tilde{\varphi}_{\mathcal{E},\boldsymbol{\mathcal{K}}}(x)$,
the exact solution $\tilde{\varphi}_{X,K}(x)$~is~\emph{not} an eigenfunction of $\partial_t$,~by virtue of
the $\Delta{t}^2$- and $\Delta{t}^3$-terms in its phase.~This circumstance~is
irrelevant for its application in physics.~Still,~$\tilde{\varphi}_{X,K}(x)$
is given by $\exp(-iM\tau)$,~where~$\tau$~stands for proper time for
a geodesic connecting $X$ with $x$,~see~\cite{Stodolsky}.~Second, the plane
waves~\eqref{eq:qft-mode-solution-y}~cannot
in general~be exact solutions of the scalar-field equation~\eqref{eq:relativistic-schroedinger-equation}
in curved spacetime.~This~is due to space-time
curvature~\cite{Emelyanov-2020,Emelyanov-2021,Emelyanov&Robertz},
which we neglect in this section.

\subsection{Wave function}

The particle's dynamics is described by $\langle x|X\rangle$ in quantum
mechanics.~Namely,~assuming 
the particle has position $\boldsymbol{X}$ and
momentum $\boldsymbol{P}$ at $t = T$,~we consider a Gaussian~wave~packet of the form
\beqa
\tilde{\varphi}_{X,P}(x) &\equiv& N\left(\frac{D^2}{\pi}\right)^\frac{3}{2}
{\int}d^3\boldsymbol{Q}\,\langle t,\boldsymbol{x}|T,\boldsymbol{X} +\boldsymbol{Q}\rangle\,
{\exp}{\left(i\boldsymbol{P}{\cdot}\boldsymbol{Q}
-D^2\boldsymbol{Q}^2\right)}\,,
\eeqa
where $D$ denotes momentum variance and $N$ is a normalisation
factor~determined from~\eqref{eq:non-relativistic-normalisation-condition}.
This wave packet is, accordingly,~given via the convolution of the propagator with the initial
wave packet~\cite{Sakurai&Napolitano}.~Alternatively,~we also have
\beqa\label{eq:non-relativistic-wp}
\tilde{\varphi}_{X,P}(x) &=& N{\int}\frac{d^3\boldsymbol{K}}{(2\pi)^3}\,
\tilde{\varphi}_{X,K}(x)\,
{\exp}{\left({-}\frac{(\boldsymbol{P}-\boldsymbol{K})^2}{4D^2}\right)}\,,
\eeqa
which follows from a Gaussian superposition of the plane waves in the Riemann frame.

\subsection{Free fall}

Direct experiments with thermal
neutrons~\cite{McReynolds,Dabbs&Harvey&Paya&Horstmann,Koester} showed that neutrons
fall down~with~the free-fall acceleration.~The precision of
these~experiments is many orders~of~magnitude~smaller 
than those with macroscopic objects.~For~instance,~the MICROSCOPE 
experiment~found~no relative acceleration
at the $10^{-15}$ level between macroscopic masses of various compositions
\cite{Touboul&etal,Berge}.~Besides,~free-fall experiments in atom interferometry
achieved the $10^{-12}$~level~for~the
E\"{o}tv\"{o}s
parameter
for\,${}^{85}\text{Rb}$~and\,${}^{87}\text{Rb}$~\cite{Asenbaum&etal},\,see also~\cite{Schlippert&etal,Albers&etal},\,that measures relative acceleration between
the atoms.

According to the quantum-mechanics operator formalism,~the position operator applied~to the
wave function reduces to the coordinate multiplication.~Thus,~its expected~value~gives~the
centre-of-mass position of the wave function,~following from Born's statistical interpretation.
We therefore obtain for the quantum-particle position that
\beqa\label{eq:qm-center-of-mass}
\langle \hat{\boldsymbol{x}}(t)\rangle &\equiv& {\int_t}d^3\boldsymbol{x}\,
\overline{\tilde{\varphi}_{X,P}(x)}
\,\boldsymbol{x}\, \tilde{\varphi}_{X,P}(x)
\nonumber\\[1mm]\label{eq:free-fall-trajectory-position}
&=& \boldsymbol{X}+\frac{\boldsymbol{P} \Delta{t}}{M} - \frac{g_\oplus \Delta{t}^2}{2}\,
\boldsymbol{e}_z\,
\eeqa
is the quantum-particle trajectory in the Earth's uniform
gravitational field.~This result~is~in
agreement with classical physics,~where the
$g_\oplus$-dependent~term~in~\eqref{eq:free-fall-trajectory-position}~is due~to~the~last~term
on the right-hand side of~\eqref{eq:non-relativistic-wave-function-phase}.

The quantum-particle momentum is~given by the momentum-operator expectation value:
\beqa
\langle \hat{\boldsymbol{p}}(t)\rangle &\equiv& {\int_t}d^3\boldsymbol{x}\,
\overline{\tilde{\varphi}_{X,P}(x)}
\,(-i\nabla)\,\tilde{\varphi}_{X,P}(x)
\nonumber\\[1mm]\label{eq:free-fall-trajectory-momentum}
&=& \boldsymbol{P} - Mg_\oplus \Delta{t}\, \boldsymbol{e}_z\,.
\eeqa
This result also agrees with classical physics,~where~the $g_\oplus$-dependent
term~in~\eqref{eq:free-fall-trajectory-momentum}~comes~from gravitational time
dilation~\cite{Czarnecka&Czarnecki},~entering the right-hand side
of~\eqref{eq:non-relativistic-wave-function-phase}
in the form $Mg_\oplus z \Delta{t}$.\;\;\;\;\;\;

Finally, the quantum-particle energy~reads
\beqa
\langle \hat{H}(t)\rangle &\equiv& {\int_t}d^3\boldsymbol{x}\,
\overline{\tilde{\varphi}_{X,P}(x)}
\,(+i\partial_t)\, \tilde{\varphi}_{X,P}(x)
\nonumber\\[1mm]\label{eq:free-fall-trajectory-energy}
&=& \frac{\boldsymbol{P}^2}{2M} + \frac{3D^2}{2M} + M\big(\phi_\oplus + g_\oplus Z\big)\,,
\eeqa
which differs from the classical result by the term depending on the momentum variance.~This
can,~however,~be eliminated by shifting~the~rest energy from $Mc^2$ to $Mc^2 + 3D^2/2M$,~resulting
in the redefinition of the (Lagrangian) rest mass.\,Furthermore,~the~quantum-particle~energy~is
independent of time,~albeit~\eqref{eq:qft-mode-solution-x} is not an eigenfunction
of $t$.~All $\Delta{t}$-dependent~terms~entering
\eqref{eq:non-relativistic-wave-function-phase} cancel each other in
the integral~\eqref{eq:free-fall-trajectory-energy}.~The energy conservation requires
the $\Delta{t}^3$-term~in
\eqref{eq:non-relativistic-wave-function-phase}, known as the Kennard phase~\cite{Kennard1927,Kennard1929}.~It has been observed
in~\cite{Amit&etal}, see also~\cite{Marletto&Vedral,Dobkowski&etal,Asenbaum&Overstreet}.\;\;\;\;\;\;

\subsection{Quantum interference induced by gravity}

An interference experiment proposed in~\cite{Colella&Overhauser} was designed to
measure a relative phase~shift
gained by two beams of thermal neutrons during their propagation at different altitudes~with respect to~the
Earth's surface.~The phase shift accordingly reads
\beqa\label{eq:cow-phase-shift}
\delta(\Delta{z}) &=& - \frac{M^2g_\oplus}{2\pi \hbar^2}\,\Delta{z}\,\lambda\,L\,,
\eeqa
where $\lambda$ is a de-Broglie wavelength of the neutrons and $L$ is a horizontal distance
covered,~see \cite{Sakurai&Napolitano,Nauenberg}~for~further~details.~In 1975,\,this theoretical
result was empirically confirmed~\cite{Colella&Overhauser&Werner},~which is
known~in~the literature as the Colella-Overhauser-Werner experiment.

The wave-function phase depends on the altitude $z$ above the Earth's surface.~Specifically,
its phase changes by shifting the altitude from $z$ to $z + \Delta{z}$ as follows:
\beqa\label{eq:our-phase-shift}
\Delta\text{Arg}(\tilde{\varphi}_{X,P}) &=& - \frac{Mg_\oplus}{\hbar}\,\Delta{t}\,\Delta{z}\,.
\eeqa
This phase shift is due to the term in~\eqref{eq:non-relativistic-wave-function-phase}~describing
gravitational time dilation.~Taking~into account that 
$\Delta{t} = L/V$ with $V = P/M$ and $P = 2\pi\hbar/\lambda$,~we find that~\eqref{eq:our-phase-shift}
agrees with~\eqref{eq:cow-phase-shift}.\;\;

\subsection{Einstein's principle}

An interference experiment with thermal neutrons was performed in~1983~by~making~use of
an accelerated interferometer~\cite{Bonse&Wroblewski}.~The observed phase shift
relative to the accelerated~device appeared to be in agreement with Einstein's
principle~\cite{Einstein-1908},~namely quantum interference~cannot be used to
distinguish between homogeneous gravity and uniform acceleration~\cite{Nauenberg}.

We wish to consider a uniformly accelerated frame parameterised by coordinates~$x_R$,~that
moves relative to a local Minkowski frame.~Specifically,~we first consider
\beqa\label{eq:local-minkowski-line-element}
ds^2\big|_\text{Universe} &\approx& 
ds^2\big|_\text{Minkowski} \;=\;
\eta_{ab}\, dy^a dy^b\,,
\eeqa
where the approximation sign is to underline that the observable Universe is not globally~flat,
however, herein we neglect the local universe curvature~\cite{Petrov}.~A
uniformly~accelerated~frame~in Minkowski spacetime is known in the literature as Rindler
spacetime\,\cite{Birrell&Davies},\,such~that~$y = y(x_R)$ are given by
\bsubeqs\label{eq:rindler-coordinates}
\beqa
y^0(x_R) &=& \left(\frac{c^2}{a} + z_R\right){\sinh}{\left(\frac{at_R}{c}\right)}\,,
\\[2mm]
y^1(x_R) &=& x_R\,,
\\[3.5mm]
y^2(x_R) &=& y_R\,,
\\[2mm]
y^3(x_R) &=& \left(\frac{c^2}{a} + z_R\right){\cosh}{\left(\frac{at_R}{c}\right)} - \frac{c^2}{a}\,,
\eeqa
\esubeqs
where $a$ stands for proper acceleration.~We then have in terms of $x_R$ that
\beqa\label{eq:rindler-spacetime}
ds^2\big|_\text{Universe} &\approx&
{\left(1+2\,\frac{\phi(\boldsymbol{x}_R)}{c^2}\right)}(cdt_R)^2
- {\left(1-2\,\frac{\psi(\boldsymbol{x}_R)}{c^2}\right)}(d\boldsymbol{x}_R)^2\,,
\eeqa
where the ``gravitational potentials" read
\bsubeqs\label{eq:gravitational-potentials-r}
\beqa
\phi(\boldsymbol{x}_R) &=& +\frac{c^2}{2}\bigg(1 + \frac{az_R}{c^2}\bigg)^2-\frac{c^2}{2}\,,
\\[2mm]
\psi(\boldsymbol{x}_R) &=& 0\,.
\eeqa
\esubeqs
In contrast to Schwarzschild spacetime,~there is no analog of gravitational length contraction
in Rindler spacetime.~This circumstance may have impact on the applicability of the Einstein
principle if one takes into account the finiteness of the speed of light, $c < \infty$~\cite{Emelyanov-2022b}.

As Rindler spacetime is a patch of Minkowski spacetime,~the plane-wave
modes~\eqref{eq:qft-mode-solution-y}~must be considered, following from
the principle of general covariance,~cf.~\cite{Fulling,Davies,Unruh}.~We~thus~have
\beqa
K{\cdot}(y(x_R)-y(X_R)) &\approx& Mc^2t_R + \frac{\boldsymbol{K}^2}{2M}\,t_R
- \boldsymbol{K}{\cdot}\boldsymbol{x}_R
\nonumber\\[1mm]
&&
+\,Mt_R\left(a z_R + \frac{(at_R)^2}{6}\right)
- K_z\,\frac{at_R^2}{2}\,,
\eeqa
in the non-relativistic limit,~where $X_R = 0$ has been set for the sake of simplicity.~Comparing
this result with~\eqref{eq:non-relativistic-wave-function-phase},~we find that
there is no physical difference between uniform acceleration and homogeneous
gravity if $c \to \infty$,~in accord with the Bonse-Wroblewski experiment~\cite{Bonse&Wroblewski}.\;\;\;\;

\section{Quantum field theory in Schwarzschild spacetime}
\label{sec:qft}

\subsection{Field quantisations in Schwarzschild spacetime}
\label{sec:field-quantisations}

The application of quantum field theory to the description of particle physics assumes~that
the observable Universe can be approximated by Minkowski spacetime.~It is justifiable~within
the framework of general relativity due to Einstein's equivalence principle.~This principle~assures
the existence of local Minkowski frames.~In such local frames,~special
relativity replaces general relativity,~where the Poincar\'{e} group plays a pivotal
role.~Even~in~Schwarzschild~spacetime,
approximating the Earth's gravitational field, one may consider
\bsubeqs\label{eq:mfq}
\beqa\label{eq:field-quantisation-particle-physics}
\hat{\Phi}(x) &=& \hat{a}(x) + \hat{a}^\dagger(x)\,,
\eeqa
where, in terms of Riemann normal coordinates from the previous section, we have
\beqa\label{eq:field-quantisation-particle-physics-annihilation-operator}
\hat{a}(x) &\approx& \frac{1}{(2\pi)^3}{\int}
\frac{d^3\boldsymbol{p}}{2\omega_{\boldsymbol{p}}}\,e^{-ip{\cdot}y(x)}\,\hat{a}_{\boldsymbol{p}}
\quad \textrm{with} \quad
\hat{a}_{\boldsymbol{p}}^\dagger|\Omega\rangle \;=\; |\boldsymbol{p}\rangle\,.
\eeqa
\esubeqs
We use the approximation sign
in~\eqref{eq:field-quantisation-particle-physics-annihilation-operator} to underline that
plane waves are non-exact solutions of the scalar-field equation.~This approximation
appears to be adequate for the description~of 
high-energy phenomena in particle accelerators and the
low-energy phenomena,~including~the effects of the Earth's gravitational field, as shown in the
previous section.

The field quantisation in Minkowski spacetime is global in theory, while local in practice.
It is generally agreed,~nevertheless,~that field quantisation in a curved spacetime needs global
isometry group of the spacetime to introduce the concept of a quantum particle.~Moreover,~it
is also generally agreed that the concept of observer's time plays a pivotal role by choosing~a
Hilbert-space representation of a field operator
algebra~\cite{Birrell&Davies,Gibbons&Hawking}.~This in turn implies~that
the concept of a quantum particle is ambiguous due to relativity of time in general
relativity.~This section is to explore consequences of this hypothesis and their coherence with
well-established laws in particle physics.

Schwarzschild
spacetime has four Killing vectors -- its isometry group is four dimensional.
Its global isometry corresponds to invariance under time translations and~spatial~rotations~to
leave the coordinate-frame origin unaltered.~The Schwarzschild-time translation is generated
by the Killing vector $\partial_t$.~It is commonly assumed that positive- and negative-frequency
modes defined with respect to $\partial_t$ are relevant to particle physics.~Namely,~this
assumption gives\;\;\;\;
\bsubeqs\label{eq:hfq}
\beqa\label{eq:global-field-quantisation-scalar-field}
\hat{\Phi}(x) &=& \hat{n}(x) + \hat{n}^\dagger(x) + \hat{h}(x) + \hat{h}^\dagger(x)\,,
\eeqa
where there are then twice as many independent mode functions in Schwarzschild
spacetime than in local Minkowski frames~\cite{DeWitt-1975,Hawking1975,Hawking1976,Wald,Castineiras&Silva&Matsas,Egorov&Smolyakov&Volobuev}:
\beqa\label{eq:n-operator}
\hat{n}(x) &\approx& {\sum\limits_{l\,=\,0}^\infty}\;{\sum\limits_{m\,=\,-l}^{m\,=\,+l}}\;
{\int\limits_{M}^{\infty}}
\frac{d\omega}{\sqrt{2\omega}}
{\left(\frac{\omega}{\sqrt{\omega^2-M^2}}\right)^\frac{1}{2}}N_{\omega lm}(x)\,\hat{n}_{\omega lm}\,,
\\[1mm]\label{eq:h-operator}
\hat{h}(x) &\approx& {\sum\limits_{l\,=\,0}^\infty}\;{\sum\limits_{m\,=\,-l}^{m\,=\,+l}}\;
{\int\limits_0^\infty}
\frac{d\omega}{\sqrt{2\omega}}\,H_{\omega lm}(x)\,\hat{h}_{\omega lm}\,,
\eeqa
\esubeqs
where,~from the time-translation and spherical (relative to $|\boldsymbol{x}| = 0$) symmetries,~one assumes
\bsubeqs
\beqa\label{eq:nwml}
N_{\omega lm}(x) &=& \frac{i}{\sqrt{2\pi}}\,e^{-i\omega t}\,
\frac{r\,\mathcal{N}_{\omega l}(r)}{\left(r + \frac{1}{4}R_S\right)^2}\,Y_{lm}(\Omega_{\boldsymbol{x}})\,,
\\[1mm]\label{eq:hwml}
H_{\omega lm}(x) &=& \frac{i}{\sqrt{2\pi}}\,e^{-i\omega t}\,
\frac{r\,\mathcal{H}_{\omega l}(r)}{\left(r + \frac{1}{4}R_S\right)^2}\,Y_{lm}(\Omega_{\boldsymbol{x}})\,,
\eeqa
\esubeqs
where $Y_{lm}(\Omega_{\boldsymbol{x}})$ denotes the spherical harmonics,
$\mathcal{N}_{\omega l}(r)$ and $\mathcal{H}_{\omega l}(r)$~are radial modes.

The approximation signs in \eqref{eq:field-quantisation-particle-physics-annihilation-operator} 
and \eqref{eq:n-operator} with \eqref{eq:h-operator} serve to emphasise
the observable Universe~is neither Minkowski nor Schwarzschild spacetime.

\subsection{Radial-mode solutions}
\label{sec:rmss}

In terms of the confluent Heun ($Hc$) function (as defined in Maple), we have
\bsubeqs
\beqa
\mathcal{N}_{\omega l}(r_s) &=&
B_l(\omega,M)\,
\frac{Hc\Big(\scalebox{0.9}{$2\alpha,-2\beta,0,\gamma,\delta,1 - \dfrac{r_s}{R_S}$}\Big)}{e^{-ikr_s}\,
\scalebox{0.9}{$\dfrac{R_S}{r_s}$}{\Big(\scalebox{0.9}{$\dfrac{r_s}{R_S}-1$}\Big)^{\hspace{-1mm}+\beta}}}\,,
\\[1mm]
\mathcal{H}_{\omega l}(r_s) &=& 
\frac{Hc\Big(\scalebox{0.9}{$2\alpha,+2\beta,0,\gamma,\delta,1 - \dfrac{r_s}{R_S}$}\Big)}{e^{-ikr_s}\,
\scalebox{0.9}{$\dfrac{R_S}{r_s}$}{\Big(\scalebox{0.9}{$\dfrac{r_s}{R_S}-1$}\Big)^{\hspace{-1mm}-\beta}}}
+ \mathcal{A}_l(\omega,M)\,
\frac{Hc\Big(\scalebox{0.9}{$2\alpha,-2\beta,0,\gamma,\delta,1 - \dfrac{r_s}{R_S}$}\Big)}{e^{-ikr_s}\,
\scalebox{0.9}{$\dfrac{R_S}{r_s}$}{\Big(\scalebox{0.9}{$\dfrac{r_s}{R_S}-1$}\Big)^{\hspace{-1mm}+\beta}}}
\,,
\eeqa
\esubeqs
where $r_s$ is the Schwarzschild radial coordinate
($r = \sqrt{\boldsymbol{x}{\cdot}\boldsymbol{x}}$ is the isotropic one),~and
\bsubeqs
\beqa
\alpha &\equiv& - ikR_S\,,
\\[1mm]
\beta &\equiv& +i\omega R_S\,,
\\[1mm]
\gamma &\equiv& \alpha^2 + \beta^2\,,
\\[1mm]
\delta &\equiv& - \alpha^2 - \beta^2 - l(l+1)\,,
\eeqa
\esubeqs
and where, by definition,
\beqa
k &\equiv& 
\left\{
\begin{array}{ll}
\sqrt{\omega^2 - M^2}\,, & \omega \,\in\, [M,\,\infty)\,, \\[2mm]
i\sqrt{M^2-\omega^2 }\,, & \omega \,\in\, [0,\,M)\,.
\end{array}
\right.
\eeqa

For the computation of the reflection coefficient $\mathcal{A}_l(\omega,M)$ and
\begin{figure}
\begin{center}
\includegraphics[scale=0.55]{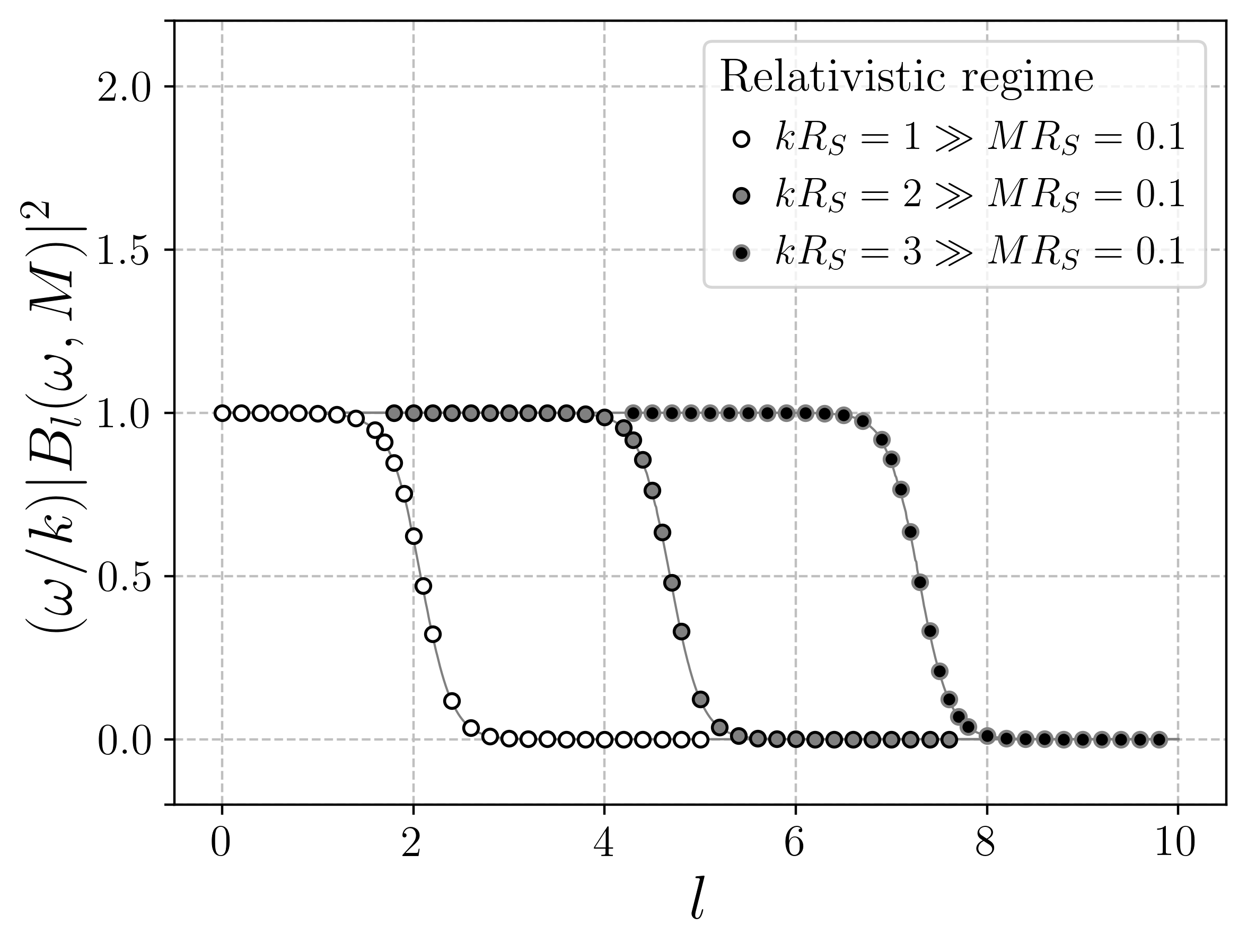}\hfill
\includegraphics[scale=0.55]{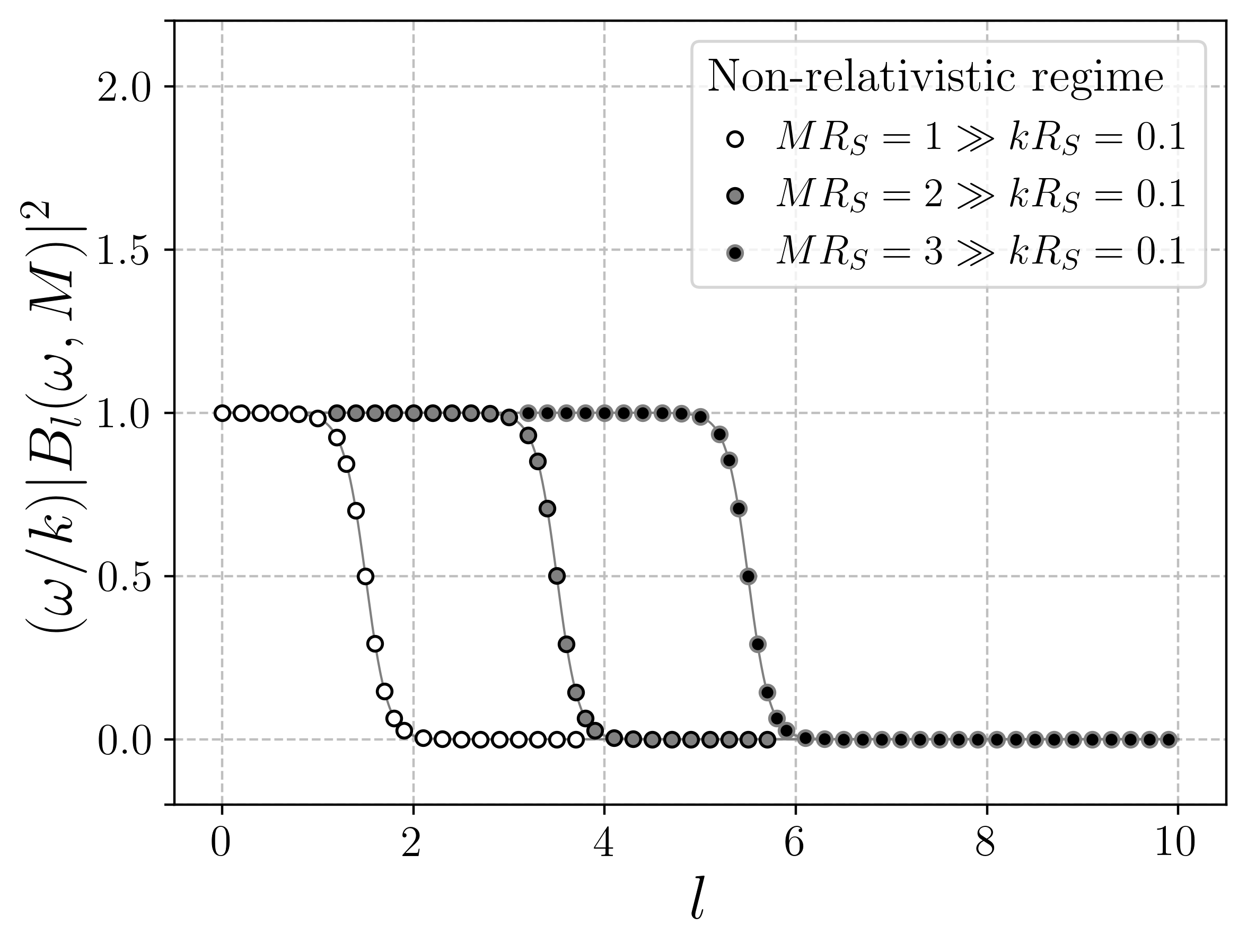}
\end{center}
\caption{Numerical computations of $(\omega/k)\,|B_{l}(\omega,M)|^2$ as a function of $l$
for various values of $\omega R_S$ and $MR_S$.~We compute
$\mathcal{A}_l(\omega,M)$ by evaluating the confluent Heun
functions entering $\mathcal{H}_{\omega l}(r_s)$
and their derivatives with respect to $r_s$ at $r_s = 10^3 R_S$.~We next confirm that the values of
$\mathcal{A}_l(\omega,M)$~are essentially independent of $r_s$ by computing some of those also
at $r_s = 10^4 R_S$.~Our numerical results shown here and below are however
based on the computations of $\mathcal{A}_l(\omega,M)$ at $r_s = 10^3 R_S$,~because
it requires less computational resources.~Left:~We first assume the relativistic
regime,~i.e.~$k \gg M$.~Our
numerics
agree~with the DeWitt approximation,
see (144) in~\cite{DeWitt-1975}.~Right:~We next consider~the~non-relativistic regime,~i.e.
$M \gg k$.~In
this case,~$|B_{l}(\omega,M)|^2$ approximately equals
$(k/\omega)\,\theta(l_\text{max} - l)$~with $l_\text{max} \equiv (3\sqrt{2}/2)\,MR_S$.
}\label{fig:1}
\end{figure}
the transmission coefficient $B_l(\omega,M)$, we need asymptotic forms
of the radial modes at spatial infinity~\cite{DeWitt-1975}.~These read\;\;\;\;\;
\bsubeqs
\beqa\label{eq:asymptotic-radial-modes-n}
\mathcal{N}_{\omega l}(r_s) &\xrightarrow[r_s \,\rightarrow\, \infty]{}& i^l
{\Big(e^{-ikr_s - i\eta \ln 2kr_s}
+ A_{l}(\omega,M)\,e^{+ikr_s + i\eta \ln 2kr_s}\Big)}\,,
\\[2mm]\label{eq:asymptotic-radial-modes-h}
\mathcal{H}_{\omega l}(r_s) &\xrightarrow[r_s \,\rightarrow\, \infty]{}&
i^l{\mathcal{B}_{l}}{(\omega,M)}\,e^{+ikr_s + i\eta \ln 2kr_s}\,,
\eeqa
where
\beqa
\eta &\equiv& \frac{R_S}{2k}\,(\omega^2 + k^2)\,.
\eeqa
\esubeqs
Using the constancy of the Wronskian for the radial-field equation
for various combinations~of $\mathcal{N}_{\omega l}(r)$, $\mathcal{H}_{\omega l}(r)$
and their complex conjugate,~we find that
the reflection coefficients~$A_l(\omega,M)$ and $\mathcal{A}_l(\omega,M)$
and the transmission coefficients~$B_l(\omega,M)$ and $\mathcal{B}_l(\omega,M)$
are related for $\omega \geq M$ as follows~\cite{DeWitt-1975}:
\bsubeqs
\beqa
|B_{l}(\omega,M)|^2 &=& (k/\omega)\big(1 - |A_{l}(\omega,M)|^2\big)\,,
\\[2mm]\label{eq:b-curlyb}
B_{l}(\omega, M) &=& (k/\omega)\,\mathcal{B}_{l}(\omega,M)\,,
\\[2mm]
|A_{l}(\omega,M)| &=& |\mathcal{A}_{l}(\omega,M)|\,.
\eeqa
\esubeqs
Our numerics for $|B_l(\omega,M)|^2$ are shown in Fig.~\ref{fig:1}.~We however
leave aside the computation~of
$|B_l(\omega,M)|^2$ for $\omega \in [0,M)$.~This is because we are
interested here in~local~physics at distances being much bigger than
$R_S$.~At such distances,~$\mathcal{H}_{\omega l}(r)$ exponentially decays for
$\omega \in [0,M)$.~In this case, the particle interpretation for the modes
$H_{\omega l m}(x)$ is no longer justifiable.

Apart from numerical computations of both $\langle n(x)|n(X)\rangle$ and $\langle h(x)|h(X)\rangle$,~we
also wish~to obtain as much analytic information about these propagators as possible.~This is particularly
needed for the analysis of numerical results.~For this reason,~we wish to consider approximate
solutions for $\mathcal{N}_{\omega l}(r)$ and $\mathcal{H}_{\omega l}(r)$,~which solve the radial-field
equation up to the leading order in $R_S$.~These approximate
radial-mode solutions are given by
\bsubeqs\label{eq:app-radial-mode-solutions}
\beqa\label{eq:app-radial-mode-solutions-n}
\mathcal{N}_{\omega l}^{(1)}(r) &=&
\frac{i^l\,\Gamma(l+1-i\eta)}{(-1)^{l+1} (2)_{2l}}\,
\frac{e^{+\pi\eta/2}\,M_{-i\eta,l+\frac{1}{2}}(+2ikr)}
{r{\left(r - \frac{1}{4}R_S\right)^{1/2}}
{\left(r + \frac{1}{4}R_S\right)^{-3/2}}}\,,
\\[2mm]\label{eq:app-radial-mode-solutions-h}
\mathcal{H}_{\omega l}^{(1)}(r) &=& 
i^l \mathcal{B}_{l}(\omega,M)\,
\frac{e^{-\pi\eta/2}\,W_{+i\eta,l+\frac{1}{2}}(-2ikr)}
{r{\left(r - \frac{1}{4}R_S\right)^{1/2}}
{\left(r + \frac{1}{4}R_S\right)^{-3/2}}}\,,
\eeqa
\esubeqs
where $M_{\kappa,\mu}(z)$ and $W_{\kappa,\mu}(z)$ are Whittaker's
functions,~$\Gamma(z)$ and $(z)_\nu$ are the gamma~function and~the~Pochhammer symbol.~The
index ``$(1)$"~means we deal with the first-order~solutions~in
the~Schwarzschild
radius~$R_S$.~However,~\eqref{eq:app-radial-mode-solutions-n} fulfils the
asymptotic condition~\eqref{eq:asymptotic-radial-modes-n} with 
\beqa\label{eq:a1}
A_l^{(1)}(\omega,M) &=& (-1)^{l+1}\frac{(l)_{1-i\eta}}{(l)_{1+i\eta}}\,.
\eeqa
This differs from $A_l(\omega,M)$.~It~is~because
the radial-mode solution~\eqref{eq:app-radial-mode-solutions-n}~is oblivious~to~higher-
order terms in $R_S$,~which also contribute~to the reflection
coefficient,~see~\cite{Landau&Lifshitz} for~more~details
about scattering theory.~In the non-relativistic limit, though, we have
\beqa\label{eq:n-to-n1}
\mathcal{N}_{\omega l}(r) &\xrightarrow[c \,\to\, \infty]{}& \mathcal{N}_{\omega l}^{(1)}(r)\,,
\eeqa
whereas $\eta \to (Mc)^2R_S/2k$.~This follows from the radial-field equation by taking into account
the asymptotic condition~\eqref{eq:asymptotic-radial-modes-n}
and that $R_S \propto 1/c^2 \to 0$ in the limit $c \to \infty$. 

\subsection{Hawking particles}
\label{sec:hps}

There are three quantum states which are usually considered in 
Schwarzschild~spacetime
\cite{Sciama&Candelas&Deutsch,Boulware,Unruh,Hartle&Hawking}.~The state choice
depends on the type of a spherically
symmetric compact object.~In case of a black hole formed via gravitational
collapse -- only future horizon is present,~--~one considers the Unruh state
$|U\rangle$ defined by
\bsubeqs\label{eq:u-state}
\beqa
\langle U|(\hat{n}_{\omega lm})^\dagger\hat{n}_{\omega'l'm'}|U\rangle &=& 0\,,
\\[1mm]
\langle U|(\hat{h}_{\omega lm})^\dagger\hat{h}_{\omega'l'm'}|U\rangle &=& 
\frac{1}{e^{4\pi\omega R_S}-1}\,\delta(\omega-\omega')\,\delta_{ll'}\,\delta_{mm'}\,.
\eeqa
\esubeqs
This is a thermal state with respect to
$\hat{h}(x)$~and~$\hat{h}^\dagger(x)$ operators,~characterised by Hawking's
temperature $1/4\pi R_S$~\cite{Hawking1974,Hawking1975} -- the Unruh state is therefore a many-Hawking-particle state.~It
is necessary~for energy-momentum tensor to be non-singular on the future horizon.

\begin{figure}
\begin{center}
\includegraphics[scale=0.515]{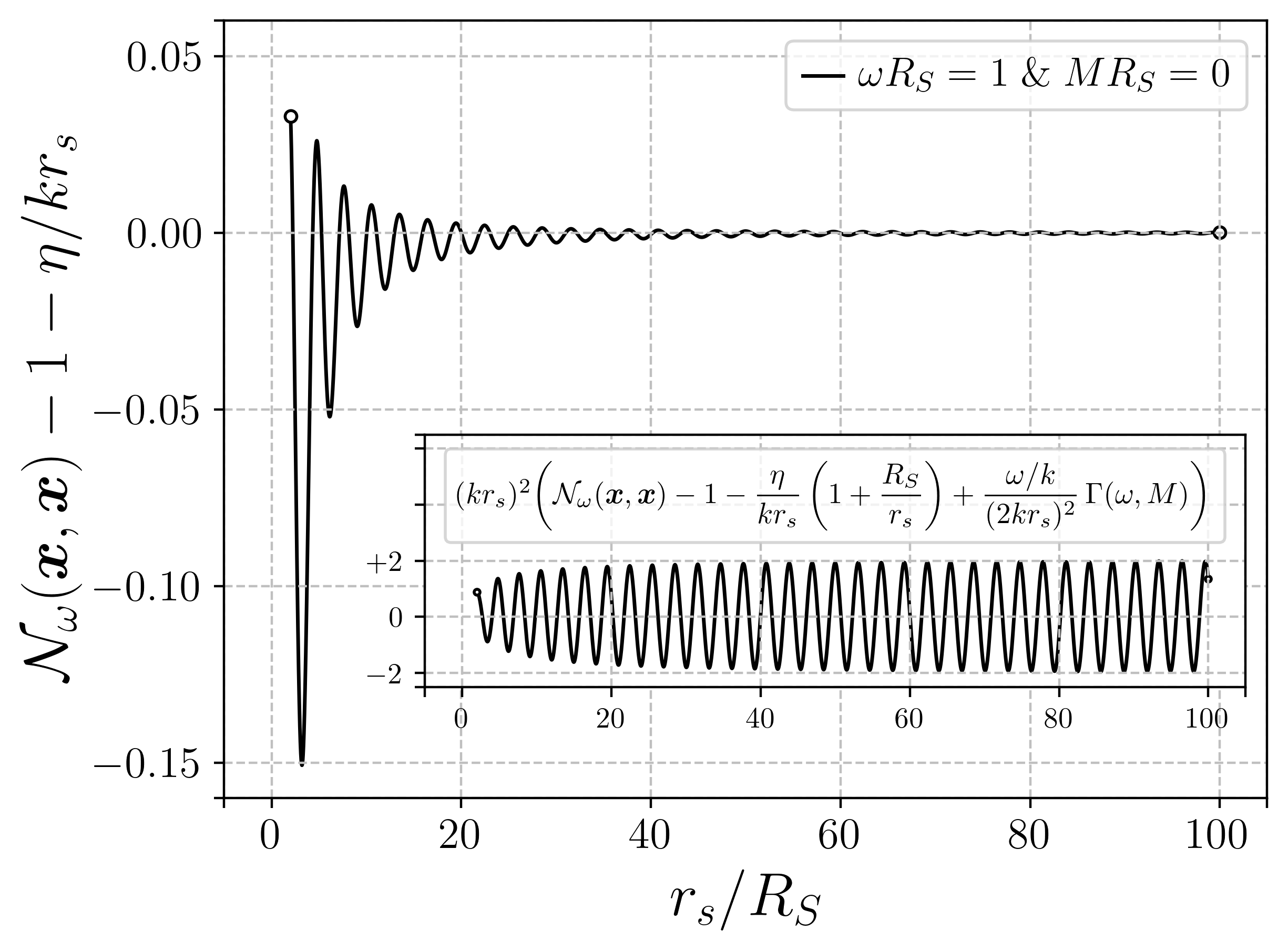}\hfill
\includegraphics[scale=0.515]{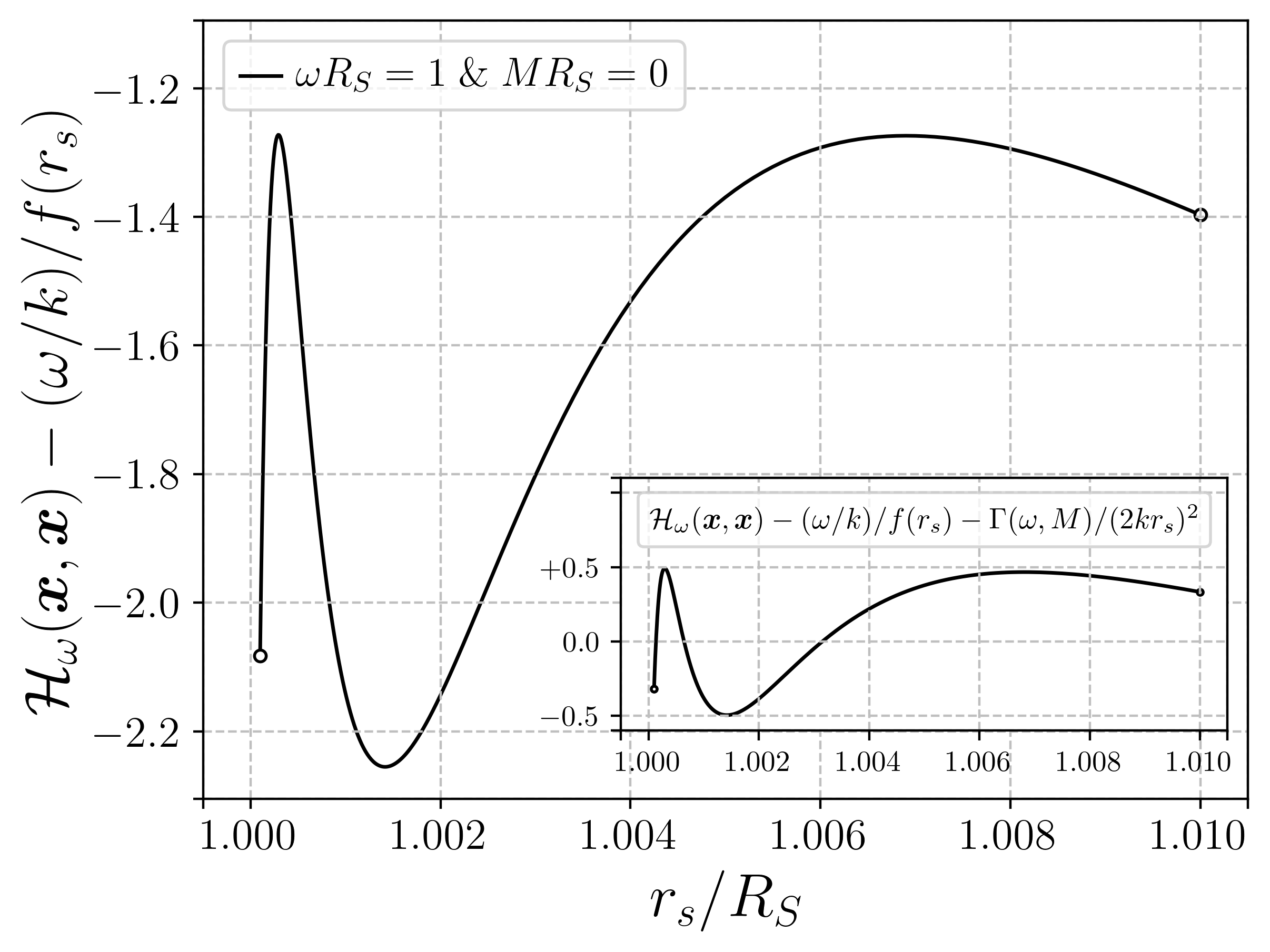}\vspace{2mm}
\includegraphics[scale=0.515]{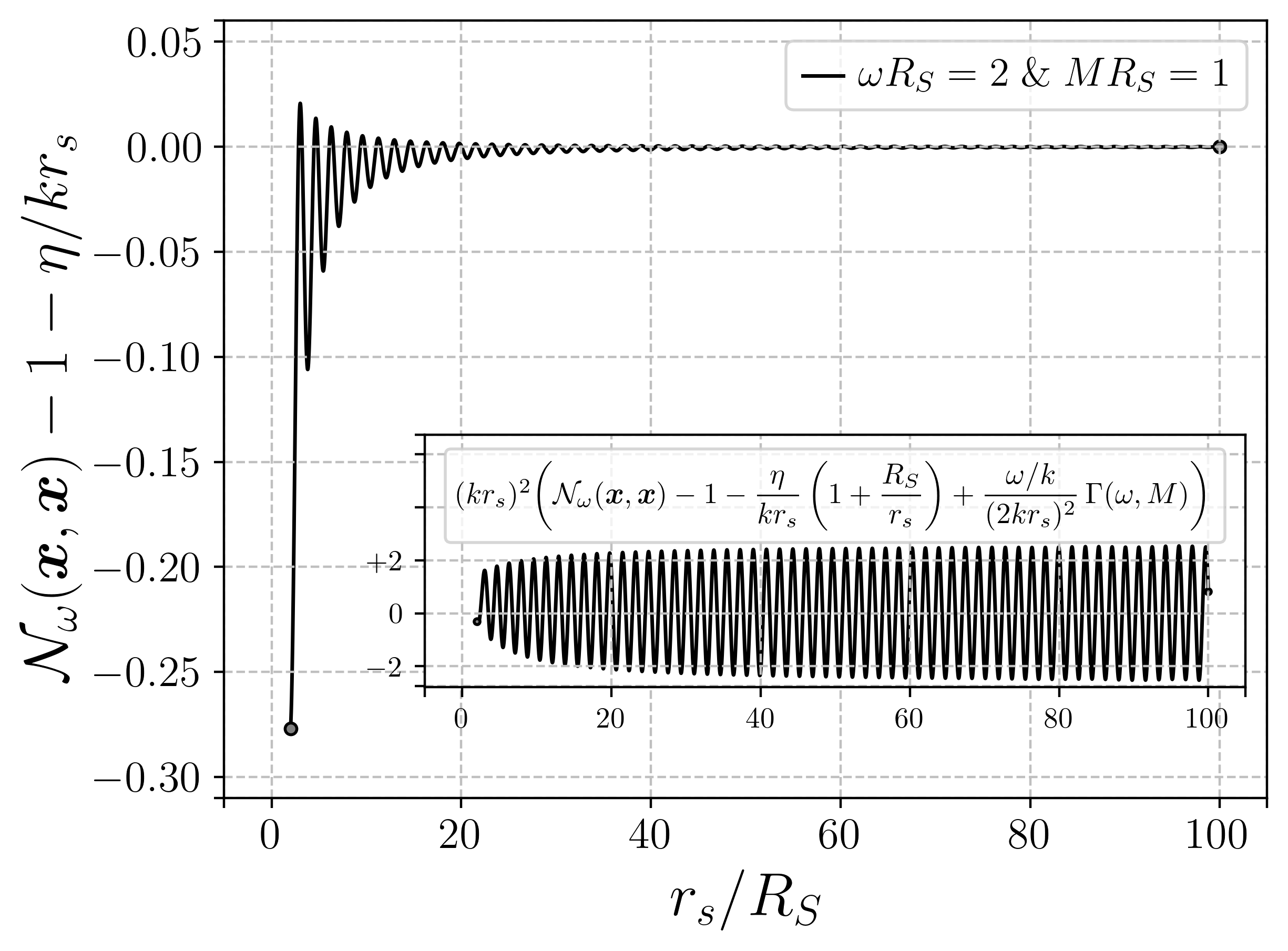}\hfill
\includegraphics[scale=0.515]{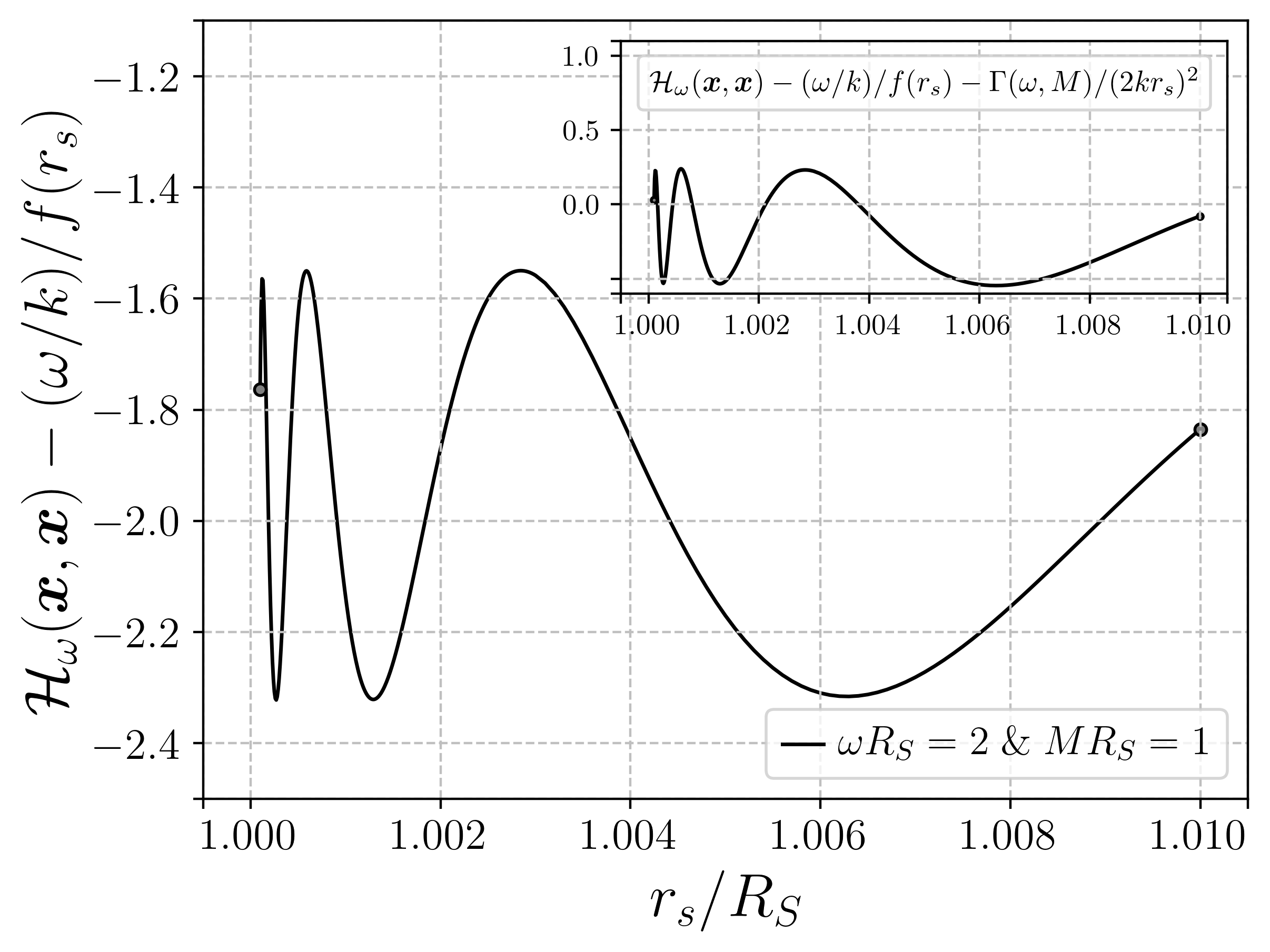}\vspace{2mm}
\includegraphics[scale=0.515]{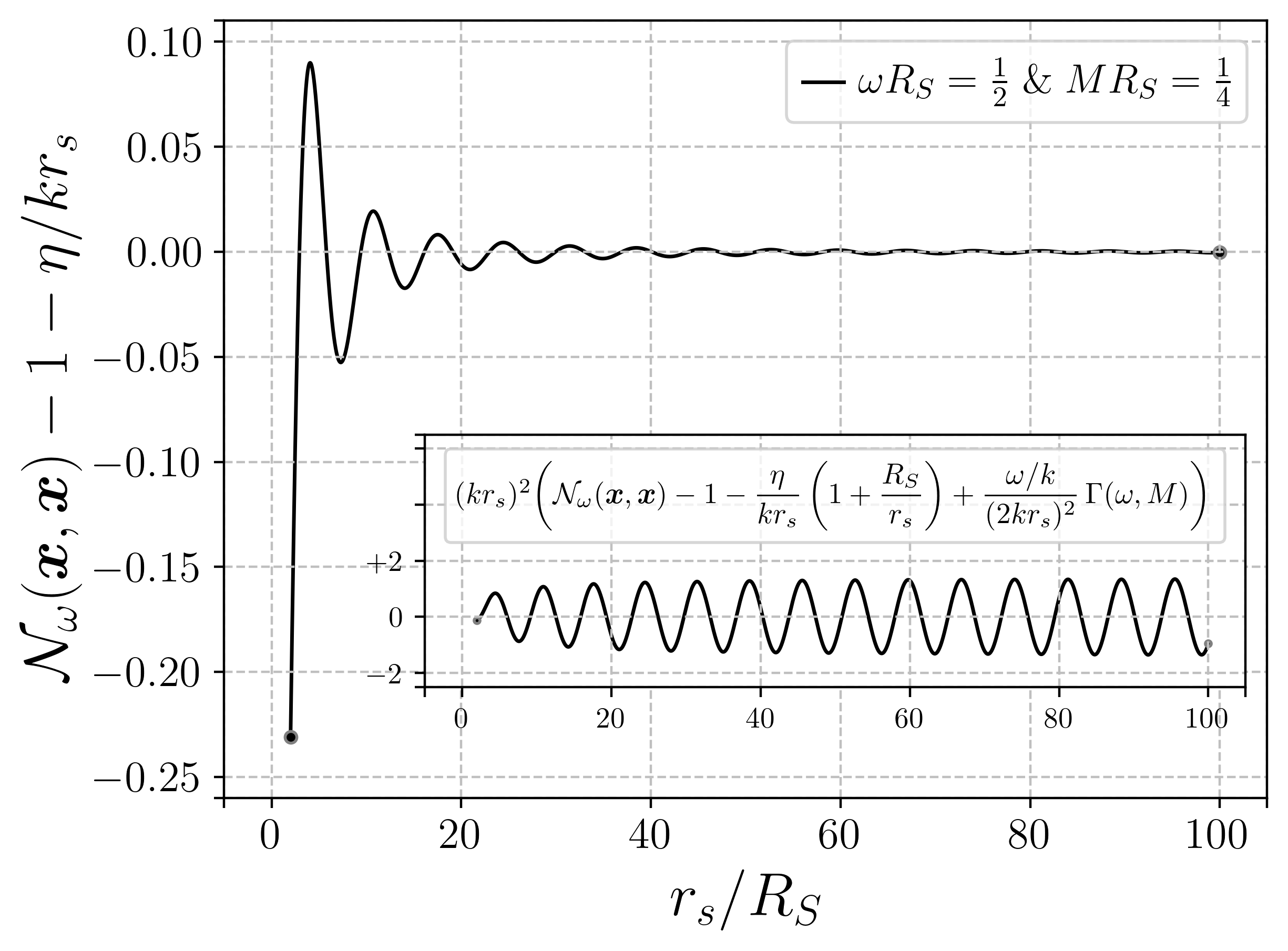}\hfill
\includegraphics[scale=0.515]{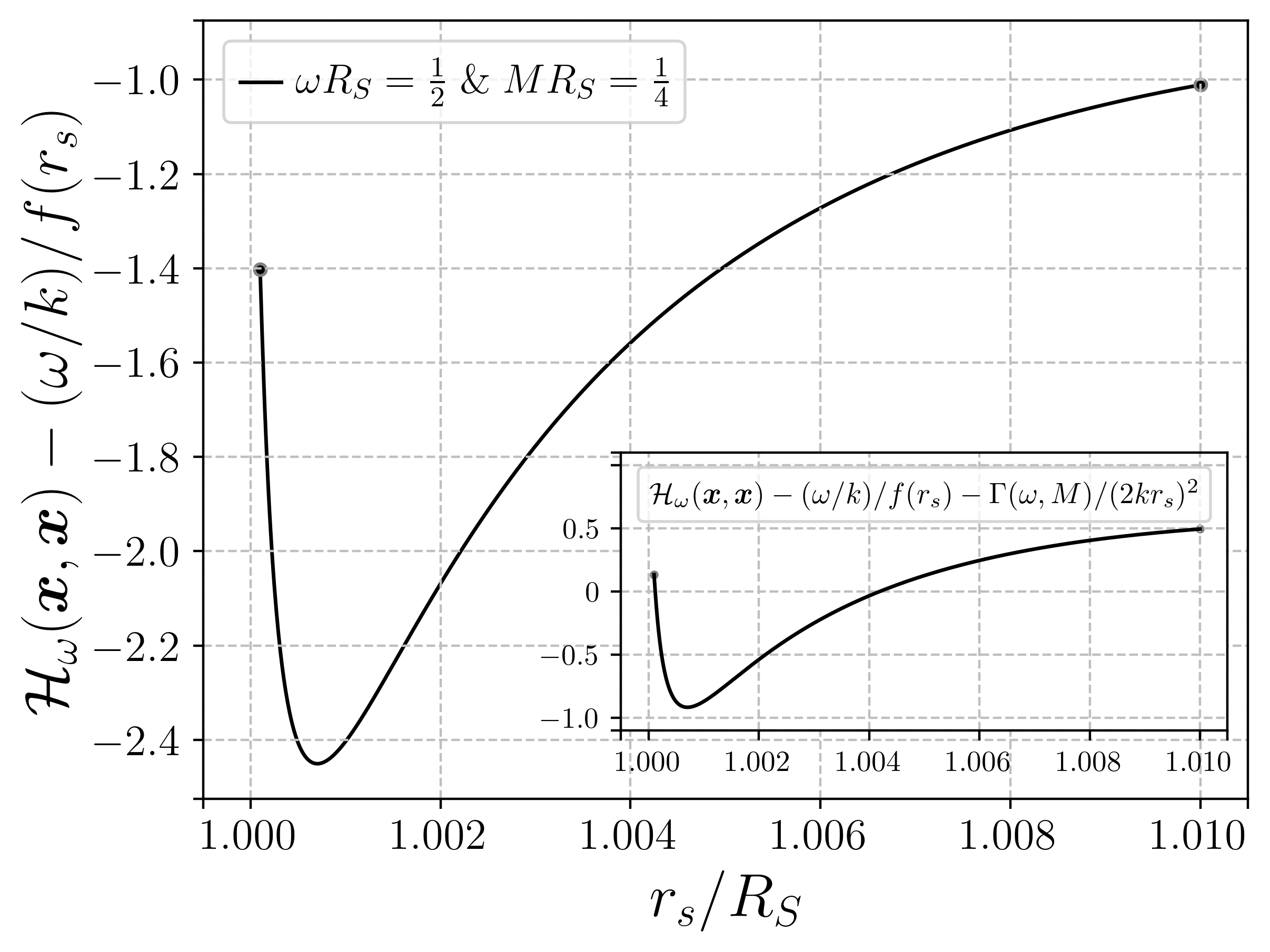}
\end{center}
\caption{Left column:~Numerical computations of $\mathcal{N}_{\omega}(\boldsymbol{x},\boldsymbol{x})$ 
for various values of $\omega$ and $M$,~while~the Schwarzschild radial coordinate
$r_s \in [2,100]{\times}R_S$.~Our numerics support \eqref{eq:g_nk_asymp} at $r_s \to \infty$.~Shortly
we shall derive this spatial-infinity asymptotic of $\mathcal{N}_{\omega}(\boldsymbol{x},\boldsymbol{x})$ by use
of $\mathcal{N}_{\omega l}^{(1)}(r)$ given in~\eqref{eq:app-radial-mode-solutions-n}.~It~will
also reveal the functional origin of the oscillations shown in the subplots.~Right 
column:~Numerical computations of
$\mathcal{H}_{\omega}(\boldsymbol{x},\boldsymbol{x})$ for the same $\omega$ and $M$,~while
$r_s \in [1.0001,1.01]{\times}R_S$.~Our~numerics~verify 
\eqref{eq:g_hk_asymp} at $r_s \to R_S$.~This~asymptotic of 
$\mathcal{H}_{\omega}(\boldsymbol{x},\boldsymbol{x})$
at $r_s \to R_S$ also agrees with~\cite{Candelas} for~$M = 0$.}\label{fig:3}
\end{figure}

\begin{figure}
\begin{center}
\includegraphics[scale=0.515]{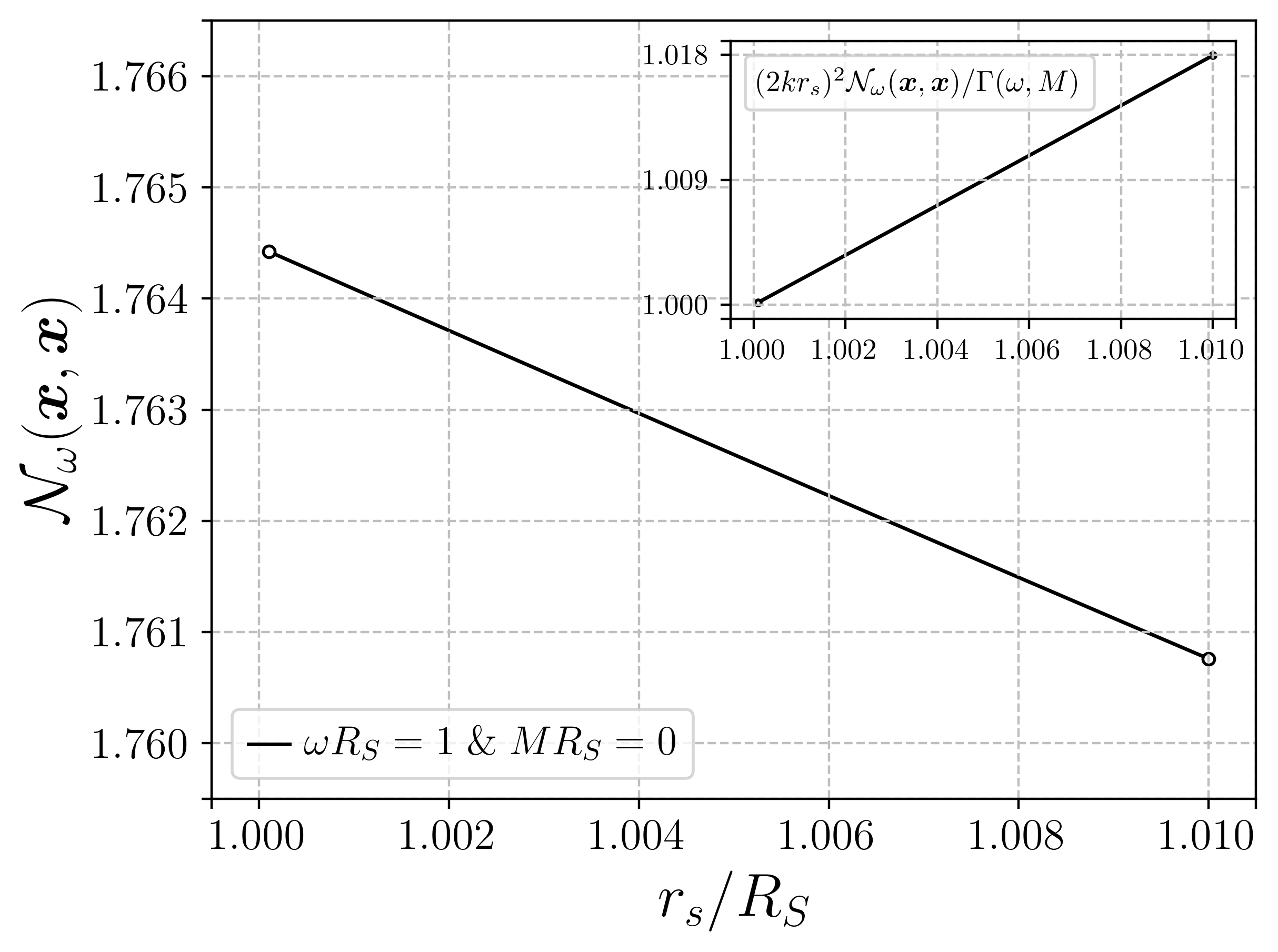}\hfill
\includegraphics[scale=0.515]{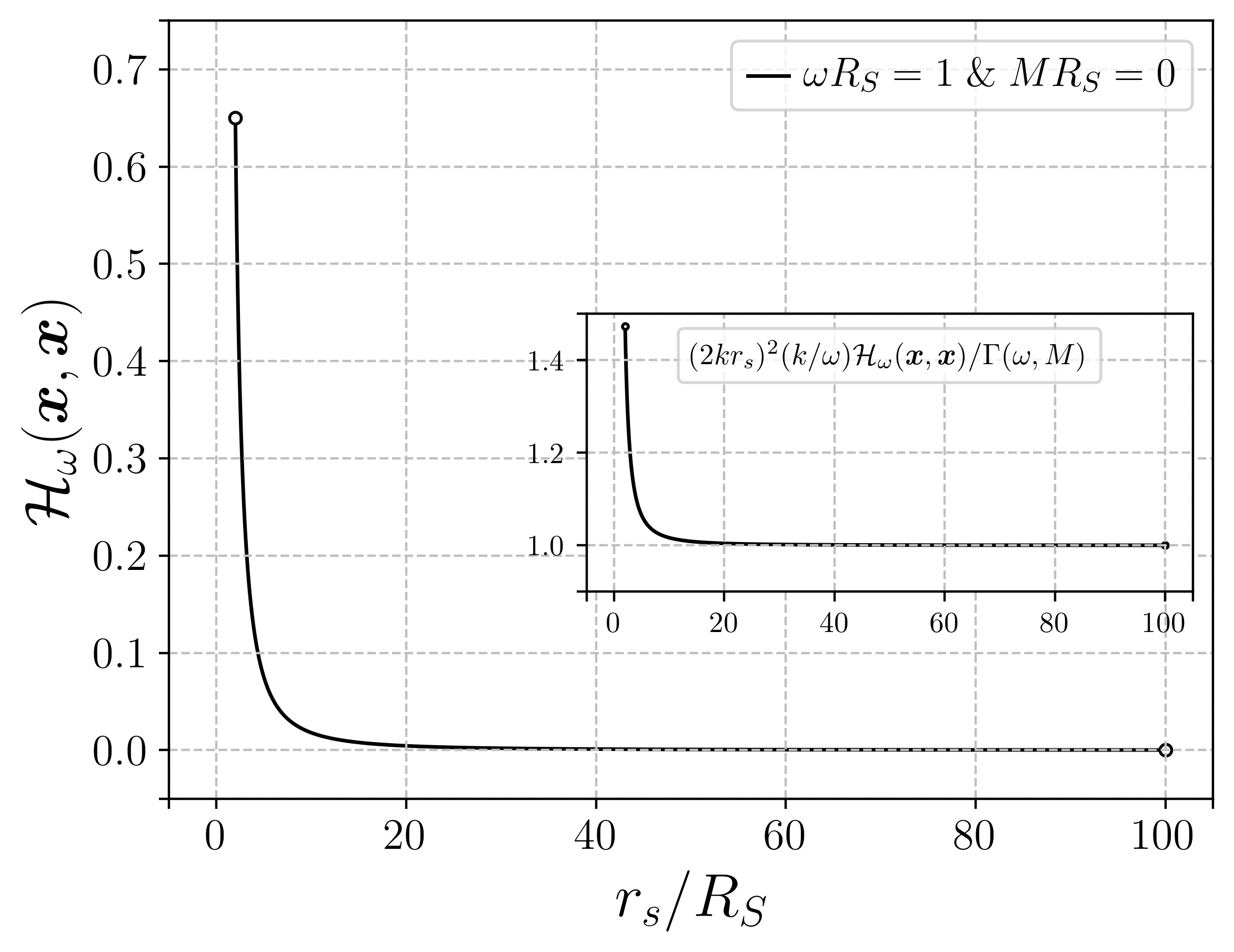}\vspace{2mm}
\includegraphics[scale=0.515]{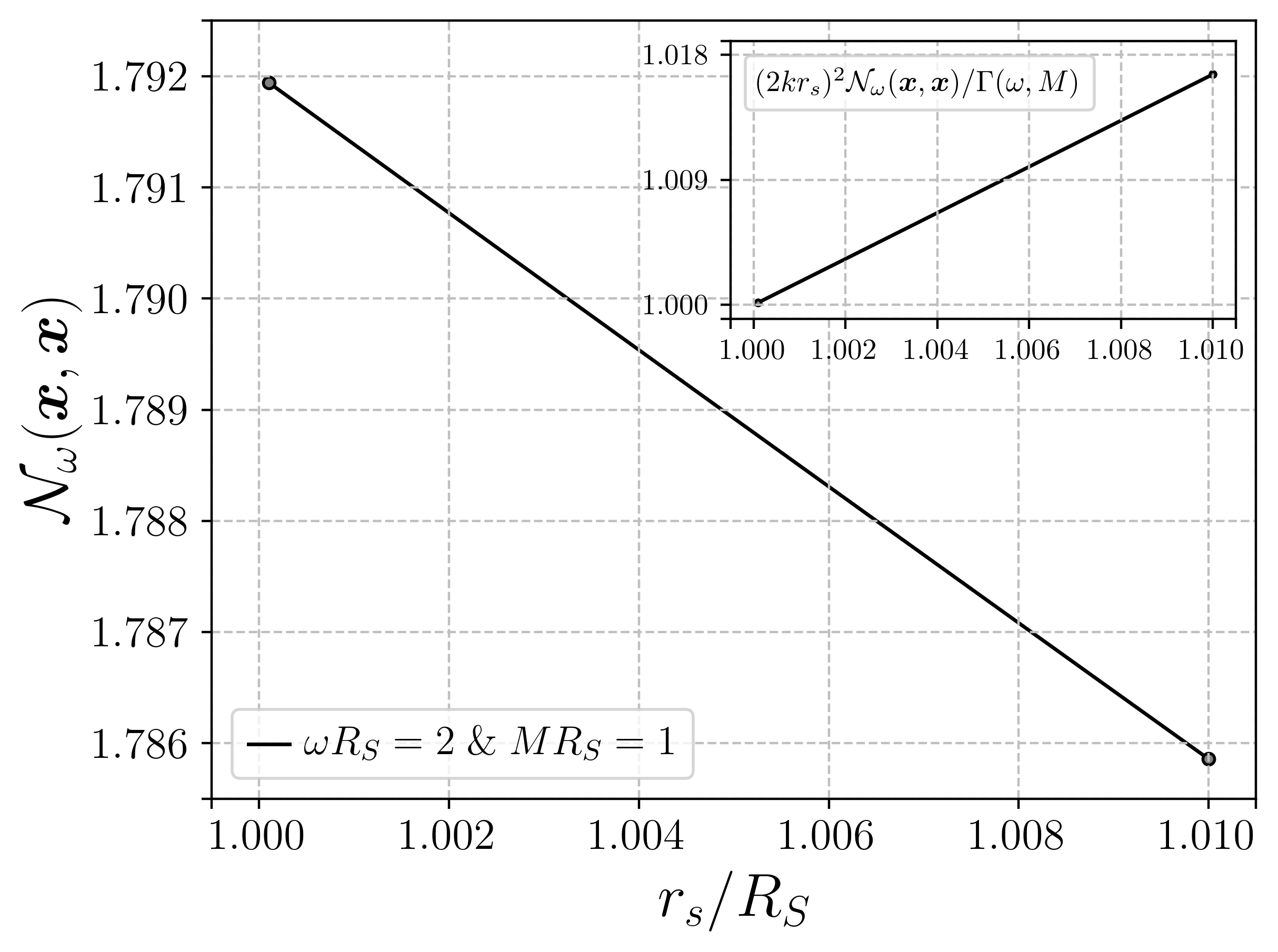}\hfill
\includegraphics[scale=0.515]{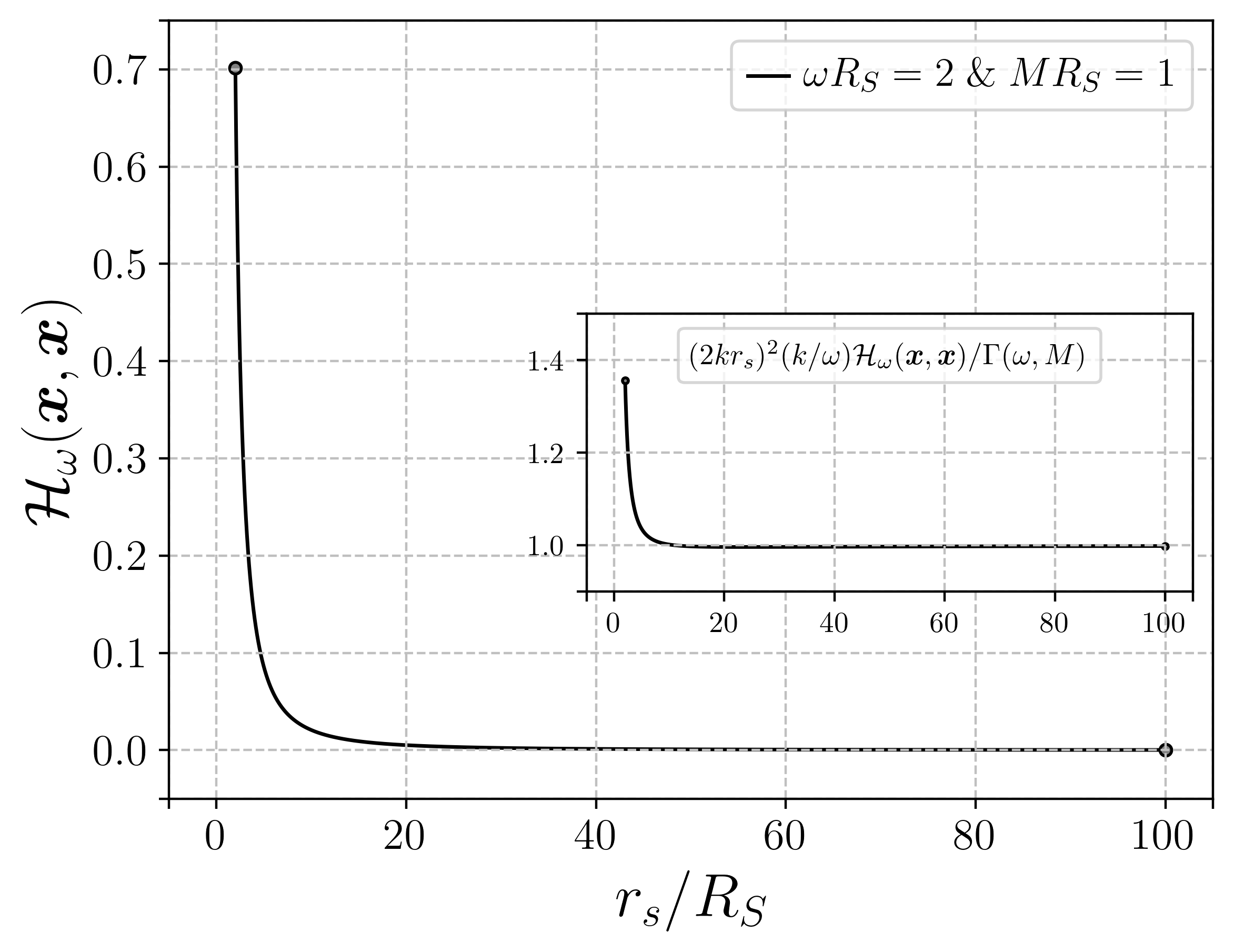}\vspace{2mm}
\includegraphics[scale=0.515]{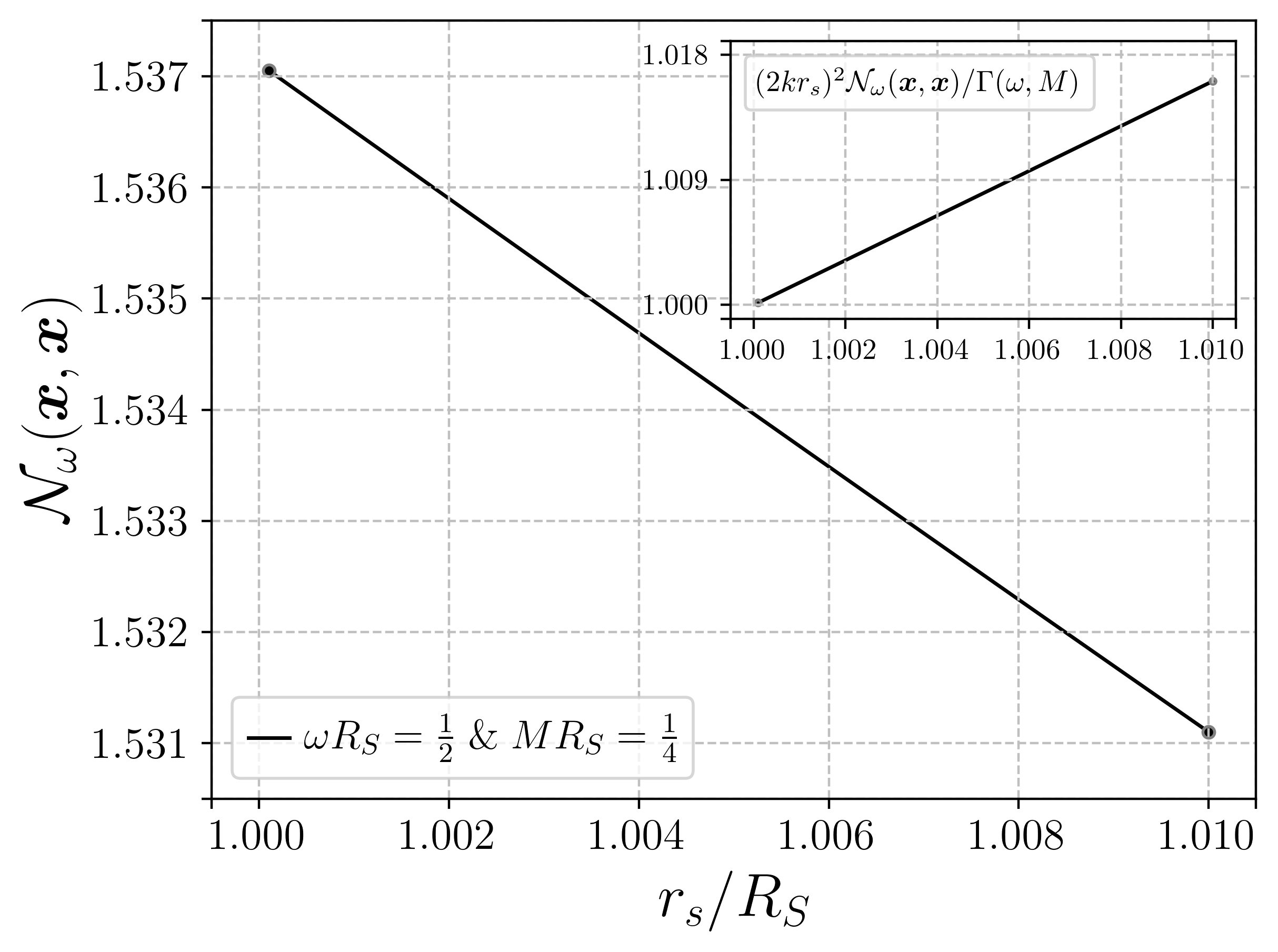}\hfill
\includegraphics[scale=0.515]{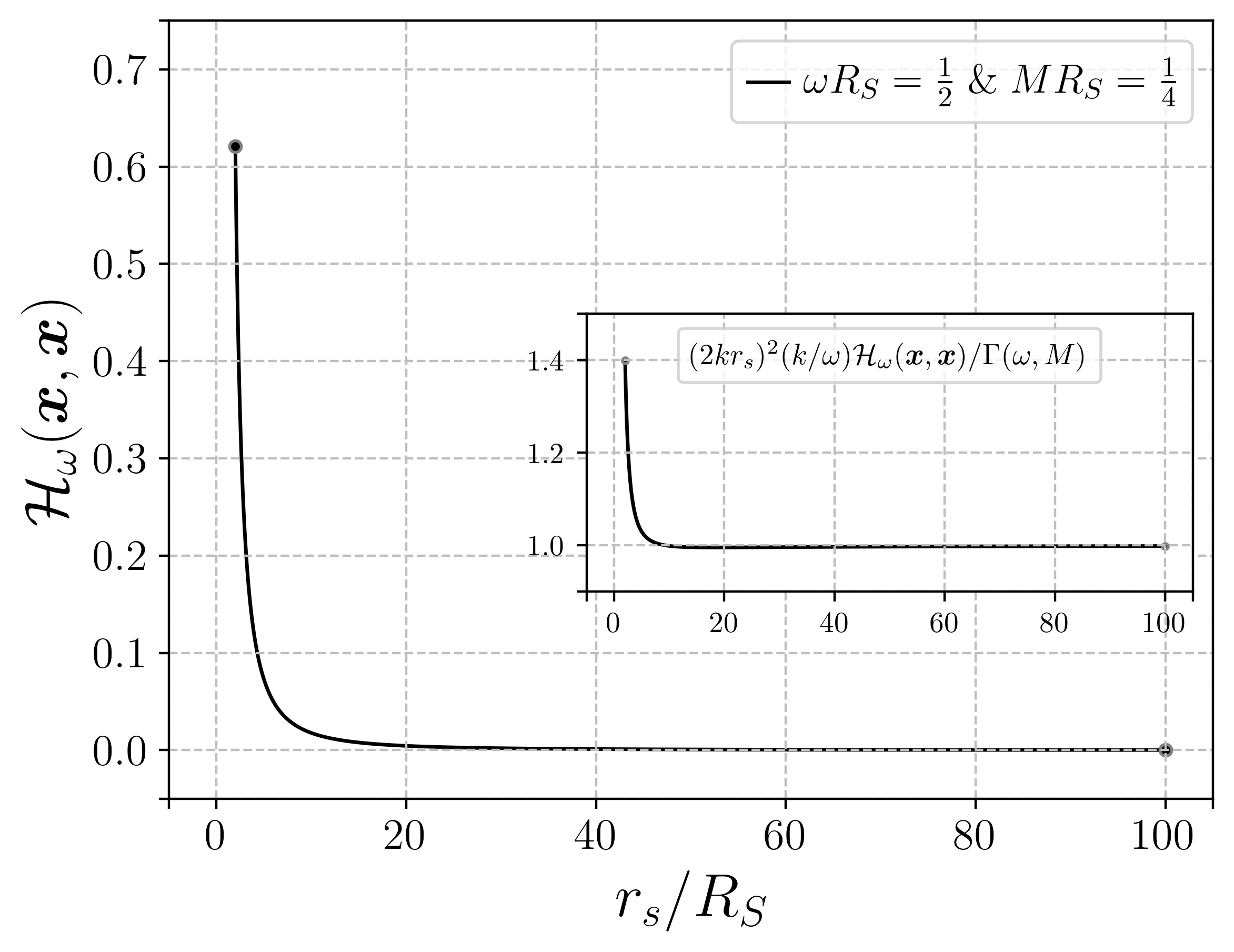}
\end{center}
\caption{Left column:~Numerical computations of $\mathcal{N}_{\omega}(\boldsymbol{x},\boldsymbol{x})$ 
for the same $\omega$ and $M$~as~in~Fig.~\ref{fig:3},~while
$r_s \in [1.0001,1.01]{\times}R_S$.~Our numerics support \eqref{eq:g_nk_asymp} at $r_s \to R_S$.~Right
column:~Numerical results for $\mathcal{H}_{\omega}(\boldsymbol{x},\boldsymbol{x})$ 
with $r_s \in [2,100]{\times}R_S$,~which confirm \eqref{eq:g_hk_asymp} at $r_s \to \infty$.~In contrast
to $\mathcal{N}_{\omega}(\boldsymbol{x},\boldsymbol{x})$,~which
approaches unity at $r_s \to \infty$ as shown in~Fig.~\ref{fig:3},~$\mathcal{H}_{\omega}(\boldsymbol{x},\boldsymbol{x})$ 
vanishes as $(1/r_s)^{2}$ at spatial~infinity.~This circumstance
particularly implies that the radial modes $N_{\omega lm}(x)$ and $H_{\omega lm}(x)$
differently behave~in the regime in which Newton's mechanics successfully works.~This has impact
on how the one-particle states $|n(x)\rangle$ and $|h(x)\rangle$ propagate
at $|\boldsymbol{x}| \gg R_S$,~as will be shown in the
subsequent sections.
}\label{fig:4}
\end{figure}

In fact,~we obtain for the massive scalar~field conformally coupled to gravity~that~the~trace
of its energy-momentum tensor reads
\beqa
\langle U|\hat{\Theta}_\mu^\mu(x)|U\rangle &=& M^2\,
\langle U| \hat{\Phi}^2(x) |U\rangle
\nonumber\\[1mm]\label{eq:trace-in-u}
&=& 
\frac{M^2}{2\pi^2}
{\int\limits_0^\infty}\frac{k^2dk}{2\omega}\,\mathcal{N}_{\omega}(\boldsymbol{x},\boldsymbol{x}) +
\frac{M^2}{2\pi^2}{\int\limits_0^\infty}\frac{kd\omega}{2}\,
\mathcal{H}_{\omega}(\boldsymbol{x},\boldsymbol{x})\,{\coth}{\big(2\pi \omega R_S\big)}\,,
\eeqa
where by definition
\bsubeqs\label{eq:g_nk_ghk_xx}
\beqa
\mathcal{N}_{\omega}(\boldsymbol{x},\boldsymbol{x}) &\equiv& \frac{1}{4k^2r_s^2}
\sum\limits_{l\,=\,0}^{\infty}\,(2l+1)\, |\mathcal{N}_{\omega l}(r_s)|^2\,,
\\[1mm]
\mathcal{H}_{\omega}(\boldsymbol{x},\boldsymbol{x}) &\equiv& \frac{1}{4k\omega r_s^2}
\sum\limits_{l\,=\,0}^{\infty}\,(2l+1)\, |\mathcal{H}_{\omega l}(r_s)|^2\,.
\eeqa
\esubeqs
By use of the asymptotic conditions and numerical computations, we obtain
\bsubeqs
\beqa\label{eq:g_nk_asymp}
\mathcal{N}_{\omega}(\boldsymbol{x},\boldsymbol{x}) &\to& 
\left\{
\begin{array}{ll}
1 + \dfrac{\eta}{k r_s}\,, & r_s \,\to\, \infty\,, \\[4mm]
\dfrac{1}{(2kr_s)^2}\,\Gamma(\omega,M)\,, & r_s \,\to\, R_S\,, \\[2mm]
\end{array}
\right.
\eeqa
and
\begin{figure}
\begin{center}
\includegraphics[scale=0.545]{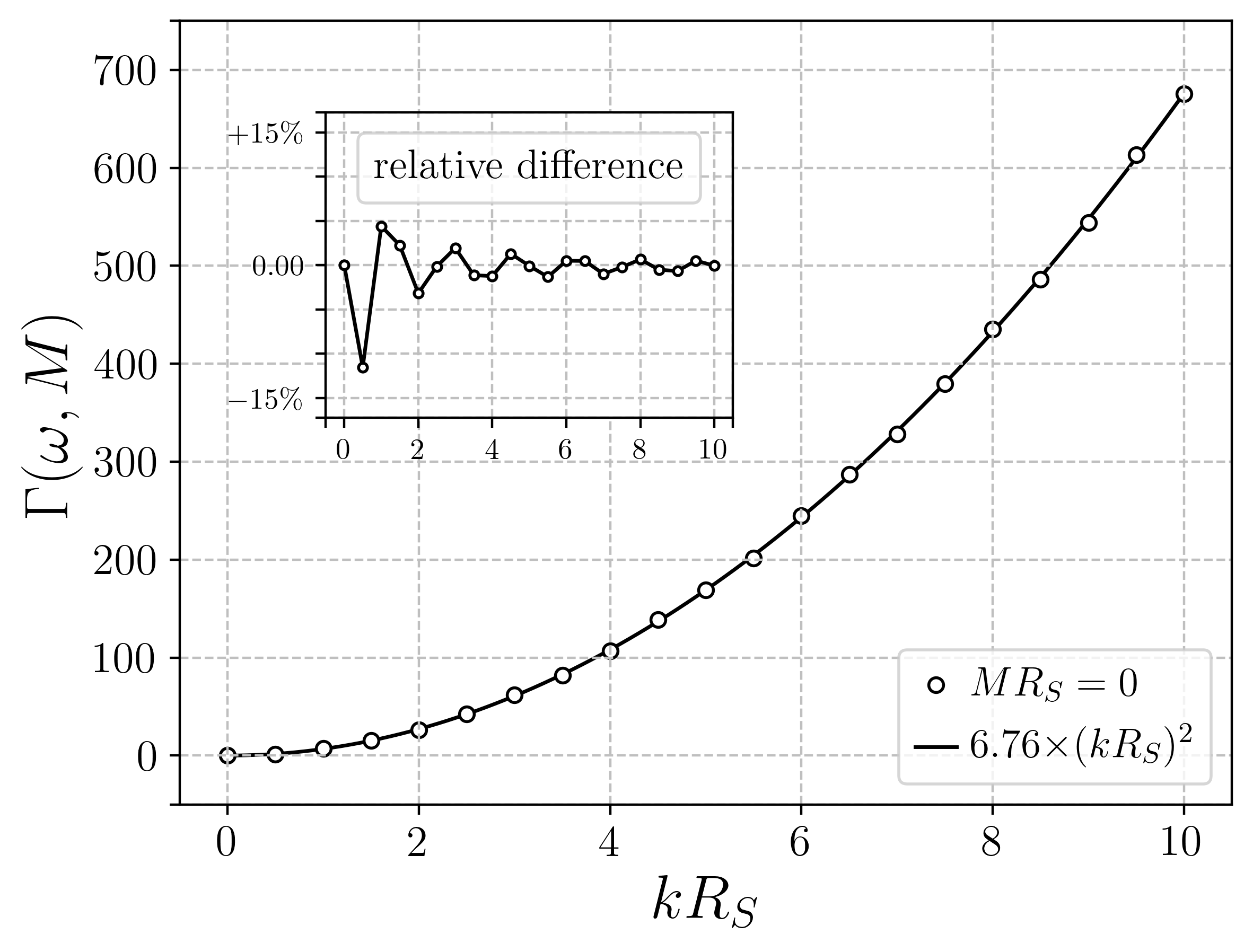}\hfill
\includegraphics[scale=0.545]{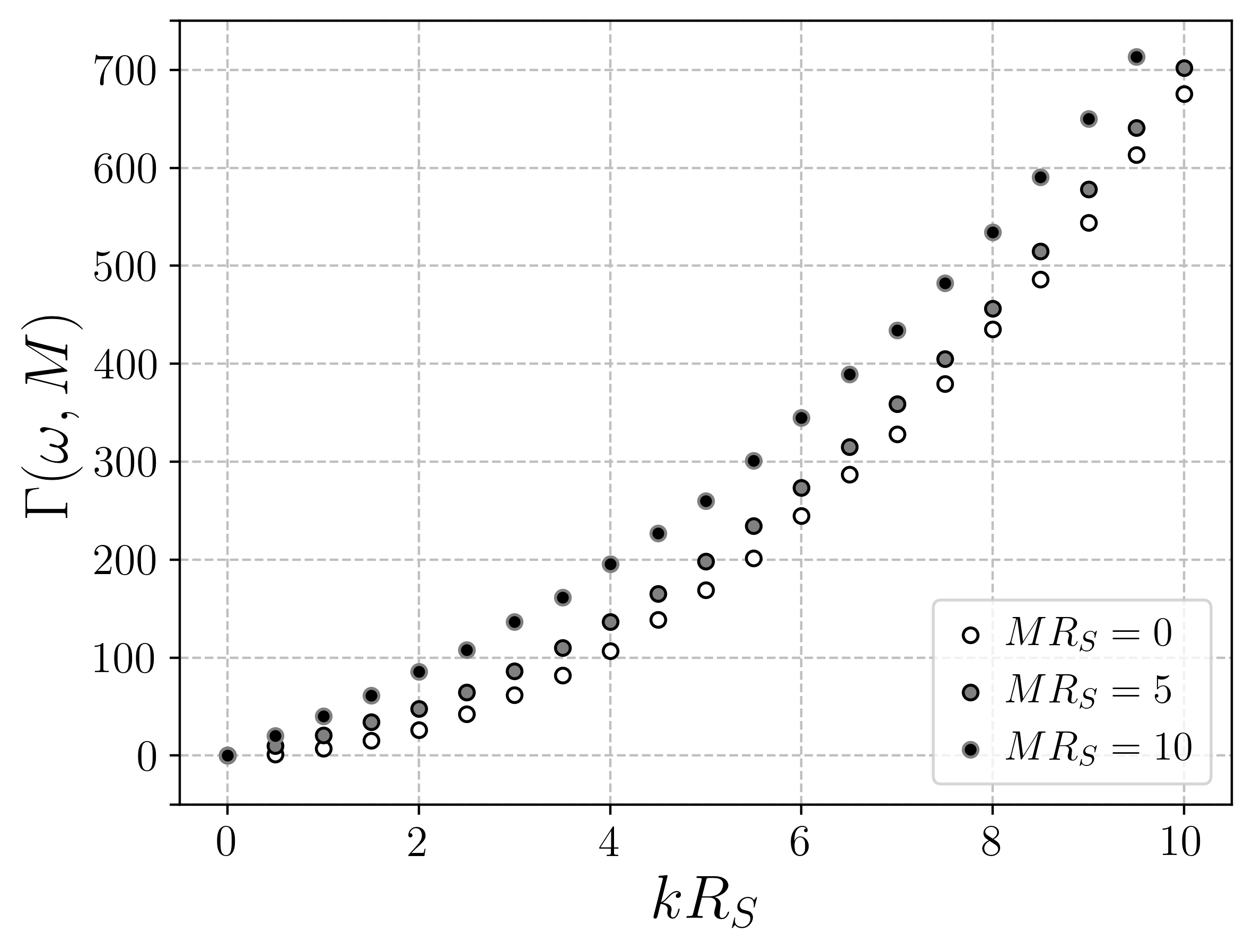}
\end{center}
\caption{Numerical computations of the gray-body factor $\Gamma(\omega,M)$ as a function of
$kR_S$ for various values of $MR_S$.\,We compute the
transmission probability $|B_l(\omega,M)|^2$~by using~the~method~outlined in the
caption of Fig.~\ref{fig:1}.~Left:~Numerical results for $\Gamma(\omega,M)$ with
$MR_S = 0$,~in accord~with~DeWitt's 
approximation~\cite{DeWitt-1975}.~Right:~Numerical results
for $\Gamma(\omega,M)$ with~$MR_S \in \{5,10\}$,~suggesting~$\Gamma(\omega,M)$
increases with growing $MR_S$ and vanishes~if~$kR_S \to 0$,~which agrees
with~\eqref{eq:a1} as $kR_S \to 0$~if~$c \to \infty$.
}\label{fig:2}
\end{figure}
\beqa\label{eq:g_hk_asymp}
\mathcal{H}_{\omega}(\boldsymbol{x},\boldsymbol{x}) &\to& 
\left\{
\begin{array}{ll}
\dfrac{\omega/k}{(2kr_s)^2}\,\Gamma(\omega,M)
\,, & r_s \,\to\, \infty\\[4mm]
\dfrac{\omega/k}{f(r_s)}\,, & r_s \,\to\, R_S\,, \\[2mm]
\end{array}
\right.
\eeqa
\esubeqs
where lapse function in terms of the Schwarzschild radial coordinate reads
\beqa
f(r_s) &\equiv& 1 + 2\phi(r(r_s)) \;=\; 1 - \frac{R_S}{r_s}
\eeqa
and the gray-body factor is defined as follows:
\beqa
\Gamma(\omega,M) &\equiv& \sum\limits_{l \,=\,0}^{\infty}(2l+1)|B_l(\omega,M)|^2\,.
\eeqa
Our numerics~for~$\Gamma(\omega,M)$~are presented in Fig.~\ref{fig:2}.~Our~numerics validating
\eqref{eq:g_nk_asymp} and~\eqref{eq:g_hk_asymp}~are presented in Figs.~\ref{fig:3} and~\ref{fig:4}.

Both integrals in~\eqref{eq:trace-in-u} diverge as $\omega^2$ at $\omega \to \infty$.~It is because
quantum fields are operator-valued distributions whose products at the same space-time
point are typically singular.~The standard approach
to deal with this problem is
renormalisation theory~\cite{Birrell&Davies}.~Since this applies to any state,~we 
treat $|\Omega\rangle$ which locally reduces to the Minkowski
quantum vacuum~as
defined in theoretical particle physics~\cite{Weinberg}.~We find in this state that
\beqa\label{eq:trace-in-m}
\langle \Omega|\hat{\Theta}_\mu^\mu(x)|\Omega\rangle &=& {M^2}\lim\limits_{X \,\to\, x}\,
\langle \Omega| \hat{\Phi}(x)\hat{\Phi}(X) |\Omega\rangle
\nonumber\\[1mm]
&=& {M^2}\lim\limits_{X \,\to\, x}\, 
{\int}{\frac{d^3\boldsymbol{p}}{(2\pi)^3}}\,
\frac{1}{2\omega_{\boldsymbol{p}}}\,e^{-ip{\cdot}(y(x)-y(X))}
+ \textrm{finite terms}
\nonumber\\[1mm]
&=& {M^2}\lim\limits_{X \,\to\, x}\, \frac{-1}{8\pi^2\sigma(x,X)} + 
\frac{M^2}{16\pi^2}\ln\sigma(x,X) + \textrm{finite terms}\,,
\eeqa
where\,$\sigma(x,X)$~is~geodesic distance~\cite{DeWitt1965}.~The singular terms~in 
$\langle \Omega| \hat{\Phi}(x)\hat{\Phi}(X) |\Omega\rangle$\,at~$X \to x$~are
common for locally Minkowski states.~These singularities
can be eliminated by making use~of the
Hadamard renormalisation~\cite{Decanini&Folacci}.~It is based on
subtracting the reciprocal and logarithmic divergences with respect to $\sigma(x,X) \to 0$
in~\eqref{eq:trace-in-m}.~We approximately obtain
by use of~\cite{Emelyanov-2018}~that
\beqa\label{eq:trace-in-m-int-rep}
\langle \Omega|\hat{\Theta}_\mu^\mu(x)|\Omega\rangle 
&\xrightarrow[r_s \,\to\, \infty]{r_s \,\to\, R_S}& 
\frac{M^2}{2\pi^2}
{\int\limits_0^\infty}\frac{kd\omega}{2}\,
\Omega_\omega(r_s)\,{\coth}{\big(\pi\omega/\kappa(r_s)\big)}
+ \text{finite terms}\,
\eeqa
in the near- and far-horizon regions,~where
\beqa\label{eq:omega-function}
\Omega_\omega(r_s) &\equiv&
\frac{\left|K_{i\omega/\kappa(r_s)+1}\hspace{-0.8mm}\left(
\scalebox{0.7}{$\dfrac{f^\frac{1}{2}(r_s)M}{\kappa(r_s)}$}\right)\right|^2 -
\left|K_{i\omega/\kappa(r_s)}\hspace{-0.8mm}\left(
\scalebox{0.7}{$\dfrac{f^\frac{1}{2}(r_s)M}{\kappa(r_s)}$}\right)\right|^2}
{(k/\omega)f(r_s)\left|\Gamma(1 + i\omega/\kappa(r_s))\right|^2\big(\kappa(r_s)/f^\frac{1}{2}(r_s)M\big)^2}\,,
\eeqa
where $K_\nu(z)$ is~the modified Bessel function,~and
\beqa
\kappa(r_s) &\equiv& f'(r_s)/2
\eeqa
is surface gravity.~In fact,~we have from 6.794.3 on p.~751~in~\cite{Gradshteyn&Ryzhik} that
\eqref{eq:trace-in-m-int-rep} with~\eqref{eq:omega-function}~is~equal~to
\eqref{eq:trace-in-m}~with~$\sigma(x,X)$~in which 
$t = T + 0$ and~$\boldsymbol{x} = \boldsymbol{X}$.~Furthermore,
we obtain
\beqa\label{eq:omega_asymp}
\Omega_\omega(r_s) &\to&
\left\{
\begin{array}{ll}
\theta(\omega - M)
{\left(1 + \dfrac{\eta}{k r_s}\right)}\,, & r_s \,\to\, \infty\,, \\[4mm]
\dfrac{\omega/k}{f(r_s)}\,, & r_s \,\to\, R_S\,, \\[2mm]
\end{array}
\right.
\eeqa
where $\theta(z)$ is the Heaviside function.~Comparing~\eqref{eq:omega_asymp}
with~\eqref{eq:g_nk_asymp} at $r_s \to \infty$ and with~\eqref{eq:g_hk_asymp}
at $r_s \to R_S$, we observe that $\Omega_\omega(r_s)$ approximately interpolates
between $\mathcal{N}_\omega(\boldsymbol{x},\boldsymbol{x})$~in~the~far-
horizon region and
$\mathcal{H}_\omega(\boldsymbol{x},\boldsymbol{x})$ in the near-horizon region.

The singularities in $\langle U|\hat{\Theta}_\mu^\mu(x)|U\rangle$ and
$\langle \Omega|\hat{\Theta}_\mu^\mu(x)|\Omega\rangle$ accordingly match at spatial infinity,~at
least up to the leading order in $1/r_s$.~This is due to the operators
$\hat{n}(x)$ and $\hat{n}^\dagger(x)$.~In~the~near-horizon region,~these 
match owing to the operators $\hat{h}(x)$ and $\hat{h}^\dagger(x)$,~at
least~up~to~the~leading order in $1/f(r_s)$.~As a consequence,~their difference
is finite at the event horizon,~at~least~in the massless limit, $M \to 0$.~The cancelation
of $1/f(r_s) \to \infty$ in the $r_s \to R_S$ limit needs~that
the Unruh state is characterised 
by the Hawking temperature $\kappa(R_S)/2\pi = 1/4\pi R_S$.~This~is~in 
accord with~\eqref{eq:u-state}.~In
the absence of future horizon,~however,~the Boulware state $|B\rangle$,~which~is vacuous with respect
to both $\hat{n}(x)$ and $\hat{h}(x)$,~is admissible as $r_s > R_S$ implies $1/f(r_s) < \infty$.\;\;\;\;

The line element of Schwarzschild spacetime approximately approaches the line element~of
Minkowski spacetime far away from~the~horizon and of Rindler spacetime nearby the horizon.
In fact,~the gravitational potentials~\eqref{eq:gravitational-potentials} vanish at
$r \to \infty$,~while if
$\boldsymbol{x} \to (\boldsymbol{x} + R_S\boldsymbol{e}_z)/4$,
where $|\boldsymbol{x}| \ll R_S$,
the line element~\eqref{eq:schwarzschild-spacetime} approaches~\eqref{eq:rindler-spacetime},
assuming
$z_R \to z_R - 1/a$ and $a = 1/2R_S$.
Rindler spacetime is a patch of Minkowski spacetime,~in~which stationary observers move~at~a 
constant proper acceleration,~equaling $\kappa(r_s)$ in our case.~The Schwarzschild-time
coordinate~$t$ is then approximated by
the Rindler time $t_R$ at the horizon and a local Minkowski time,~$t_M$,~at
spatial infinity~\cite{Emelyanov-2018}:
\beqa
\Delta{t}_M &\approx&  \frac{f^\frac{1}{2}(r_s)}{\kappa(r_s)}\,{\sinh}\big(\kappa(r_s)\Delta{t}\big)\,,
\eeqa
where we have neglected terms vanishing as $(f(r_s))^{3/2}$ at $r_s \to R_S$ and
as $(f^\prime(r_s))^{2}$ at $r_s \to \infty$.
This explains the emergence
of~the~effective temperature parameter $\kappa(r_s)/2\pi$
in~\eqref{eq:trace-in-m-int-rep},~see~\cite{Sciama&Candelas&Deutsch}.
This effective temperature approaches Hawking's temperature $1/4\pi R_S$~at~the~future~horizon
and asymptotically vanishes
at spatial infinity.~If there is more than one black hole,~then~the effective
temperature~in~\eqref{eq:trace-in-m-int-rep}
acquires different values depending on the Schwarzschild~radius of a given black hole.~This
illustrates the local character of the state $|\Omega\rangle$.

The Unruh state $|U\rangle$ is thermal with respect to 
$\hat{h}(x)$~and~$\hat{h}^\dagger(x)$ operators at the prescribed temperature.~It~is~necessary 
for the energy-momentum tensor $\langle U|\hat{\Theta}_\nu^\mu(x)|U\rangle$ to be
non-singular on the event horizon of a prescribed black hole.~This 
aspect of $|U\rangle$ must thus manifest~itself~at spatial infinity.~In particular,~we obtain by
use of~\eqref{eq:asymptotic-radial-modes-n} and~\eqref{eq:asymptotic-radial-modes-h} that
\beqa\label{eq:he}
\langle U|\hat{\Theta}_t^{\boldsymbol{x}}(x)|U \rangle &\xrightarrow[r \,\to\, \infty]{}&
+ \frac{1}{8\pi^2r^2}\,
{\int\limits_0^\infty}\frac{dk}{\omega}\,
\frac{\omega^2\,\Gamma(\omega,M)}{e^{4\pi\omega R_S} - 1}\,\frac{\boldsymbol{x}}{|\boldsymbol{x}|}\,,
\eeqa
where we have taken into account that $H_{\omega lm}(x)$ exponentially vanishes
at spatial infinity for $\omega < M$.~This quantity describes an outward
(spherically symmetric) flux of energy,~in~accord
with~the~Hawking effect~\cite{Hawking1975},~see also~\cite{DeWitt-1975}.~This energy flux is
carried by Hawking particles.~In other words,~the one-particle state
$|h(x)\rangle = \hat{h}^\dagger(x)|B\rangle$ defined in~\eqref{eq:ohps}
can be interpreted~as~a one-Hawking-particle state at the point $x$.

\subsection{Propagator $\langle n(x) | n(X) \rangle$ in the far-horizon region}
\label{sec:nxnX}

It proves useful to get the scalar-field mode $N_{\omega lm}(x)$ in
Cartesian coordinates.~It suffices~to obtain that in perturbation theory up to the leading
order in the Schwarzschild radius~$R_S$.~We obtain from
\beqa
N_{\omega lm}^{(1)}(x) &=&
\frac{i}{\sqrt{2\pi}}\,e^{-i\omega t}\,
\frac{r\,\mathcal{N}_{\omega l}^{(1)}(r)}{\left(r + \frac{1}{4}R_S\right)^2}\,Y_{lm}(\Omega_{\boldsymbol{x}})\,,
\eeqa
which follows from~\eqref{eq:nwml} with~\eqref{eq:app-radial-mode-solutions-n},~by
making use of 6.6.3.4 on p.\,347 in~\cite{Prudnikov&Brychkov&Marichev-3},~that
\beqa
N_{\boldsymbol{k}}^{(1)}(x) &\equiv& \frac{(2\pi)^\frac{3}{2}}{k}\,
{\sum\limits_{l\,=\,0}^\infty}\;{\sum\limits_{m\,=\,-l}^{m\,=\,+l}}\,i^l\,
N_{\omega lm}^{(1)}(x)\, \overline{Y_{lm}(\Omega_{\boldsymbol{k}})}\,,
\nonumber\\[1mm]
&=&
e^{-ik{\cdot}x+\pi\eta/2}\,\Gamma(1-i\eta)\,
\frac{M\hspace{-0.3mm}\big(i\eta,\,1,\,
i(kr - \boldsymbol{k}{\cdot}\boldsymbol{x})\big)}{\left(1 - (R_S/4r)^2\right)^\frac{1}{2}}\,,
\eeqa
where $M(a,b,z)$ is the Kummer function.~This~non-exact solution matches the exact~solution
used in quantum mechanics to describe a~charged
particle elastically scattered~by Coulomb's potential,~assuming the non-relativistic limit,~i.e.~$M \gg k$,
the denominator set to unity
and the Sommerfeld parameter numerically given by
$M^2R_S/2k$,~see~\cite{Landau&Lifshitz} for more details.~In~this
case,
$N_{\boldsymbol{k}}(x)$ can be interpreted at $r \gg R_S$ as modelling
a non-relativistic~particle elastically scattered by Newton's
potential.

The interpretation of the field quantisation~\eqref{eq:hfq}
in terms of scattering, e.g.,~see~\cite{Wald},~differs 
from that of the setup we have dealt with in Sec.~\ref{sec:qm}.~Namely, a quantum particle
of mass~$M$~at the Earth's surface has no asymptotic descriptions of scattering theory.~The
quantum particle in free fall is not asymptotically free.~The field
quantisation~\eqref{eq:mfq} is based on the application~of quantum field theory
to the description of scattering processes in collider physics.~These are,
however,~owing to non-gravitational interactions.~In general relativity, non-spinning particles
move along geodesics.~In particular,~geodesics which satisfy appropriate~conditions for initial
position and momentum may model scattering processes due to gravitational interaction.~In
terms of Riemann normal coordinates, all geodesics passing through
$y = Y$ turn locally into straight world lines.~These in turn correspond to trajectories
of asymptotic states entering~the 
$S$-matrix in collider physics.~In non-inertial coordinates,~these trajectories
may correspond~to free fall or scattering via gravitational interaction,~explaining
why~the~field~quantisation~\eqref{eq:mfq}
also agrees with particles' motion which is not asymptotically free.~The field 
quatisation~\eqref{eq:hfq} appears,~in contrast,~to be designed relying on scattering
theory,~see~\cite{Wald}.~This~thus~needs~the consideration of asymptotic
regions in space and time which in practice do not~exist,~bearing
in mind the observable Universe is a non-Schwarzschild spacetime.

The scattering modes $N_{\omega lm}(x)$ may locally give rise to a propagator
which still matches~\eqref{eq:qm-propagator}
in the $c \to \infty$ limit -- this is what we intend to study here.~If affirmative, then the one-particle
state $|n(x)\rangle$ defined in~\eqref{eq:onps} has properties matching those of observable
particles, particularly considered in Sec.~\ref{sec:qm}.~We~have from~\eqref{eq:onps}
and~\eqref{eq:n-operator}~that
\bsubeqs
\beqa\label{eq:Nk-function}
\langle n(x)|n(X)\rangle &=& \frac{1}{2\pi^2}
{\int\limits_0^\infty}\frac{k^2dk}{2\omega}\,e^{-i\omega\Delta{t}}\,\mathcal{N}_{\omega}(\boldsymbol{x},\boldsymbol{X})\,,
\eeqa
where by definition
\beqa\label{eq:spatial-Nk-function}
\mathcal{N}_{\omega}(\boldsymbol{x},\boldsymbol{X}) &\equiv& \frac{1}{4k^2}
\sum\limits_{l \,=\, 0}^{+\infty}\,(2l+1)\,\frac{r\,\mathcal{N}_{\omega l}(r)}{\left(r+\frac{1}{4}R_S\right)^2}
\,\frac{R\,\overline{\mathcal{N}_{\omega l}(R)}}{\left(R+\frac{1}{4}R_S\right)^2}\,
P_l\left(\scalebox{0.9}{$\dfrac{\boldsymbol{x}{\cdot}\boldsymbol{X}}{rR}$}\right),
\eeqa
\esubeqs
where~$P_\nu(z)$ is the Legendre polynomial and we recall that $r = |\boldsymbol{x}|$
and,~accordingly,~$R = |\boldsymbol{X}|$. Note that
$\mathcal{N}_{\omega}(\boldsymbol{x},\boldsymbol{X})$
reduces to $\mathcal{N}_{\omega}(\boldsymbol{x},\boldsymbol{x})$ from~\eqref{eq:g_nk_ghk_xx}~if
$\boldsymbol{X} = \boldsymbol{x}$.~We,~first,~obtain at $R_S \to 0$~by use
of~\eqref{eq:app-radial-mode-solutions-n} that
\beqa
\mathcal{N}_{\omega}(\boldsymbol{x},\boldsymbol{X}) &\xrightarrow[R_S \,\to\, 0]{}&
j_0\big(k|\Delta\boldsymbol{x}|\big)\,,
\eeqa
where $j_\nu(z)$ is the spherical Bessel function,~and
we have taken into account 13.18.8, 10.27.6, 10.47.3 and 10.60.2 in~\cite{Olver&etal}.~This gives
\beqa
\langle n(x)|n(X)\rangle &\xrightarrow[c \,\to\, \infty]{}& \frac{e^{-iMc^2\Delta{t}}}{2Mc}\, 
\langle x|X\rangle|_{G \,\to\, 0}\,.
\eeqa
Thus,~the one-particle state $|n(x)\rangle$ models~a quantum particle of mass $M$,~which
freely~moves
as in classical mechanics in the absence of gravity (Newton's constant $G \to 0$).

To determine how $|n(x)\rangle$ moves in the presence of gravity,~$G > 0$,~we,~second,~approximate
$\mathcal{N}_{\omega l}(r)$ entering~\eqref{eq:spatial-Nk-function} 
by $\mathcal{N}_{\omega l}^{(1)}(r)$ given in~\eqref{eq:app-radial-mode-solutions-n}.~Up to the leading
order in the Schwarzschild radius $R_S$, we have
\beqa\label{eq:Nk-(1)-function-def}
\mathcal{N}_{\omega}^{(1)}(\boldsymbol{x},\boldsymbol{X}) &=&
\sum\limits_{l \,=\, 0}^{+\infty}\,
\frac{(1)_l\,|(1-i\eta)_l|^2}{(-1)^l\,l!\,(1)_{2l}\,(2)_{2l}}
\frac{M_{-i\eta,l+\frac{1}{2}}(2ikr)M_{-i\eta,l+\frac{1}{2}}(2ikR)}{e^{-\pi\eta}\,|\Gamma(1-i\eta)|^{-2}(2ikr)\,(2ikR)}
P_l\left(\scalebox{0.9}{$\dfrac{\boldsymbol{x}{\cdot}\boldsymbol{X}}{rR}$}\right).
\eeqa
Making use of~\cite{Hostler&Pratt}, we obtain
\beqa\label{eq:Nk-(1)-function-res}
\mathcal{N}_{\omega}^{(1)}(\boldsymbol{x},\boldsymbol{X}) &=& \left.
\frac{\big(\partial_{\xi_{+}} - \partial_{\xi_{-}}\big)
M_{i\eta,\frac{1}{2}}(-i\xi_{+})M_{i\eta,\frac{1}{2}}(-i\xi_{-})}
{e^{-\pi\eta}\,|\Gamma(1-i\eta)|^{-2}(\xi_{+}-\xi_{-})}\right|_{\scalebox{0.8}{$
\xi_\pm \,=\, k(r+R \pm |\Delta\boldsymbol{x}|)$}}\,.
\eeqa
In fact,~in the limit $\eta \rightarrow 0$~or~$R_S \rightarrow 0$,\,the right-hand sides of
\eqref{eq:Nk-(1)-function-def} and \eqref{eq:Nk-(1)-function-res} coincide.~This~can be shown by taking
into account 13.6.9 on p.~328 in~\cite{Olver&etal}~and 5.10.3.3 on p.~621
in~\cite{Prudnikov&Brychkov&Marichev-2}.~Moreover,~assuming that
$\boldsymbol{x}{\cdot}\boldsymbol{X} = - rR$,~the sum
in~\eqref{eq:Nk-(1)-function-def} can also be exactly evaluated for~any~$\eta > 0$ with the help of
6.6.2.8 on p.~347~in~\cite{Prudnikov&Brychkov&Marichev-3},~which matches that of
\eqref{eq:Nk-(1)-function-res} if $\boldsymbol{x}$~and~$\boldsymbol{X}$ are collinear~in opposite directions.
\begin{figure}
\begin{center}
\includegraphics[scale=0.5425]{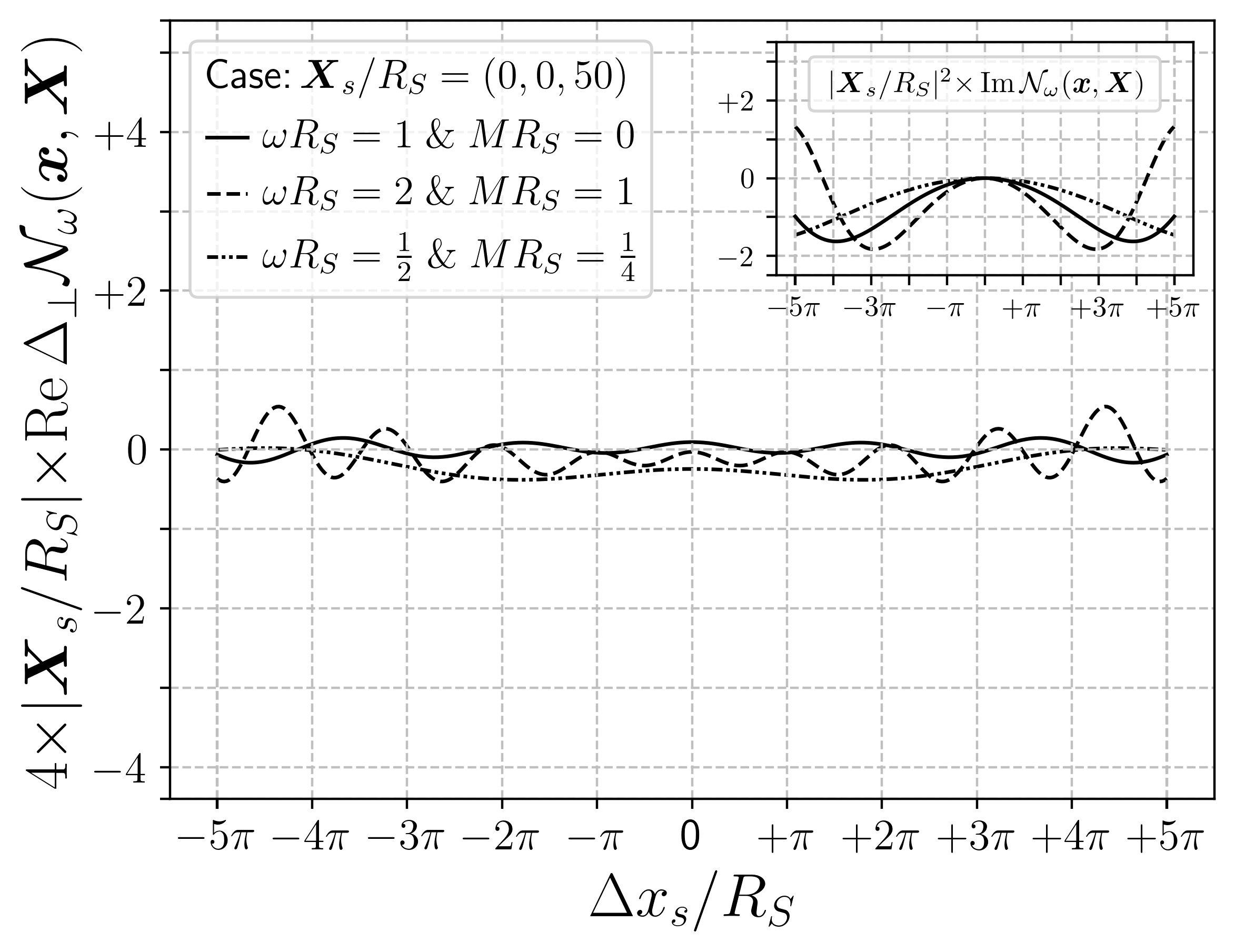}\hfill
\includegraphics[scale=0.5425]{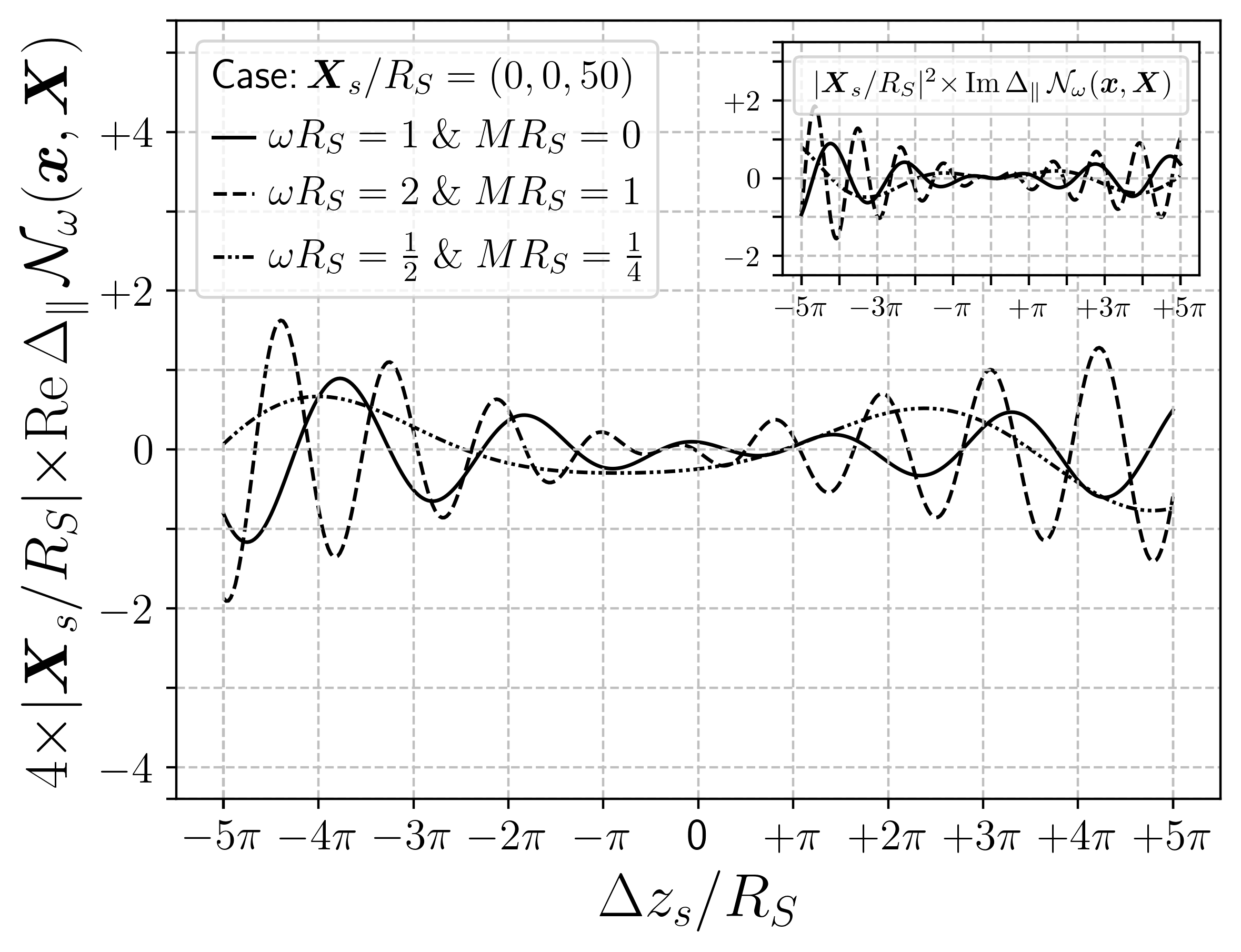}\vspace{2mm}
\includegraphics[scale=0.5425]{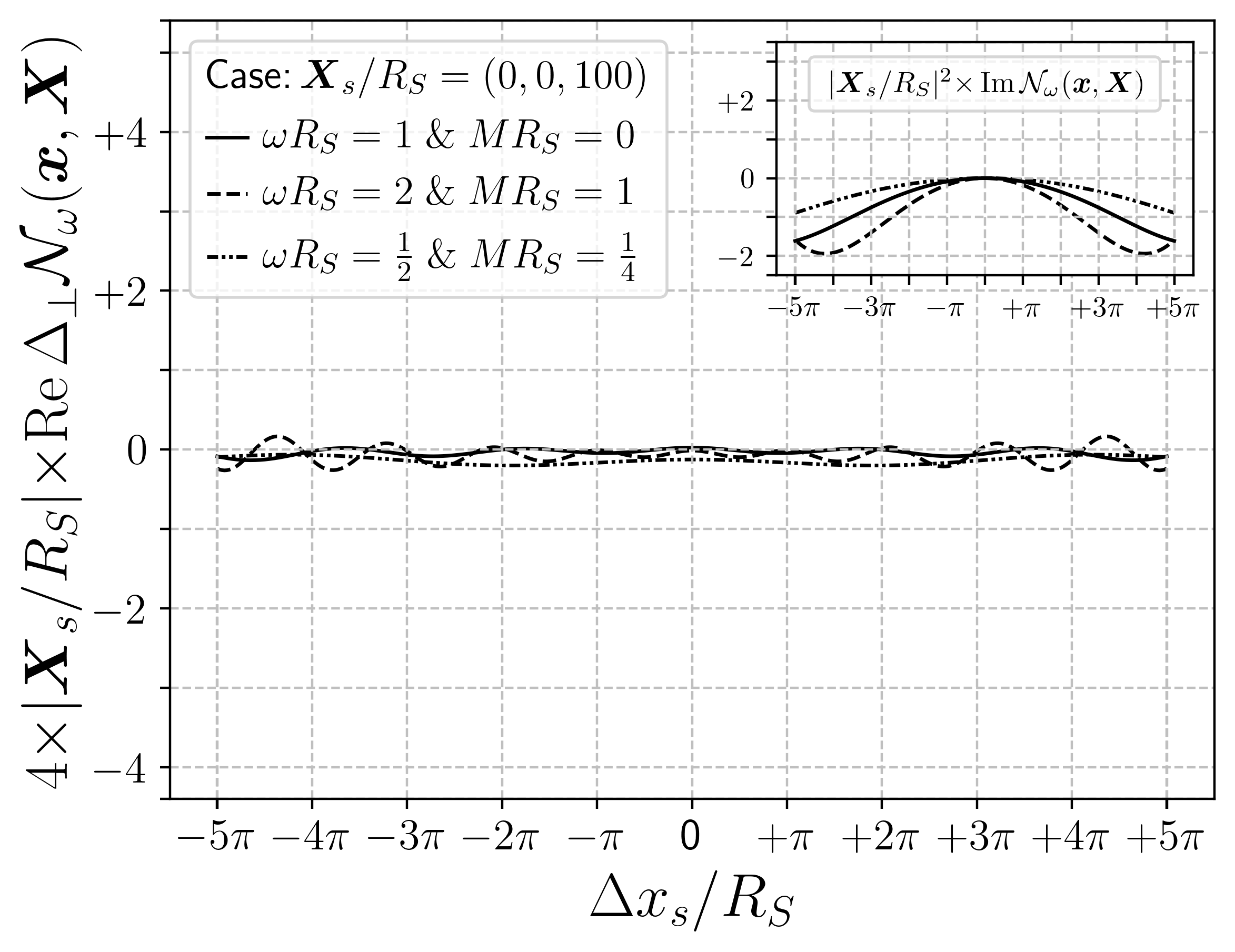}\hfill
\includegraphics[scale=0.5425]{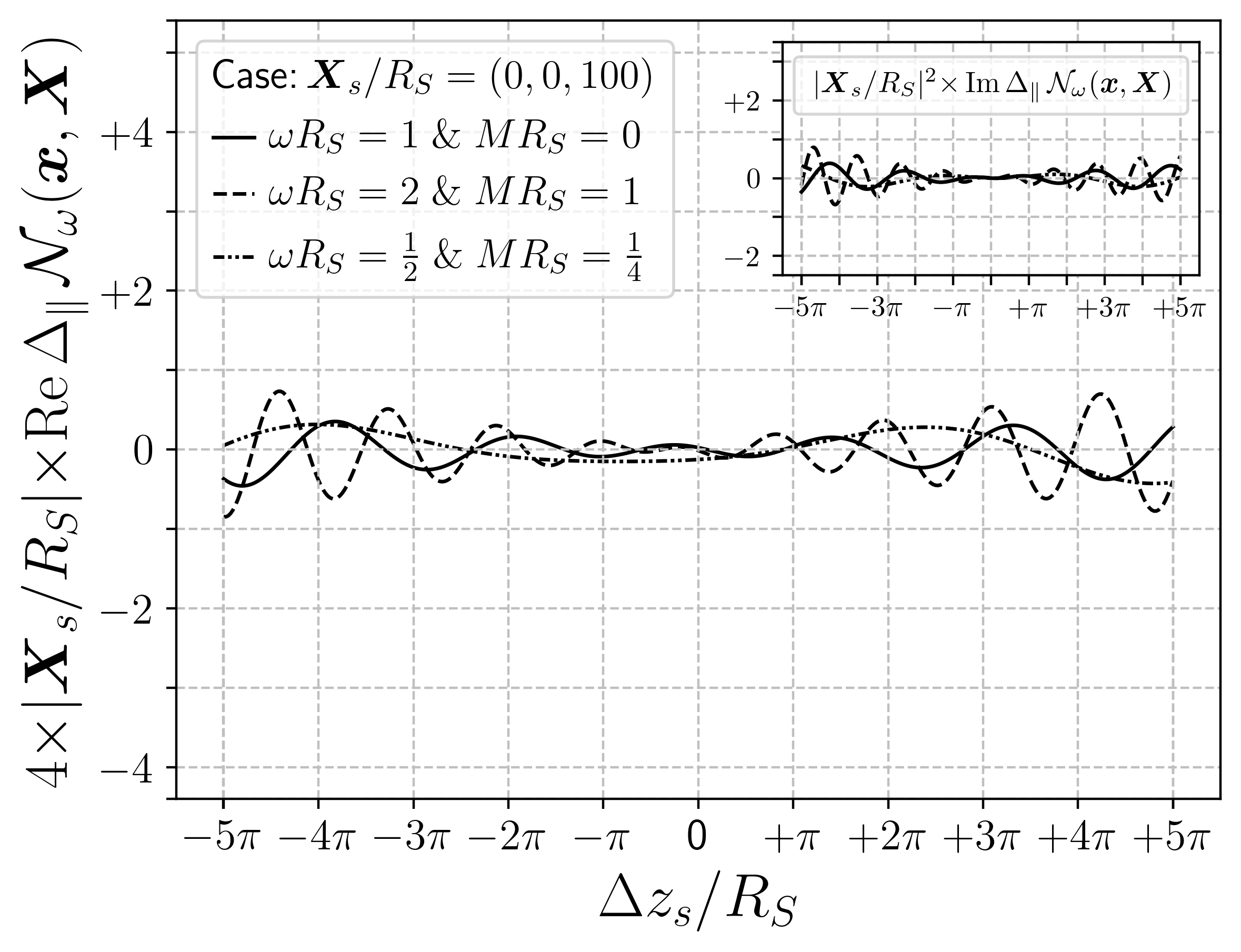}\vspace{2mm}
\includegraphics[scale=0.5425]{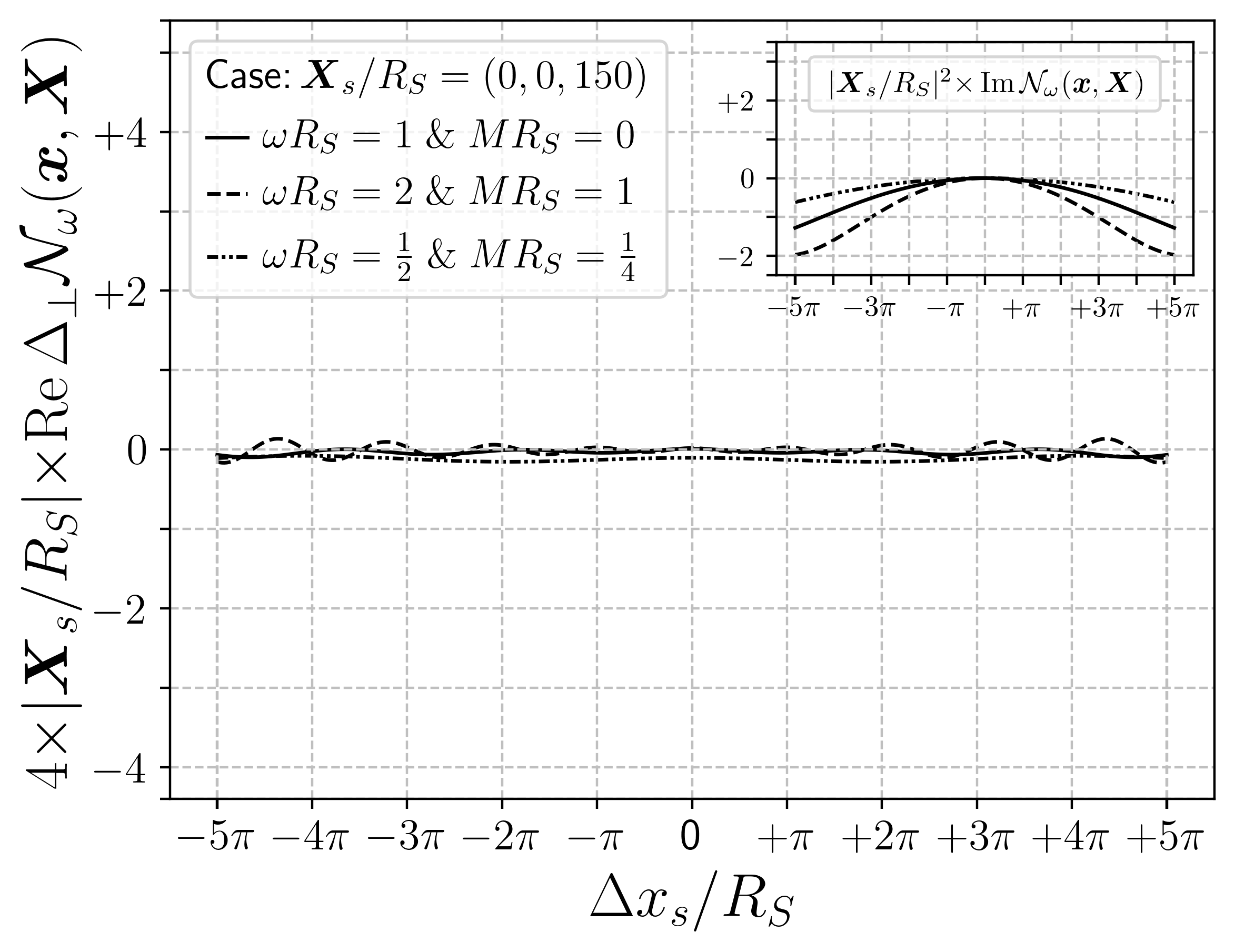}\hfill
\includegraphics[scale=0.5425]{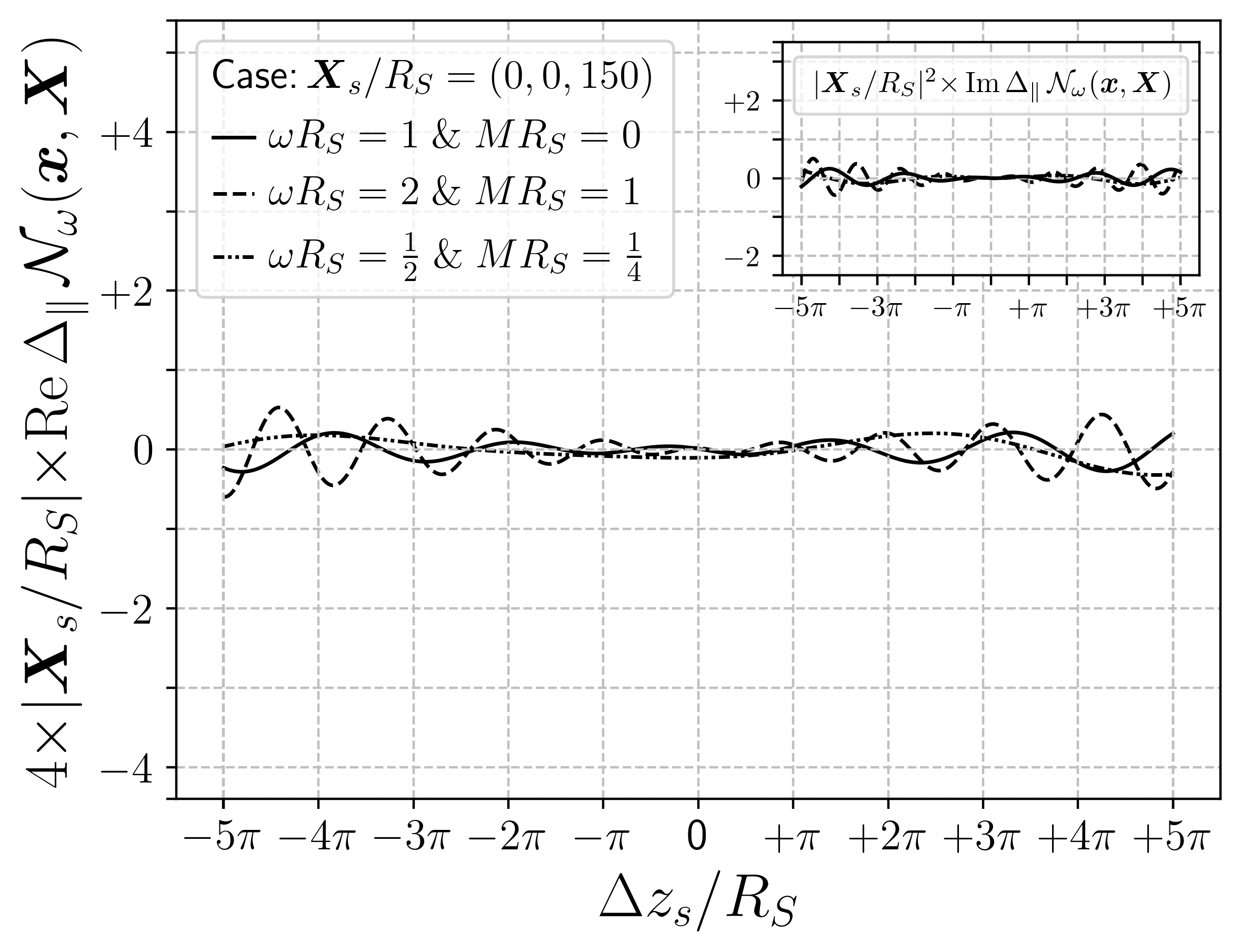}
\end{center}
\caption{Numerical computations of $\mathcal{N}_\omega(\boldsymbol{x},\boldsymbol{X})$
for various values of $\omega$ and $M$,~whereas
$\boldsymbol{x} = \Delta\boldsymbol{x} + \boldsymbol{X}$
with $\boldsymbol{X} = (0,0,Z)$ and $\Delta\boldsymbol{x}$ being either $(\Delta{x},0,0)$ or
$(0,0,\Delta{z})$.~For a given value of $Z$ in units~of~$R_S$, we thus compute how $\mathcal{N}_\omega(\boldsymbol{x},\boldsymbol{X})$ changes by varying $\Delta\boldsymbol{x}$ either 
perpendicularly or parallelly~to~$\boldsymbol{X}$.
Left column:~$\Re\Delta_\perp\mathcal{N}_\omega(\boldsymbol{x},\boldsymbol{X})$
with $\Delta\boldsymbol{x} = (\Delta{x},0,0)$ and $Z_s/R_S \in \{50,100,150\}$,~while
the subplots show $\Im\mathcal{N}_\omega(\boldsymbol{x},\boldsymbol{X})$.~Right
column:~$\Re\Delta_\parallel\,\mathcal{N}_\omega(\boldsymbol{x},\boldsymbol{X})$
with $\Delta\boldsymbol{x} = (0,0,\Delta{z})$ and the same values~of~$Z_s$ and
the subplots display $\Im\Delta_\parallel\,\mathcal{N}_\omega(\boldsymbol{x},\boldsymbol{X})$.
}\label{fig:5}
\end{figure}

With this result at hand,~we next consider~\eqref{eq:Nk-(1)-function-res} 
with $\boldsymbol{x} = \Delta\boldsymbol{x} + \boldsymbol{X}$ and
$|\boldsymbol{X}| \gg R_S$,~whereas
$|\Delta\boldsymbol{x}| \ll |\boldsymbol{X}|$, and define for
$\Delta\boldsymbol{x}\perp\boldsymbol{X}$ and
$\Delta\boldsymbol{x}\parallel\boldsymbol{X}$,~respectively,
\bsubeqs\label{eq:delta-n}
\beqa
\Delta_\perp\mathcal{N}_\omega(\boldsymbol{x},\boldsymbol{X}) &\equiv&
\mathcal{N}_\omega(\boldsymbol{x},\boldsymbol{X}) -
{\left(1 + \frac{R_S}{2|\boldsymbol{X}_s|}\right)} j_0\big(k|\Delta\boldsymbol{x}_s|\big)
- \frac{\omega^2R_S}{2k^2|\boldsymbol{X}_s|} \cos k|\Delta\boldsymbol{x}_s|\,,
\\[2mm]
\Delta_\parallel\,\mathcal{N}_\omega(\boldsymbol{x},\boldsymbol{X}) &\equiv&
\mathcal{N}_\omega(\boldsymbol{x},\boldsymbol{X}) - j_0\big(k|\Delta\boldsymbol{x}_s|\big)
- \frac{\eta\cos k|\Delta\boldsymbol{x}_s|}{k|\boldsymbol{X}_s|}
- i\omega \Gamma(\omega,M)
\frac{\sin k|\Delta\boldsymbol{x}_s|}{4k^3|\boldsymbol{X}_s|^2}\,,
\eeqa
\esubeqs
where the index ``$s$" refers to the Schwarzschild-Cartesian coordinates.~Our
numerical results given in Fig.~\ref{fig:5} show that
\eqref{eq:Nk-(1)-function-res} properly
approximates~\eqref{eq:spatial-Nk-function} up to the leading~order~in~$1/|\boldsymbol{X}|$.
Accordingly,~we find
\beqa
\langle n(x)|n(X)\rangle &\xrightarrow[c \,\to\, \infty]{}& 
\frac{e^{-iMc^2\Delta{t}}}{2Mc}\, 
\langle x|X\rangle
+ {\textrm{O}}{\left(\frac{1}{\boldsymbol{X}^2}\right)}\,,
\eeqa
where we have taken into account that $R_S \to 0$ in the limit $c \to \infty$,~whereas
$\omega^2R_S > 0$.~Thus, this result
agrees with~\eqref{eq:qm-propagator} to the leading order in $1/|\boldsymbol{X}|$.

A few remarks are in order.~First,~$\mathcal{N}_\omega^{(1)}(\boldsymbol{x},\boldsymbol{X})$
given in~\eqref{eq:Nk-(1)-function-res} is insufficient to unambiguously determine 
terms in $\mathcal{N}_\omega(\boldsymbol{x},\boldsymbol{X})$
with $\boldsymbol{x} \sim \boldsymbol{X}$,~approaching zero faster than
$1/|\boldsymbol{X}|$~at~$|\boldsymbol{X}| \to \infty$. Among of such terms is
$g \propto R_S/\boldsymbol{X}^2$.~This~is because
$\mathcal{N}_\omega^{(1)}(\boldsymbol{x},\boldsymbol{X})$~is~an~approximate
solution which merely~takes~the~leading-order correction with 
respect to $R_S/|\boldsymbol{X}|$ into account.~Second,
$\mathcal{N}_\omega^{(1)}(\boldsymbol{X},\boldsymbol{X})$
at $|\boldsymbol{X}| \to \infty$ has the Taylor-series term
$\eta\sin(2k|\boldsymbol{X}|)/2|k\boldsymbol{X}|^2$.~This qualitatively 
accounts for the oscillations
shown in the subplots~in Fig.~\ref{fig:3},~left column, while~their~phase~and
amplitude
should gain corrections depending~on higher-order terms in $R_S$.~Such
corrections must vanish at $c \to \infty$,~according
to~\eqref{eq:n-to-n1}.~Replacing~\eqref{eq:spatial-Nk-function} by 
\eqref{eq:Nk-(1)-function-res} in~\eqref{eq:Nk-function} at
$c \to \infty$~gives~\eqref{eq:qm-propagator} with the term $g \propto R_S/\boldsymbol{X}^2$ included.~However, $\eta\sin(2k|\boldsymbol{X}|)/2|k\boldsymbol{X}|^2$ represents
an extra term which is still regular at $k \to 0$ in~\eqref{eq:Nk-function},~whereas higher-order
Taylor-series terms diverge.~This gives an additive correction to~\eqref{eq:qm-propagator} of the order~of
$g_\oplus M^2/\hbar^2$ if $\Delta{t} \ll MR_\oplus^2/\hbar \sim 10^{21}\,(M/M_n)\,\text{s}$,
where $M_n$ is neutron's mass and the universe age is roughly $10^{17}\,\text{s}$.~At the Earth's surface,~this 
correction to~\eqref{eq:qm-propagator}
is negligible~if $\Delta{t} \ll (\hbar/Mg_\oplus^2)^{1/3}$.~This is~particularly~violated~in~the~classical
limit.~And,~finally,
\eqref{eq:axaX} implies in terms of isotropic coordinates that
\beqa
\langle a(x)|a(X) \rangle &=& \frac{1}{2\pi^2}{\int\limits_0^\infty}{\frac{p^2dp}{2\omega_{\boldsymbol{p}}}}\,
e^{-i\omega_{\boldsymbol{p}}\sqrt{1+2\phi(\boldsymbol{x})}\,\Delta{t}}\,
j_0\Big(p\sqrt{1 - 2\psi(\boldsymbol{x})}\,|\Delta\boldsymbol{x}|\Big)
+ {\textrm{O}}{\left(\frac{1}{\boldsymbol{X}^2}\right)}\,,
\eeqa
where we have used
$y(x) \approx y(X) + (\sqrt{1 + 2\phi(\boldsymbol{x})}\Delta{t},\,\sqrt{1 - 2\psi(\boldsymbol{x})}
\Delta\boldsymbol{x})$~approximately holding
up to $1/|\boldsymbol{X}|$~in the limit $|\boldsymbol{X}| \to \infty$.~In contrast to
the integral representation of $\langle n(x)|n(X) \rangle$,
the integral representation of $\langle a(x)|a(X) \rangle$ explicitly involves 
both gravitational time dilation and gravitational length contraction.~It is the former
general-relativity effect which
gives rise to both $\phi_\oplus$- and $g_\oplus$-dependent terms in~\eqref{eq:qm-propagator}.

\subsection{Propagator $\langle h(x) | h(X) \rangle$ in the far-horizon region}
\label{sec:hxhX}

We finally wish to explore how a Hawking particle propagates far away from a spherically
symmetric compact object.~The probability amplitude for $|h(X)\rangle$
to evolve into $|h(x)\rangle$ reads
\bsubeqs
\beqa\label{eq:Hk-function}
\langle h(x)|h(X)\rangle &=& \frac{1}{2\pi^2}
{\int\limits_0^\infty}\frac{k^2dk}{2\omega}\,e^{-i\omega\Delta{t}}\,\mathcal{H}_{\omega}(\boldsymbol{x},\boldsymbol{X})\,,
\eeqa
where by definition
\beqa\label{eq:spatial-Hk-function}
\mathcal{H}_{\omega}(\boldsymbol{x},\boldsymbol{X}) &\equiv& \frac{1}{4k\omega}
\sum\limits_{l \,=\, 0}^{+\infty}\,(2l+1)\,\frac{r\,\mathcal{H}_{\omega l}(r)}{\left(r+\frac{1}{4}R_S\right)^2}
\,\frac{R\,\overline{\mathcal{H}_{\omega l}(R)}}{\left(R+\frac{1}{4}R_S\right)^2}\,
P_l\left(\scalebox{0.9}{$\dfrac{\boldsymbol{x}{\cdot}\boldsymbol{X}}{rR}$}\right),
\eeqa
\esubeqs
such that $\mathcal{H}_{\omega}(\boldsymbol{x},\boldsymbol{X})$ reduces to
$\mathcal{H}_{\omega}(\boldsymbol{x},\boldsymbol{x})$ defined in~\eqref{eq:g_nk_ghk_xx}
if $\boldsymbol{X} = \boldsymbol{x}$.
\begin{figure}
\begin{center}
\includegraphics[scale=0.5425]{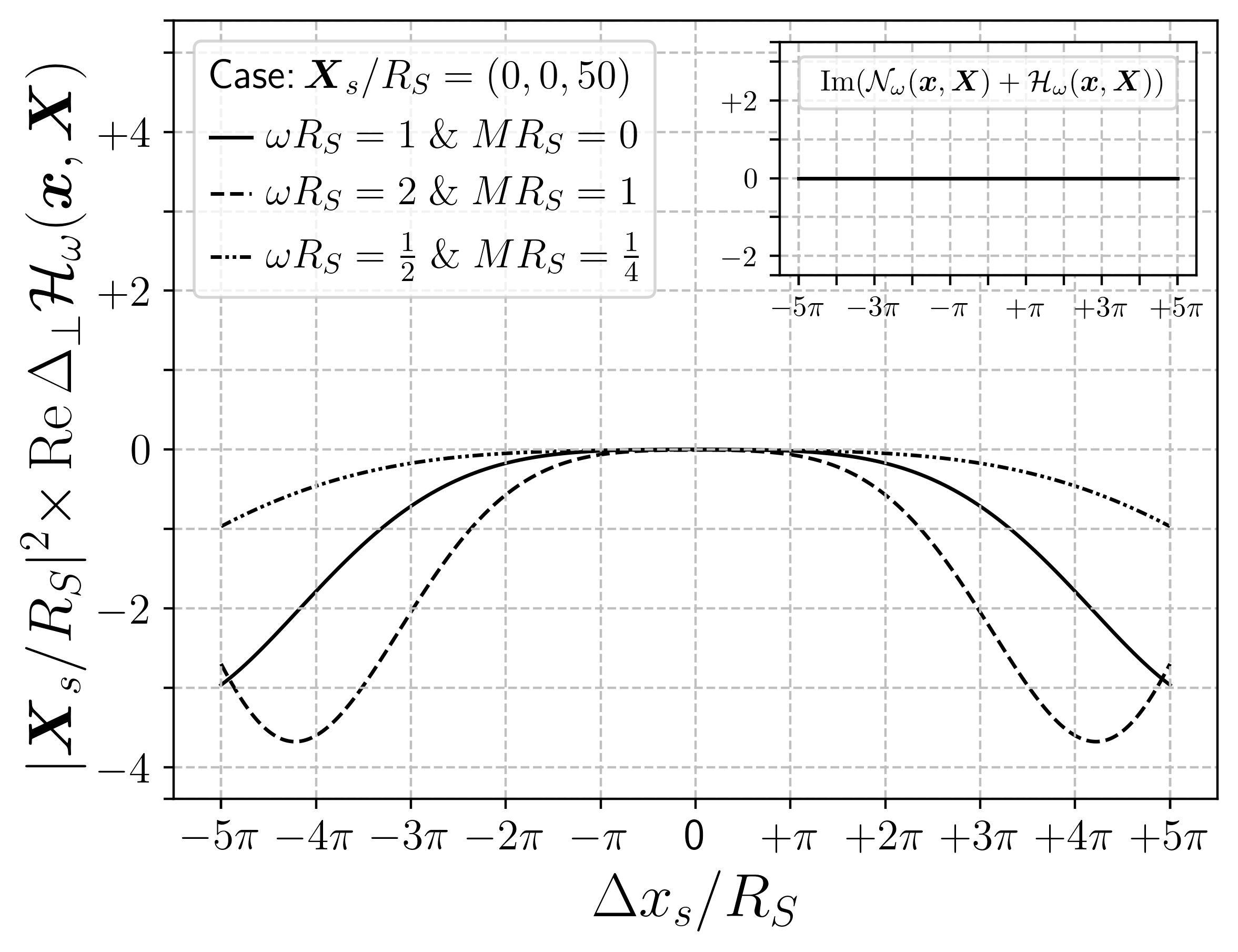}\hfill
\includegraphics[scale=0.5425]{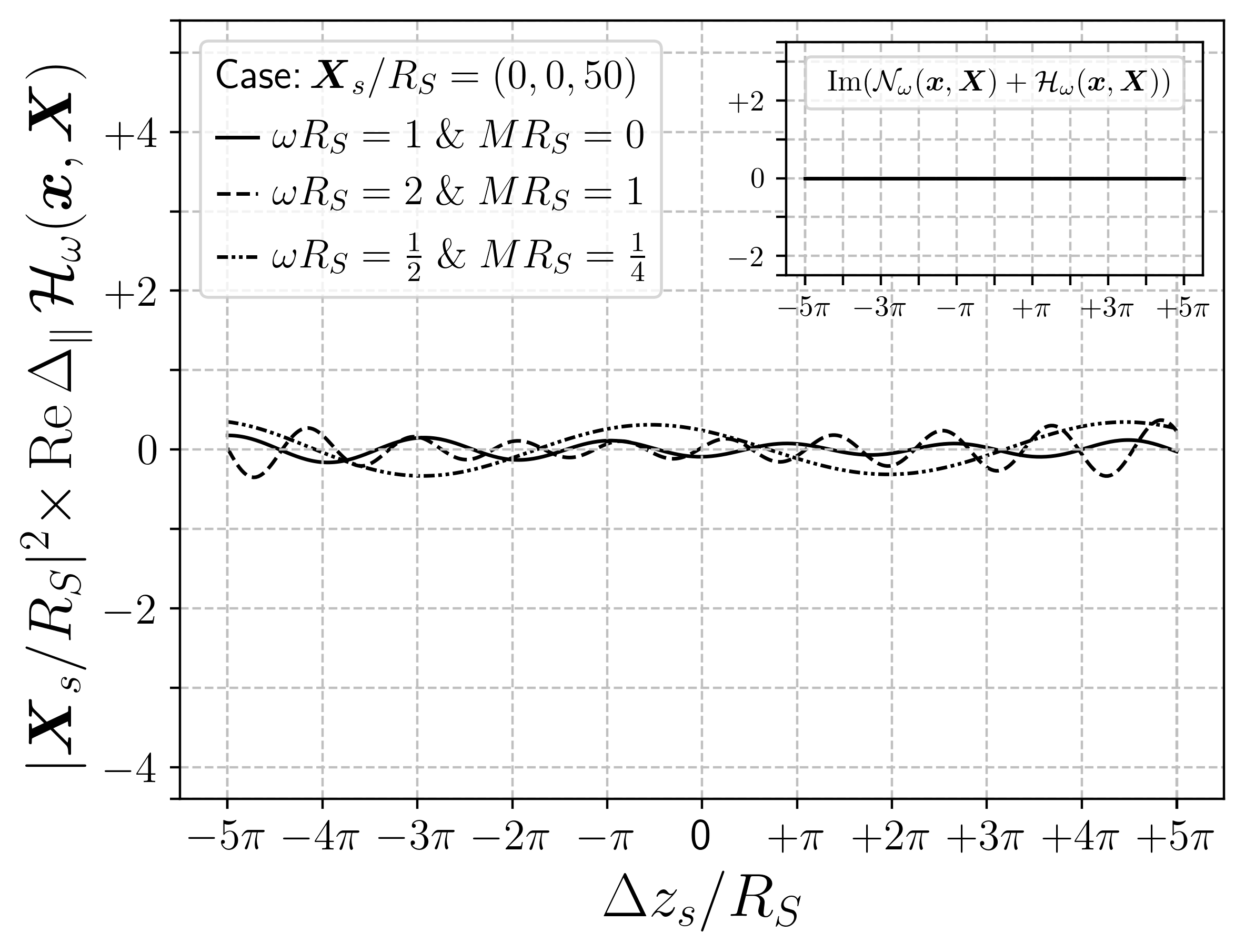}\vspace{2mm}
\includegraphics[scale=0.5425]{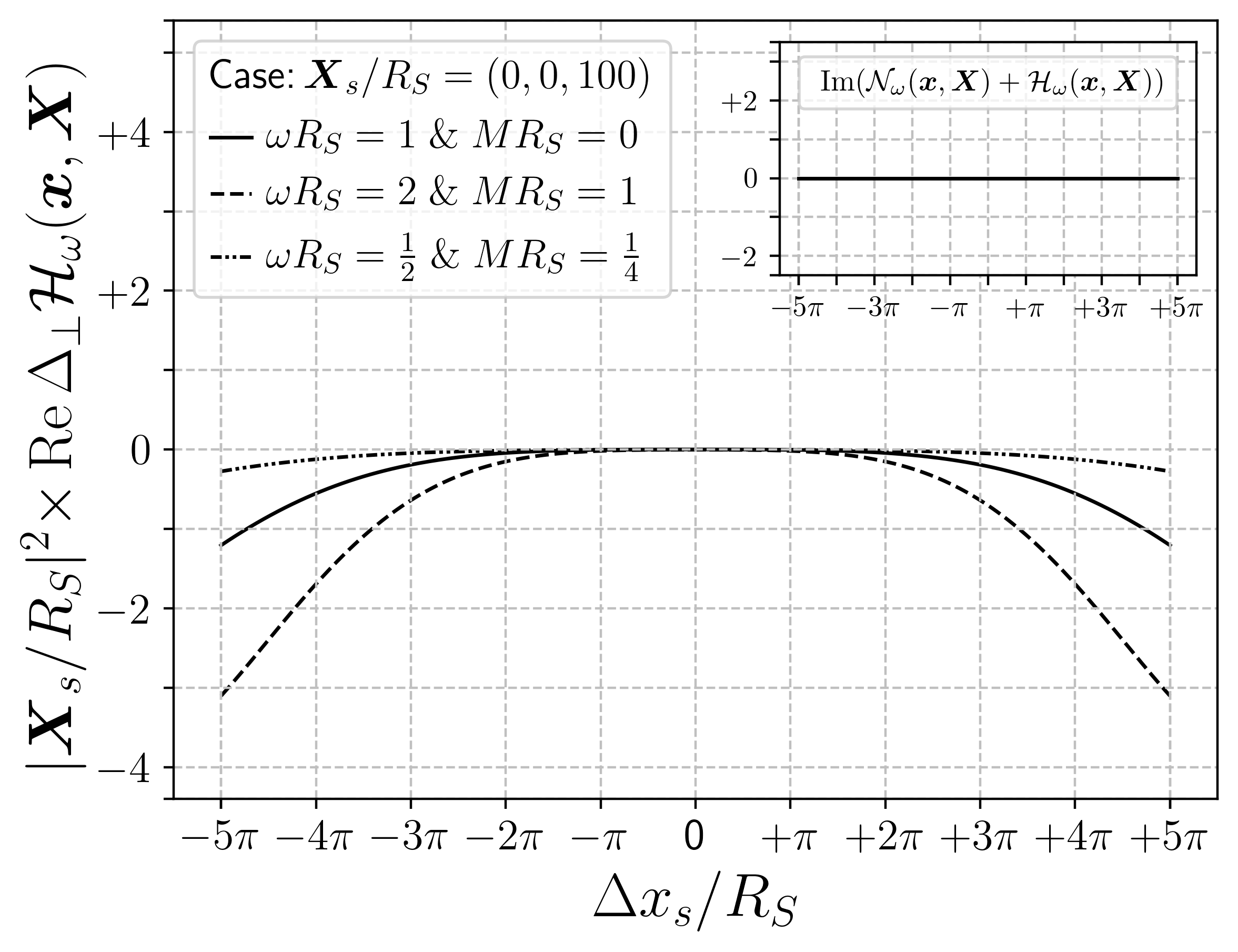}\hfill
\includegraphics[scale=0.5425]{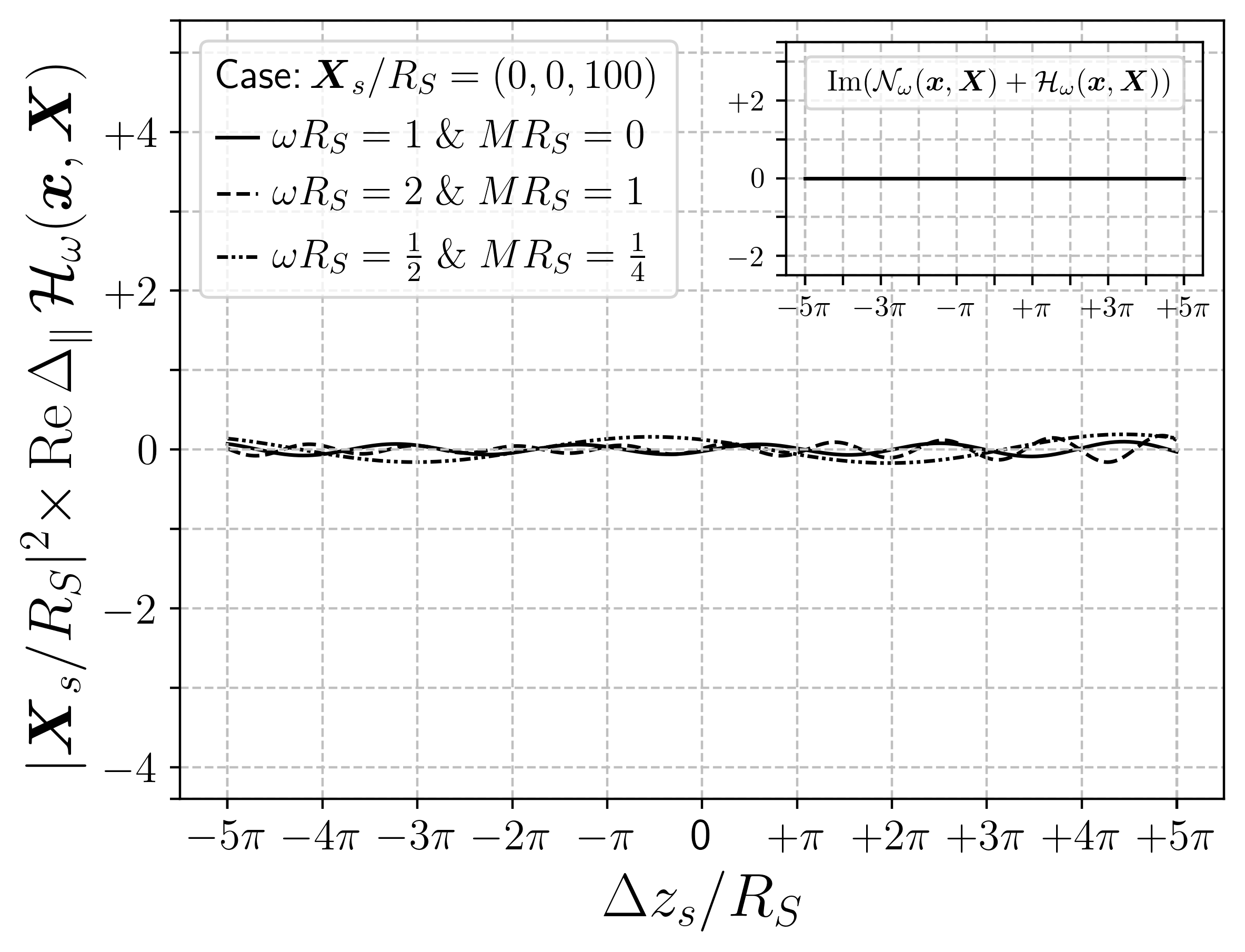}\vspace{2mm}
\includegraphics[scale=0.5425]{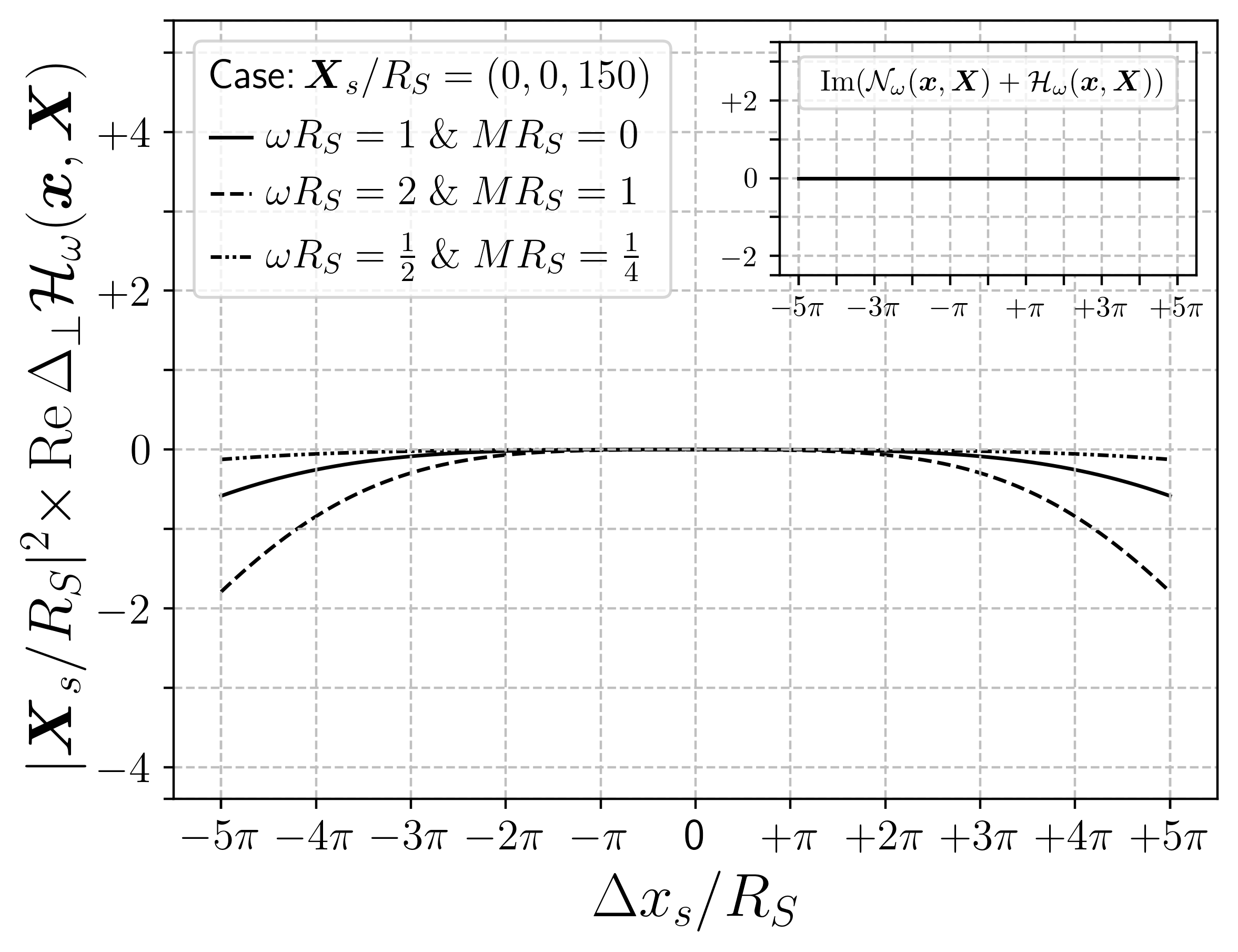}\hfill
\includegraphics[scale=0.5425]{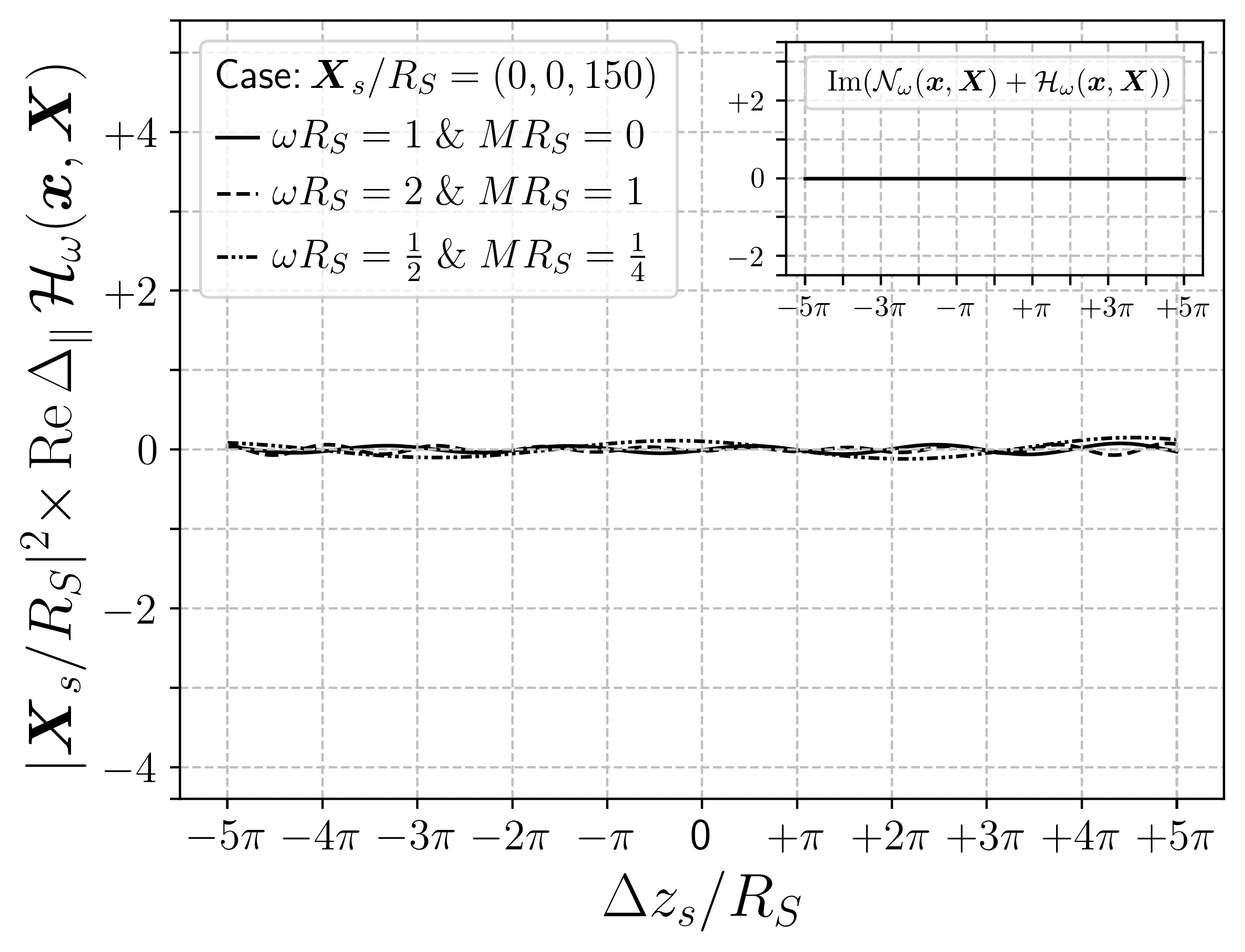}
\end{center}
\caption{Numerical computations of
$\mathcal{H}_\omega(\boldsymbol{x},\boldsymbol{X})$ for the same values of $\omega$ and $M$,
$\boldsymbol{x} = \Delta\boldsymbol{x} + \boldsymbol{X}$~and~$\boldsymbol{X}$
as in Fig.~\ref{fig:5}.~Left
column:~$\Re\Delta_\perp\mathcal{H}_\omega(\boldsymbol{x},\boldsymbol{X})$
for $\Delta\boldsymbol{x} \perp \boldsymbol{X}$.~Right
column:~$\Re\Delta_\parallel\mathcal{H}_\omega(\boldsymbol{x},\boldsymbol{X})$
for $\Delta\boldsymbol{x} \parallel \boldsymbol{X}$. 
Both $\Re\Delta_\perp\mathcal{H}_\omega(\boldsymbol{x},\boldsymbol{X})$
and $\Re\Delta_\parallel\mathcal{H}_\omega(\boldsymbol{x},\boldsymbol{X})$ decrease with 
increasing values of $|\boldsymbol{X}|$.~In 
both cases,~we find that
$\Im\mathcal{H}_\omega(\boldsymbol{x},\boldsymbol{X}) = -\Im\mathcal{N}_\omega(\boldsymbol{x},\boldsymbol{X})$ with accuracy of 15 digits after the comma.~It~corresponds
to the number of digits we have kept by our numerical computations.
}\label{fig:6}
\end{figure}

Using the approximation\,\eqref{eq:app-radial-mode-solutions-h}
and our numerical results for\,\eqref{eq:spatial-Hk-function},\,we assume
$|\Delta\boldsymbol{x}| \ll |\boldsymbol{X}|$ and define
for
$\Delta\boldsymbol{x}\perp\boldsymbol{X}$ and
$\Delta\boldsymbol{x}\parallel\boldsymbol{X}$, respectively,
\bsubeqs\label{eq:chw}
\beqa
\Delta_\perp\mathcal{H}_\omega(\boldsymbol{x},\boldsymbol{X}) &\equiv&
\Re\mathcal{H}_\omega(\boldsymbol{x},\boldsymbol{X}) - \frac{\omega/k}{|2k\boldsymbol{X}_s|^2}\,\Gamma(\omega,M) \,,
\\[2mm]
\Delta_\parallel\,\mathcal{H}_\omega(\boldsymbol{x},\boldsymbol{X}) &\equiv&
\Re\mathcal{H}_\omega(\boldsymbol{x},\boldsymbol{X}) -
\omega\Gamma(\omega,M)\frac{\cos k|\Delta\boldsymbol{x}_s|}{4k^3|\boldsymbol{X}_s|^2}
-\frac{4R_S}{|\boldsymbol{X}_s|}
\Re\Delta_\parallel\,\mathcal{N}_\omega(\boldsymbol{x},\boldsymbol{X}) 
\,,
\eeqa
\esubeqs
where we have taken into account that
\beqa
\Im\big(\mathcal{N}_\omega(\boldsymbol{x},\boldsymbol{X}) +
\mathcal{H}_\omega(\boldsymbol{x},\boldsymbol{X}) \big) &=& 0\,,
\eeqa
which can be proved by using the way the confluent~Heun function
transforms under $\omega \to -\omega$ and $k \to -k$.~Our numerical results
for~\eqref{eq:chw}~are shown in Fig.~\ref{fig:6}.~We~thus find for
$\boldsymbol{x} = \Delta\boldsymbol{x} + \boldsymbol{X}$ with 
$|\Delta\boldsymbol{x}| \ll |\boldsymbol{X}|$
and $\boldsymbol{X} \propto \boldsymbol{e}_z$ -- local outward radial direction -- that
\beqa
\langle h(x)|h(X)\rangle &=&
 \frac{1}{16\pi^2}\,\frac{1}{\boldsymbol{X}^2}
{\int\limits_0^\infty}\frac{dk}{k}\,\Gamma(\omega,M)\,
e^{-i\omega\Delta{t} + ik\Delta{z}}
+ {\textrm{O}}{\left(\frac{1}{\boldsymbol{X}^3}\right)}\,.
\eeqa
In contrast to $\langle n(x)|n(X)\rangle$,~which approaches the Minkowski-spacetime
propagator at spatial infinity, $\langle h(x)|h(X)\rangle \to 0$ in the far-horizon
region.~This agrees with our expectation~in~Sec.~\ref{sec:introduction}.
Moreover,~the Hawking-particle motion~is suppressed in directions
perpendicular to $\boldsymbol{e}_z$.~This 
contradicts to quantum particles' dynamics following from~\eqref{eq:qm-propagator},~assuming
$c \to \infty$.

A few remarks are in order.~First,~this form of $\langle h(x)|h(X)\rangle$ agrees~with the
Hawking~effect
\eqref{eq:he} if $|B\rangle$ is replaced by $|U\rangle$.~Second,~our numerical computations~of
$\Gamma(\omega,M)$ shown in Fig.~\ref{fig:2} suggest that
$\Gamma(\omega,M) \to 0$ if $c \to \infty$ as this assumes both $\omega R_S \to 0$
and $kR_S \to 0$,~resulting~in $\langle h(x)|h(X)\rangle \to 0$ if $c \to \infty$.~This is consistent 
with the fact that there is only one type~of~$|\boldsymbol{x}\rangle$ in quantum mechanics.~Third,~if the 
compact object is a black hole,~then
the computations~of $\mathcal{N}_\omega(\boldsymbol{x},\boldsymbol{x})$
and $\mathcal{H}_\omega(\boldsymbol{x},\boldsymbol{x})$ 
made in Sec.~\ref{sec:hps} suggest
that $\langle h(x)|h(X)\rangle$ may approximately~turn 
into the Minkowski-spacetime propagator in the near-horizon region~if 
$|B\rangle$ is replaced~by~$|U\rangle$.
If so,~it would validate the application of such concepts as geodesic and 
classical action~to~a 
Hawking particle assumed in~\cite{Parikh&Wilczek}.~This~is~certainly~the~case~for
$|a(x)\rangle = \hat{a}^\dagger(x)|\Omega\rangle$,~although~no black-hole evaporation is present in
this case.~However,~in the far-horizon~region,~$\langle n(x)|n(X)\rangle$
asymptotically approaches the Minkowski-spacetime propagator.~Therefore,~such 
concepts as geodesic and classical action are still not applicable to a Hawking 
particle at spatial infinity, even~in the presence of event horizon.~And,~finally,~the
fact that $\langle h(x)|h(X)\rangle$ 
differs~from~the Minkowski-spacetime propagator~at spatial infinity~agrees
with the canonical commutation relation.~Indeed,~$[\hat{\Phi}(x),\hat{\Phi}(X)]$ 
equals $\langle a(x)|a(X)\rangle - \langle a(X)|a(x)\rangle$,~whereas,~at~spatial~infinity, 
$\langle n(x)|n(X)\rangle \to \langle a(x)|a(X)\rangle$, as found
in~Sec.~\ref{sec:nxnX},~implying then that $\langle h(x)|h(X)\rangle \to 0$.\;\;\;\;

\section{Concluding remarks}

Classical mechanics is successful by the description of particle physics in the regime~which
is compatible with $\hbar \to 0$ and $c \to \infty$.~Classical mechanics is replaced by
quantum mechanics for particle-physics phenomena
which require $\hbar > 0$.~Quantum mechanics is in turn replaced by
quantum field theory if $\hbar > 0$ and $c < \infty$ need to be taken into
account.~However,~classical mechanics and quantum mechanics deal primarily with
particles,~while quantum field theory with a field operator algebra.~The
concept of a particle depends accordingly on the choice~of~a 
Hilbert-space representation of such an algebra.~In contrast to
quantum mechanics,~quantum
field theory allows multiple Hilbert-space
representations which may or may not be unitarily 
equivalent to each other~\cite{Haag}.

The formalism of quantum field theory is successfully used for the description of scattering
processes and decay rates in collider physics.~The Poincar\'{e} group plays a key role by choosing the unique Hilbert-space representation.~Particles' states~are accordingly linked
to~irreducible unitary representations of the Poincar\'{e} group.~This~is the isometry
group~of~Minkowski~spacetime in theory,~while of local Minkowski frames in practice.~On
the Earth's surface,~the~space-time geometry can approximately be modelled
by Schwarzschild~spacetime,~provided~that~the Earth's rotation is
ignored.~Particles' dynamics is accordingly modelled by their propagators
which have been established in quantum field theory in Minkowski
spacetime~\cite{Weinberg}.~By~relying on the principle of general covariance,~their dynamics
can be treated in terms of~a~coordinate frame which is at rest with respect to the
Earth's surface.~We have shown in Sec.~II that,~still,
the Minkowski-spacetime
propagators properly model particles'\,dynamics.\,Particularly,\,in\,the
regime $\hbar \to 0$ and $c \to \infty$,~particles' dynamics agrees with free
fall,~while,~in~the~regime~$\hbar > 0$ 
and $c \to \infty$,~that agrees with quantum interference induced by gravity.~The field
quantisation
\eqref{eq:mfq} which is used in theoretical particle physics works not only for high-energy
processes~in
collider physics,~but also for the low-energy phenomena in the Earth's gravitational
field.\;\;\;\;\;\;

The field quantisation~\eqref{eq:hfq} which relies on the isometry group of Schwarzschild spacetime
is~generally used in quantum field theory in curved spacetime.~This field quantisation 
assumes the doubling of particle types in theory.~One of these particle types is known
in the literature as~a Hawking particle.~We have shown in Sec.~III that Hawking
particles cannot be identified with particles which are coherently described by
quantum mechanics and classical~mechanics in the weak-gravity regime.~This conclusion
follows from the observation that the propagator of a Hawking
particle in~the far-horizon region of Schwarzschild spacetime differs from that following from
the path-integral formalism.~This field quantisation accordingly lacks~not~only
experimental confirmations for the moment,~but also
coherence with the well-established~laws in particle physics.~This observation
implies Hawking particles obey non-standard~mechanics. 
Therefore,~insisting on the existence of Hawking particles (and, accordingly, of the Hawking
effect) should,~at least,~influence experimental techniques designed for their detection.

\section*{
ACKNOWLEDGMENTS} A great part of this research has been accomplished over my employment
at Institute~for Theoretical Physics, Karlsruhe Institute of Technology.

\end{document}